\documentclass[cernpreprint, atlasdraft=false, texlive=2023, USenglish, texmf, orcidlogo]{atlasdoc}
 
\usepackage{atlaspackage}
\usepackage{atlasbiblatex}
 
\usepackage{atlasphysics}
\usepackage{atlasprocess}

\graphicspath{{logos/}{figures/}}
 
\usepackage{ANA-STDM-2019-22-INT1-defs}
\usepackage{ANA-STDM-2019-22-PAPER-defs}
\usepackage{PMGRefs-defs}

\AtlasTitle{Measurement of the production of a $W$ boson in association with a charmed hadron in $pp$ collisions at $\sqrt{s} = \SI{13}{\tera{eV}}$ with the ATLAS detector}

\AtlasAbstract{
The production of a $W$ boson in association with a single charm quark is studied using \SI{140}{\per\femto\barn}
of $\sqrt{s}=\SI{13}{\tera{eV}}$ proton--proton collision data collected with the ATLAS detector at the
Large Hadron Collider. The charm quark is tagged by the presence of a charmed hadron, reconstructed with a
secondary-vertex fit. The $W$ boson is reconstructed from the decay to either an electron or a muon and the
missing transverse momentum present in the event. The charmed mesons reconstructed are $D^{+} \to K^- \pi^+ \pi^+$ and
$D^{*+} \to D^{0} \pi^+ \to (K^- \pi^+) \pi^+$ and the charge conjugate decays in the fiducial regions where $p_{\text{T}}(e, \mu) > \SI{30}{\giga{eV}}$,
$|\eta(e, \mu)| < 2.5$, $p_{\text{T}}(D^{(\ast)}) > \SI{8}{\giga{eV}}$, and $|\eta(D^{(\ast)})| < 2.2$.  The integrated and normalized
differential cross-sections as a function of the pseudorapidity of the lepton from the $W$ boson decay, and of the
transverse momentum of the charmed hadron, are extracted from the data using a profile likelihood fit.
The measured total fiducial cross-sections are
$\sigma^{\mathrm{OS-SS}}_{\mathrm{fid}}(W^{-}{+}D^{+})  = 50.2\pm0.2\,\mathrm{(stat.)}\,^{+2.4}_{-2.3}\,\mathrm{(syst.)}\,\mathrm{pb}$,
$\sigma^{\mathrm{OS-SS}}_{\mathrm{fid}}(W^{+}{+}D^{-})  = 48.5\pm0.2\,\mathrm{(stat.)}\,^{+2.3}_{-2.2}\,\mathrm{(syst.)}\,\mathrm{pb}$,
$\sigma^{\mathrm{OS-SS}}_{\mathrm{fid}}(W^{-}{+}D^{*+}) = 51.1\pm0.4\,\mathrm{(stat.)}\,^{+1.9}_{-1.8}\,\mathrm{(syst.)}\,\mathrm{pb}$, and
$\sigma^{\mathrm{OS-SS}}_{\mathrm{fid}}(W^{+}{+}D^{*-}) = 50.0\pm0.4\,\mathrm{(stat.)}\,^{+1.9}_{-1.8}\,\mathrm{(syst.)}\,\mathrm{pb}$.
Results are compared with the predictions of next-to-leading-order quantum chromodynamics calculations
performed using state-of-the-art parton distribution functions.
Additionally, the ratio of charm to anti-charm production cross-sections is studied to probe the $s$-$\bar{s}$
quark asymmetry. The ratio is found to be $R_c^\pm  =  0.971\pm0.006\,\mathrm{(stat.)}\pm0.011\,\mathrm{(syst.)}$.
The ratio and cross-section measurements are consistent with the predictions obtained with parton
distribution function sets that have a symmetric $s$-$\bar{s}$ sea, indicating that any $s$-$\bar{s}$
asymmetry in the Bjorken-$x$ region relevant for this measurement is small.
}

\AtlasNote{ANA-STDM-2019-22}
 
\PreprintIdNumber{CERN-EP-2022-291}

\arXivId{2302.00336}
 
\HepDataRecord{136060}
 
\AtlasJournalRef{\PRD 108 (2023) 032012}
\AtlasDOI{10.1103/PhysRevD.108.032012}

\hypersetup{pdftitle={ATLAS document},pdfauthor={The ATLAS Collaboration}}
 
\begin{document}
 
\maketitle
 
\tableofcontents

\clearpage
\section{Introduction}
\label{sec:introduction}
 
Parton distribution functions (PDFs) describe the momentum distributions of quarks and gluons inside nucleons.
Currently, only limited information is available about the PDF of strange quarks in the proton.
The sea distributions for the three light quarks, up, down and strange, might be equal due to flavor SU(3) symmetry; alternatively the strange quark distribution might be suppressed due to its larger mass. Current knowledge of the strange PDF comes largely from measurements of deep-inelastic lepton--proton scattering~\cite{HERMES:2008pug,ZEUS:2019oro} and charged-current neutrino scattering~\cite{CCFR:1994ikl,CHORUS:2008vjb,NuTeV:2001dfo,NuTeV:2007uwm,NOMAD:2013hbk}, and from vector-boson measurements at the Large Hadron Collider (LHC)~\cite{Evans:2008zzb,STDM-2020-32,STDM-2012-20,STDM-2014-12}.
However, constraints on the strange quark and antiquark PDFs are much weaker than those on the up and down sea
quarks and antiquarks~\cite{Faura:2020oom}.
 
\begin{figure}[hbtp!]
\centering
\includegraphics[width=0.48\linewidth,trim={0 {16 cm} 0 {1 cm}},clip]{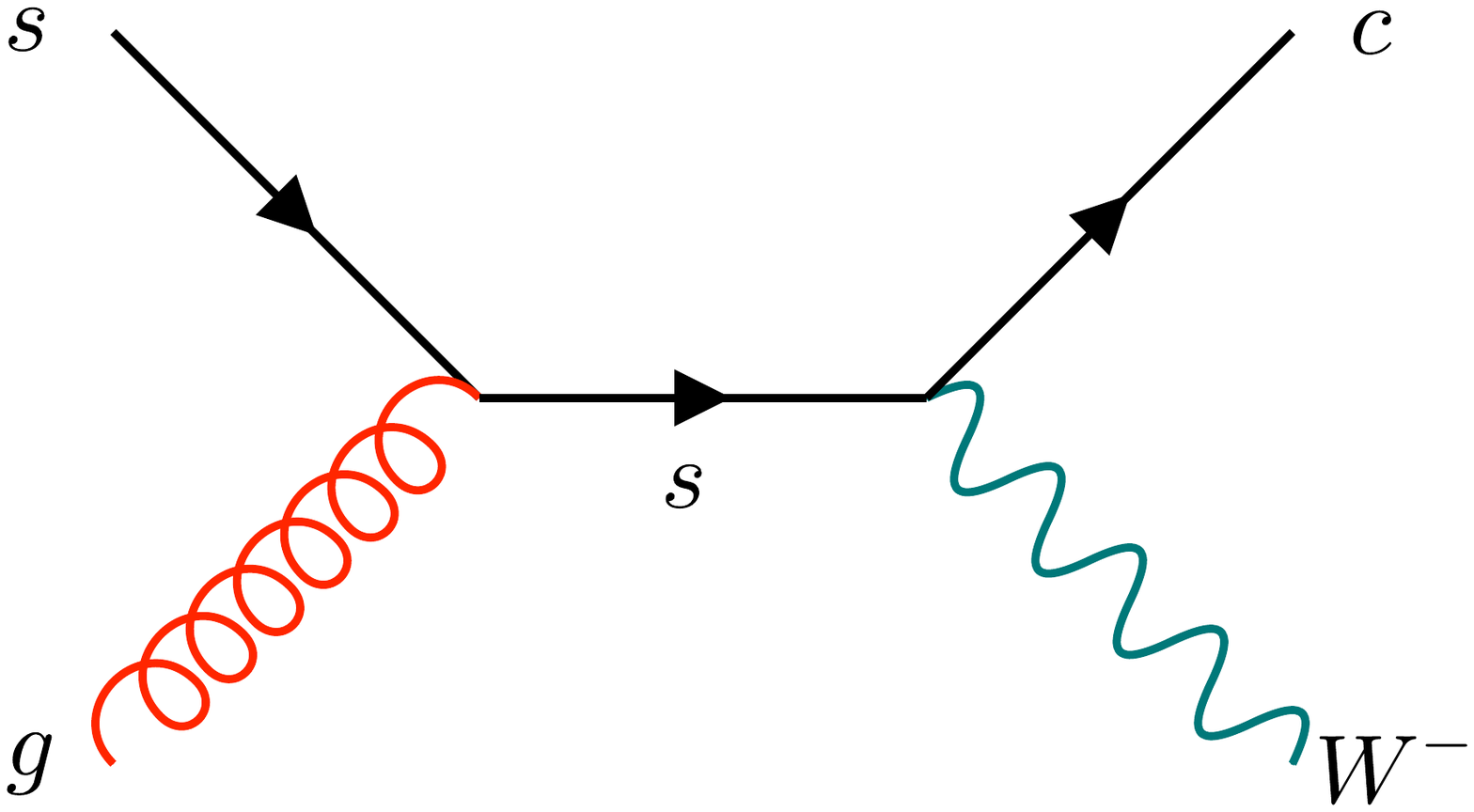}
\includegraphics[width=0.48\linewidth,trim={0 {15.8 cm} 0 {1.2 cm}},clip]{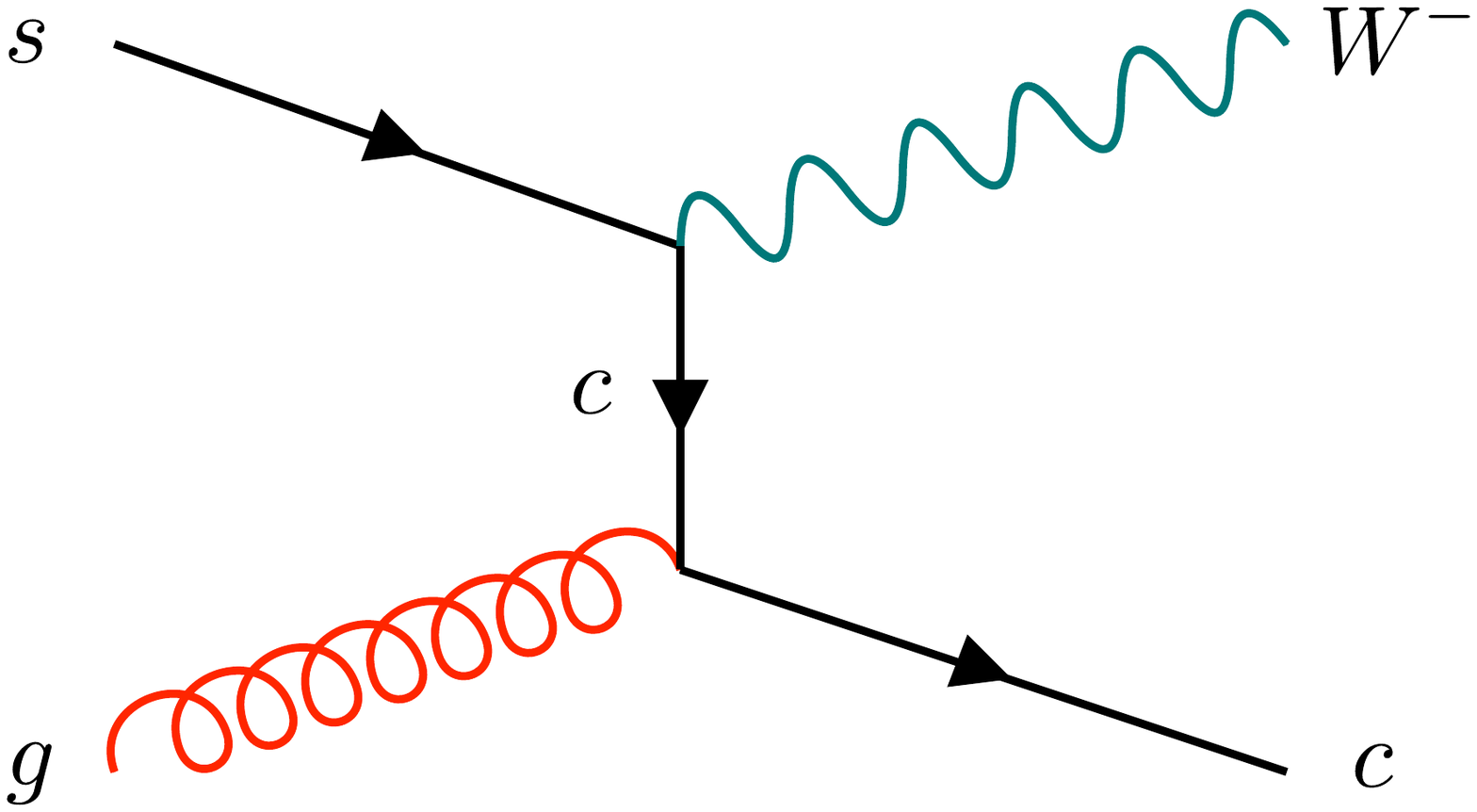}
\caption[]{The leading-order diagrams for $W^- + c$ production}
\label{fig:WcProd}
\end{figure}

In perturbative quantum chromodynamics (QCD), the production of a $W$ boson in association with a single charm quark occurs through the scattering of a gluon and a down-type quark, i.e.\ down, strange or bottom, at leading order (LO), as shown in \Fig{\ref{fig:WcProd}}. The relative contributions to the cross-section of \Wplusc production from each of the three different quarks depends on their PDFs and on the values of the three relevant terms from the Cabbibo--Kobayashi--Maskawa (CKM) mixing matrix~\cite{Cabibbo:1963yz,Kobayashi:1973fv}: $V_{cd}$, $V_{cs}$, and $V_{cb}$. At the LHC, the process $gs \to W^- c$ and its charge conjugate are dominant, while
the process $gd \to W^- c$ ($g\overline d \to W^+\overline c$) contributes only ${\sim}10\%$ (${\sim}5\%$)
to the $W^-c$ ($W^+\overline c$) rate.  The difference between the $d$ and $\overline d$ contributions  can be attributed to the presence of  valence $d$-quarks~\cite{Czakon:2020coa}.
The contribution from $b$-quark-initiated processes is negligible. The largest next-to-leading-order (NLO) contributions are the one-gluon-loop corrections to
$gs \to W^- c$ ($gs \to W^+ \overline c$); however,  various other partonic initial states
such as  $qq'$, $gg$ and $sq$ or $\overline sq$
are also present.
 
The idea of using $\Wplusc$ events to measure the strange PDF was first proposed in Refs.~\cite{Baur:1993zd,Giele:1995kr} and their production was first observed at the Tevatron~\cite{CDF:2012mhm}. At the LHC, it has been measured both by ATLAS and CMS using data taken at $\sqrt{s} = \SI{7}{\TeV}$~\cite{STDM-2010-07,STDM-2012-14,CMS-SMP-12-002} and by CMS using data taken at $\sqrt{s}=\SI{8}{\TeV}$ and $\SI{13}{\TeV}$~\cite{CMS-SMP-18-013,CMS-SMP-17-014}.  Measurements of $\Wplusc$ production
in the forward region at $\sqrt{s} = \SI{7}{\TeV}$ and $\SI{8}{\TeV}$ have also been performed by the LHCb Collaboration~\cite{LHCb:2015bwt}.
In these measurements, the charm quark or antiquark is tagged either by the presence of a jet of particles containing a secondary vertex or a semileptonic decay to a muon, or by explicit reconstruction of a \Dplus or \Dstarp meson or its charge conjugate, collectively written as \Dmeson.
 
This paper presents a measurement of $W$ boson production in association with a \Dmeson meson using \SI{140}{\ifb} of $\sqrt{s}=\SI{13}{\TeV}$ proton--proton ($pp$) collision data recorded by the ATLAS detector at the LHC. Events in which the $W$ boson decays to an electron or a muon (and the associated neutrino) are studied and the presence of the charm quark is detected through explicit charmed hadron reconstruction.  The measurement does not require the presence of a reconstructed jet.
The production of charmed hadrons
is studied using the following decay modes (and their charge conjugates):
\begin{itemize}
\item \DtoKpipi and
\item \DstarplustoKpipi.
\end{itemize}
The signal \WplusDmeson events are extracted through a profile likelihood~\cite{Cowan:2010js}  fit to the reconstructed
secondary-vertex mass distribution for the \Dplus and the mass difference \mdiff for the \Dstarp. The main backgrounds are
single-\Wboson-boson events that do not contain the requisite  \Dmeson decays and \ttbar events.
 
In \Wplusc production, at LO the \Wboson boson and charm quark always have opposite-sign electric charges,
i.e.,\ either $W^+ + \bar{c}$ or $W^- + c$.  For those processes where one of the initial-state partons is a strange or anti-strange quark,
this charge correlation remains at NLO and next-to-next-to-leading order (NNLO)~\cite{Czakon:2020coa}.\footnote{If there is a significant asymmetry between the
charm and anti-charm PDFs, there would be a contribution from processes with charm quarks in the initial state, i.e.\
$dc \to W^{-}uc$ and $d\bar{c} \to W^{-}u\bar{c}$, but this is expected to be small~\cite{Czakon:2020coa}.}
However, many of the backgrounds (e.g.\ heavy-flavor pair production or $b$-hadron production from \ttbar events)
have equal rates for the production of leptons and \Dmeson
with opposite-sign (OS) or same-sign (SS) charges.
This is exploited in the analysis by extracting the signal as the difference of the numbers of OS and SS candidates,
denoted by OS--SS, and extrapolating the background
estimate from SS candidates. The \ttbar background with events containing $W \to cs$ decays is not charge symmetric and is
measured in situ by categorizing the events according to whether  $b$-tagged jets separated in phase space from the
\Dmeson candidate are present.
 
The \WplusDmeson cross-sections, \sigmaWplusD,  are measured in a fiducial region defined by requirements for \Wboson boson
and \Dmeson meson selection. The requirements for \Wboson boson selection are a charged lepton, $\ell$ ($e$ or $\mu$), of transverse momentum
$\pT^\ell > \SI{30}{\GeV}$ and pseudorapidity $|\eta^\ell| < 2.5$. The requirements for \Dmeson meson selection
are $\pT(\Dmeson) > \SI{8}{\GeV}$ and $|\eta(\Dmeson)| < 2.2$. The total fiducial
cross-section is presented along with two differential cross-sections, in \diffpt and \diffeta.
The measurements are performed separately for events with positively and negatively charged \Wboson
bosons and the ratio $\Rc \equiv \sigmaWpplusD/\sigmaWmplusD$ is also presented.
These measurements are
compared with QCD predictions obtained using state-of-the-art  PDF sets~\cite{Alekhin:2017kpj,STDM-2020-32,Hou:2019efy,Bailey:2020ooq,PDF4LHCWorkingGroup:2022cjn,NNPDF:2017mvq,Faura:2020oom,NNPDF:2021njg}.

An important development in the theoretical study of \Wplusc production
is the recent publication of the first NNLO calculation~\cite{Czakon:2022khx}
of the process.  This calculation
includes an off-shell treatment of the $W$ boson and  is performed in a five-flavor scheme using the infrared- and colinear-safe flavored $k_t$ algorithm~\cite{Banfi:2006hf} and
neglecting $c$-quark finite-mass effects.
Nondiagonal CKM matrix elements and the dominant NLO electroweak (EW) corrections are included.
Scale uncertainties obtained using this calculation are below 2\%, significantly smaller than  PDF uncertainties for most PDF set  choices.
Such NNLO calculations will ultimately allow the incorporation  of \Wplusc measurements into NNLO PDF fits.
For \Wplusc-jet measurements,  comparisons with NNLO predictions require that cross-sections be unfolded
to jet observables calculated in the flavored $k_t$ scheme;  such unfolded results are not currently available.
Alternatively, in the
case of \WplusDmeson measurements, the charm fragmentation function could in the future be
incorporated into theory predictions using methods pioneered in \Refn{\cite{Czakon:2021ohs}}.
 
The measurements presented here are compared with QCD calculations with NLO plus parton shower accuracy.  The baseline
framework for these calculations and the QCD  scale uncertainities associated with them  is \mgamc\cite{Alwall:2014hca}.
Theoretical uncertainties associated with the choice of matching scheme are assessed using
the difference between
predictions obtained with \mgamc  and those obtained with recent calculations~\cite{Bevilacqua:2021ovq}
implemented in  the \Powhel event generator~\cite{Garzelli:2011iu}.
 
This paper is structured as follows. \Sect{\ref{sec:atlas_detector}} introduces the ATLAS detector.
The data and Monte Carlo simulation samples used in the analysis are discussed in \Sect{\ref{sec:samples}}.
\Sect{\ref{sec:object_selection}} describes the physics objects used in the analysis and their selection criteria.
The reconstruction and selection of charmed mesons are discussed in \Sect{\ref{sec:meson_selection}}. The event selection is summarized in \Sect{\ref{sec:event_selection}}.
Signal and background modeling are described in \Sect{\ref{sec:signal_and_bkg_modelling}}.  \Sect{\ref{sec:likelihood_fit}} presents the method used to extract the \WplusDmeson  differential cross-section and \Sect{\ref{sec:sys}} summarizes the relevant systematic uncertainties.
The cross-section measurements and their comparison with theoretical predictions
are presented in \Sect{\ref{sec:results}}.  Conclusions are provided in \Sect{\ref{sec:conclusions}}.
 
\FloatBarrier


\newcommand{\AtlasCoordFootnote}{
ATLAS uses a right-handed coordinate system with its origin at the nominal interaction point (IP)
in the center of the detector and the \(z\)-axis along the beam pipe.
The \(x\)-axis points from the IP to the center of the LHC ring,
and the \(y\)-axis points upwards. Cylindrical coordinates \((r,\phi)\) are used in the transverse plane,
\(\phi\) being the azimuthal angle around the \(z\)-axis.
The pseudorapidity is defined in terms of the polar angle \(\theta\) as \(\eta = -\ln \tan(\theta/2)\).
Angular distance is measured in units of \(\Delta R \equiv \sqrt{(\Delta\eta)^{2} + (\Delta\phi)^{2}}\).}
 
\section{The ATLAS detector}
\label{sec:atlas_detector}
 
The ATLAS detector~\cite{PERF-2007-01} at the LHC covers nearly the entire solid angle around the collision point.\footnote{\AtlasCoordFootnote}
It consists of an inner tracking detector surrounded by a thin superconducting solenoid, electromagnetic and hadron calorimeters,
and a muon spectrometer incorporating three large superconducting air-core toroidal magnets.
 
The inner-detector system (ID) is immersed in a \qty{2}{\tesla} axial magnetic field
and provides charged-particle tracking in the range \(|\eta| < 2.5\).
The high-granularity silicon pixel detector covers the vertex region and typically provides four measurements per track,
the first hit normally being in the insertable B-layer (IBL) installed before Run~2~\cite{ATLAS-TDR-19,PIX-2018-001}.
It is followed by the silicon microstrip tracker, which usually provides eight measurements per track.
These silicon detectors are complemented by the transition radiation tracker (TRT),
which enables radially extended track reconstruction up to \(|\eta| = 2.0\).
The TRT also provides electron identification information
based on the fraction of hits (typically 30 in total) above a higher energy-deposit threshold corresponding to transition radiation.
 
The calorimeter system covers the pseudorapidity range \(|\eta| < 4.9\).
Within the region \(|\eta|< 3.2\), electromagnetic calorimetry is provided by barrel and
endcap high-granularity lead/liquid-argon (LAr) calorimeters,
with an additional thin LAr presampler covering \(|\eta| < 1.8\)
to correct for energy loss in material upstream of the calorimeters.
Hadron calorimetry is provided by the steel/scintillator-tile calorimeter,
segmented into three barrel structures within \(|\eta| < 1.7\), and two copper/LAr hadron endcap calorimeters.
The solid angle coverage is completed with forward copper/LAr and tungsten/LAr calorimeter modules
optimized for electromagnetic and hadronic energy measurements, respectively.
 
The muon spectrometer (MS) comprises separate trigger and
high-precision tracking chambers measuring the deflection of muons in a magnetic field generated by the superconducting air-core toroidal magnets.
The field integral of the toroids ranges between \num{2.0} and \qty{6.0}{\tesla\metre}
across most of the detector.
Three layers of precision chambers cover the region \(|\eta| < 2.7\). They consist of layers of monitored drift tubes,
complemented by cathode-strip chambers in the forward region, where the background is highest.
The muon trigger system covers the range \(|\eta| < 2.4\) with resistive-plate chambers in the barrel, and thin-gap chambers in the endcap regions.
 
Interesting events are selected by the first-level trigger system implemented in custom hardware,
followed by selections made by algorithms implemented in software in the high-level trigger~\cite{TRIG-2016-01}.
The first-level trigger accepts events from the \qty{40}{\MHz} bunch crossings at a rate below \qty{100}{\kHz},
which the high-level trigger further reduces in order to record events to disk at about \qty{1}{\kHz}.
 
An extensive software suite~\cite{ATL-SOFT-PUB-2021-001} is used in data simulation, in the reconstruction
and analysis of real and simulated data, in detector operations, and in the trigger and data acquisition
systems of the experiment.

\FloatBarrier


\section{Data and Monte Carlo samples}
\label{sec:samples}

\subsection{Data set description}
\label{sec:samples:data}
 
Events are selected from $\sqrt{s}=\SI{13}{\TeV}$ $pp$ collision data
collected by ATLAS in the period between 2015 and 2018 (Run~2 of the LHC).
After data quality requirements~\cite{DAPR-2018-01} are applied to ensure that all detector
components are in good working condition, the data set amounts to an integrated luminosity of \SI{140}{\ifb}.
The uncertainty in the combined 2015--2018 integrated luminosity is \SI{0.83}{\percent}~\cite{DAPR-2021-01},
obtained using the LUCID-2 detector \cite{LUCID2} for the primary luminosity measurements, complemented
by measurements using the inner detector and calorimeters.
The absolute luminosity scale was determined using van der Meer scans during dedicated running periods in
each year and extrapolated to physics data-taking using complementary measurements from several luminosity-sensitive detectors.
 
Events were recorded by either single-electron or single-muon triggers.
The minimum \pT threshold ranged during data-taking from \SI{24}{\GeV} to \SI{26}{\GeV} for electrons and from \SI{20}{\GeV} to \SI{26}{\GeV} for muons.
Triggers with low \pT thresholds, below \SI{60}{\GeV} for electrons and below \SI{50}{\GeV} for muons, include isolation requirements. For electrons, the requirement is $\pt^{\mathrm{iso}}(\DeltaR_{\mathrm{var}} < 0.2)/\pt < 0.10$, where $\pt^{\mathrm{iso}}(\DeltaR_{\mathrm{var}} < 0.2)$
is the scalar sum of transverse momenta of tracks within a variable-size cone, $\DeltaR_{\mathrm{var}}$, around the electron. The cone size has a maximum value of $0.2$ and decreases as a function of electron's \pT as $\SI{10}{\GeV}/\pt[{\GeV}]$\cite{TRIG-2018-05}. The muon isolation criterion
is constructed by summing the \pt of ID tracks with $\pt^{\text{trk}} > \SI{1}{\GeV}$ around the muon candidate satisfying $\Delta z < \SI{6}{\milli\metre}$, with $\Delta z$ being the distance of the track from the primary vertex in the $z$-direction. This cut was found to be inefficient in events with high \pileup in 2017 and was tightened to $\Delta z < \SI{2}{\milli\metre}$, which allowed the loosening of the isolation criterion for data-taking in 2018. The muon isolation cut is then defined as $\pt^{\mathrm{iso}}(\Delta z)/\pt < 0.07$, where $\pt^{\mathrm{iso}}(\Delta z)$ is the scalar sum of transverse momenta of additional nearby tracks~\cite{TRIG-2018-01}. Triggers with higher \pT thresholds of \SI{60}{\GeV} and \SI{140}{\GeV} for electrons and \SI{50}{\GeV} for muons are added
to increase the selection efficiency.


\subsection{Simulated event samples for signal and background modeling}
\label{sec:samples:mc}
Monte Carlo (MC) simulations are used to model the signal and all backgrounds except \MJ.
Samples  produced with various MC generators are processed using a full detector simulation~\cite{SOFT-2010-01}
based on \GEANT~\cite{Agostinelli:2002hh} and then reconstructed using the same algorithms as the data.
The effect of multiple interactions in the same and neighboring bunch crossings (\pileup) is modeled by overlaying
each simulated hard-scattering event with inelastic $pp$ events generated with \PYTHIA[8.186]~\cite{Sjostrand:2007gs}
using the \NNPDF[2.3lo] set of PDFs~\cite{Ball:2012cx} and a set of tuned parameters called the A3 tune~\cite{ATL-PHYS-PUB-2016-017}.
The MC events are weighted to reproduce the distribution of the average number of interactions per bunch crossing (\avgmu)
observed in the data, scaled up by a factor of $1.03\pm 0.04$ to improve agreement between
data and simulation in the visible inelastic \pp cross-section~\cite{STDM-2015-05}. A reweighting procedure is applied
to all MC samples to correct the charmed hadron production fractions to the world-average values~\cite{Lisovyi:2015uqa,ATL-PHYS-PUB-2022-035}.
The change in the individual charmed meson production fractions is as large as 20\%, depending on the MC configuration.
An overview of all signal and background processes and the generators used to model them is given in \Tab{\ref{tab:samples:mc}},
and further information about the relevant generators configurations is provided below. Processes with more than one jet, known as
multi-leg processes, can have different numbers of jets in each event. To improve the accuracy of calculations, samples with different
jet multiplicities are often merged. In such multi-leg samples, the QCD accuracy for each jet multiplicity is specified in the table.
 
\begin{table}[htbp]
\caption{
The generator configurations used to simulate the signal and background processes.
The acronyms ME, PS and UE stand for matrix element, parton shower and underlying event, respectively.
The column \enquote{HF decay} specifies which software package is used to model the heavy-flavor decays of bottom and charmed hadrons.
For multi-leg samples where different jet multiplicities are merged, the QCD accuracy for each jet multiplicity is specified.
}
\label{tab:samples:mc}
\centering
\setlength\tabcolsep{3.2pt}
\resizebox{\textwidth}{!}{
\begin{tabular}{
l | l l l l l l
}
\toprule
Process       & ME generator & QCD accuracy   & ME PDF     & PS generator    & UE tune & HF decay  \\
\midrule
\multicolumn{7}{l}{\Wjets (background modeling)} \\
\midrule
\Wjets        & \SHERPA[2.2.11] &  0--2j@NLO+3--5j@LO & \NNPDF[3.0nnlo] & \SHERPA & Default & \SHERPA \\
\Wjets        & \AMCatNLO (CKKW-L) & 0--4j@LO & \NNPDF[3.0nlo]     & \PYTHIA[8] & A14 & \EVTGEN    \\
\Wjets        & \AMCatNLO (FxFx)  & 0--3j@NLO   & $\NNPDF[3.1nnlo]\_$luxqed & \PYTHIA[8] & A14 & \EVTGEN    \\
\midrule
\multicolumn{6}{l}{\WplusDmeson (signal modeling and theory predictions)} \\
\midrule
\WplusDmeson  & \SHERPA[2.2.11] & 0--1j@NLO+2j@LO& \NNPDF[3.0nnlo] & \SHERPA & Default & \EVTGEN \\
\WplusDmeson  & \AMCatNLO (NLO)   & NLO    & \NNPDF[3.0nnlo]    & \PYTHIA[8] & A14 & \EVTGEN    \\
\WplusDmeson  & \AMCatNLO (FxFx)  & 0--3j@NLO  & $\NNPDF[3.1nnlo]\_$luxqed & \PYTHIA[8] & A14 & \EVTGEN    \\
\midrule
\multicolumn{6}{l}{Backgrounds} \\
\midrule
\Zjets                  & \SHERPA[2.2.11] & 0--2j@NLO+3--5j@LO & \NNPDF[3.0nnlo] & \SHERPA & Default & \SHERPA \\
\ttbar                  & \POWHEGBOX[v2]  & NLO & \NNPDF[3.0nlo]  & \PYTHIA[8]      & A14    & \EVTGEN \\
Single-$t$, $Wt$        & \POWHEGBOX[v2] & NLO & \NNPDF[3.0nlo]  & \PYTHIA[8]      & A14    & \EVTGEN \\
Single-$t$, $t$-channel & \POWHEGBOX[v2] & NLO & \NNPDF[3.0nlo]  & \PYTHIA[8]      & A14    & \EVTGEN \\
Single-$t$, $s$-channel & \POWHEGBOX[v2] & NLO & \NNPDF[3.0nlo]  & \PYTHIA[8]      & A14    & \EVTGEN \\
$t\bar{t}V$             & \AMCatNLO & NLO & \NNPDF[3.0nlo]  & \PYTHIA[8]      & A14    & \EVTGEN \\
Diboson fully leptonic  & \SHERPA[2.2.2] & 0--1j@NLO+2--3j@LO & \NNPDF[3.0nnlo] & \SHERPA  & Default & \SHERPA \\
Diboson hadronic        & \SHERPA[2.2.1] & 0--1j@NLO+2--3j@LO & \NNPDF[3.0nnlo] & \SHERPA  & Default & \SHERPA \\
\bottomrule
\end{tabular}
}
\end{table}
 
\subsubsection{Background \Vjets samples}
\label{sec:vjetsMC}
Three generator configurations are used to model inclusive vector boson ($W$ or $Z$) plus jet production.
These samples are used to estimate the \WplusDmeson backgrounds and the corresponding experimental and
theory systematic uncertainties.
 
\textbf{\SHERPA:} The nominal MC generator used for this analysis is \SHERPA[2.2.11]~\cite{Bothmann:2019yzt}.
NLO-accurate matrix elements (ME) for up to two partons, and LO-accurate matrix elements for
between three and five partons, are calculated in the five-flavor scheme using the Comix~\cite{Gleisberg:2008fv} and \OPENLOOPS~\cite{Buccioni:2019sur,Cascioli:2011va,Denner:2016kdg}
libraries. The $b$- and $c$-quarks are treated as massless at matrix-element level and massive in the parton shower.
The Hessian \NNPDF[3.0nnlo] PDF set~\cite{Ball:2014uwa} is used. The default \SHERPA parton shower~\cite{Schumann:2007mg} based on
Catani--Seymour dipole factorization and the cluster hadronization model~\cite{Winter:2003tt} is used.  The samples are
generated using a dedicated set of tuned parameters developed by the \SHERPA authors and use the \NNPDF[3.0nnlo] set.
The NLO matrix elements for a given jet multiplicity are matched to the parton shower (PS) using a color-exact variant of the
MC@NLO algorithm~\cite{Hoeche:2011fd}. Different jet multiplicities are then merged into an inclusive sample using an improved CKKW
matching procedure~\cite{Catani:2001cc,Hoeche:2009rj} which is extended to NLO accuracy using the MEPS@NLO prescription~\cite{Hoeche:2012yf}.
The merging scale $Q_{\mathrm{cut}}$ is set to \SI{20}{\GeV}.
 
Uncertainties from missing higher orders in \SHERPA samples are evaluated~\cite{Bothmann:2016nao} using seven variations of the QCD
renormalization (\muR) and factorization (\muF) scales in the matrix elements by factors of $0.5$ and $2$, avoiding variations in opposite directions.
The strong coupling constant $\alphas$ is varied by $\pm 0.001$ to assess the effect of its uncertainty. Additional details of the use
of these samples are available in \Refn{\cite{PMGR-2021-01}}.
 
\textbf{\MGNLO (CKKW-L):} \Vjets production is simulated with LO-accurate matrix elements
for up to four partons with \MGNLO[2.2.2]~\cite{Alwall:2014hca}. The matrix-element calculation is interfaced with
\PYTHIA[8.186] for the modeling of the parton shower, hadronization and underlying event.
To remove overlap between the matrix element and the parton shower, the CKKW-L merging procedure~\cite{Lonnblad:2001iq,Lonnblad:2011xx}
is applied with a merging scale of $Q_{\mathrm{cut}}= \SI{30}{\GeV}$ and a jet-clustering radius parameter of $0.2$. In order to better model the
region of large jet \pt, the strong coupling \alphas is evaluated at the scale of each splitting to determine the weight.
The matrix-element calculation is performed with the \NNPDF[3.0nlo] PDF set~\cite{Ball:2014uwa} with $\alphas= 0.118$.
The calculation is done in the five-flavor scheme with massless $b$- and $c$-quarks. Cross-sections are calculated
using a diagonal CKM matrix. Heavy-quark masses are reinstated in the \PYTHIA[8] shower. The values of \muR and \muF  are set to one half of the transverse
mass of all final-state partons and leptons.
The A14 tune~\cite{ATL-PHYS-PUB-2014-021} of
\PYTHIA[8] is used with the \NNPDF[2.3lo] PDF set with $\alphas=0.13$. The decays of bottom and charmed hadrons are
performed by \EVTGEN[1.7.0]~\cite{Lange:2001uf}.
 
\textbf{\MGNLO (FxFx):} The \MGNLO[2.6.5] program~\cite{Alwall:2014hca} is used to generate weak bosons with up to three additional partons in the final state at NLO accuracy. The  scales \muR and \muF are set to one half of the transverse
mass of all final-state partons and leptons. Cross-sections are calculated using a diagonal CKM matrix. The showering and subsequent
hadronization are performed using \PYTHIA[8.240] with the A14 tune and the \NNPDF[2.3lo] PDF set with $\alphas=0.13$.
The different jet multiplicities are merged using the \fxfx  NLO matrix-element and parton-shower merging prescription~\cite{Frederix:2012ps}.
\PYTHIA[8.186] is used to model the parton shower, hadronization and underlying event.
 
The calculation uses a five-flavor scheme with massless $b$- and $c$-quarks at the matrix-element level, and massive
quarks in the \PYTHIA[8] shower. At the event-generation level, the jet transverse momentum is required to be at least \SI{10}{\GeV},
with no restriction on the absolute value of the jet pseudorapidity. The PDF set used for event generation is $\NNPDF[3.1nnlo]\_$luxqed.
The merging scale is set to $Q_{\mathrm{cut}} = \SI{20}{\GeV}$. Scale variations where \muR and \muF
are varied independently by a factor of $2$ or $0.5$ in the matrix element are included as generator event weights.
The decays of bottom and charmed hadrons are
performed by \EVTGEN[1.7.0].
 
\subsubsection{Signal \WplusDmeson signal samples}
 
Only about \SI{2}{\percent} of the events in the inclusive \Wjets samples pass the \WplusDmeson fiducial requirements. This, coupled
with the branching ratios of \SI{9.2}{\percent} (\SI{2.5}{\percent}) to the \Dplus (\Dstarp) decay mode of interest,  means that
even very large \Wjets samples provide statistically inadequate measurements of the \WplusDmeson fiducial efficiency. Filtered
signal samples are therefore used to enhance the statistical precision.  The generated events
are filtered to require the presence of  a single lepton with $\pt > \SI{15}{\GeV}$ and $|\eta|<2.7$ and either a \Dstarp or a \Dplus meson
with $\pt > \SI{7}{\GeV}$ and $|\eta|<2.3$. \EVTGEN[1.7.0] is used to force all \Dzero mesons to decay through the mode $D^0 \rightarrow K^-\pi^+$
and all \Dplus mesons to decay through the mode $D^+\rightarrow K^-\pi^+\pi^+$ (plus charge conjugates). \EVTGEN describes this
three-body \Dplus decay using a Dalitz plot amplitude that includes contributions from the $\overline K^{*0}(892)$, $\overline K^{*0}(1430)$,
$\overline K^{*0}(1680)$ and $\kappa(800)$ resonances, as measured by CLEO-c~\cite{CLEO:2008jus}.
 
These samples are used for signal modeling, for calculating the detector response matrix and fiducial efficiencies
with small statistical uncertainties, and for determining the \WplusDmeson signal mass distribution used in the statistical
analysis described in \Sect{\ref{sec:likelihood_fit}}. The \aMGNLO simulation described below is also used to calculate the theory
predictions with the up-to-date PDF sets in \Sect{\ref{sec:results}}. Three such filtered samples are used:
 
\textbf{\SHERPA[2.2.11] \WplusDmeson:} To reduce the per-event CPU time for the generation of the \WplusDmeson
signal data sets, \SHERPA[2.2.11] is configured to have lower perturbative accuracy than
for the inclusive \Vjets samples described above. Events are generated with NLO-accurate matrix elements for up to one jet,
and LO-accurate matrix elements for two partons, in the five-flavor scheme.  Other \SHERPA parameters are set to the same values
as for the baseline inclusive samples and uncertainties are evaluated using the same variations in QCD scale and \alphas\ as for the baseline.
The production cross-section for this configuration differs from that of the inclusive sample
by ${\sim}2\%$. The two configurations show no significant differences in kinematic distributions associated with the \Dmeson meson or \Wboson boson.
 
\textbf{\aMGNLO \WplusDmeson:} \MGNLO[2.9.3] is used to generate the $W+c$-jet process at NLO accuracy.
A finite charm quark mass of $m_c=\SI{1.55}{\GeV}$ is used to regularize the cross-section, and a full CKM matrix is
used to calculate the hard-scattering amplitudes. The values of \muR and \muF  are set to half of
the transverse mass of all final-state partons and leptons. The PDF set used for event generation is \NNPDF[3.0nnlo] with $\alphas=0.118$.
The matrix-element calculation is interfaced with \PYTHIA[8.244] for the modeling of the parton shower, hadronization,
and underlying event and the A14 tune is employed. Scale variations where \muR and \muF
are varied independently by a factor of 2 or 0.5 in the matrix element are included as generator event weights.
 
\textbf{\MGFX \WplusDmeson:} Events are generated using the same \PYTHIA[8] configuration as used for the
inclusive \MGFX sample, but with the event-level filtering and configuration described above.
 
\subsubsection{Top quark pair production background samples}
 
The production of \ttbar\ events is modeled using the \POWHEGBOX[v2]~\cite{Frixione:2007nw,Nason:2004rx,Frixione:2007vw,Alioli:2010xd}
generator which provides matrix elements at NLO in the strong coupling constant \alphas
with the \NNPDF[3.0nlo] PDF and the \hdamp\ parameter\footnote{The \hdamp\ parameter
controls the transverse momentum \pt\ of the first additional emission beyond the leading-order Feynman diagram
in the parton shower and therefore regulates the  high-\pt\ emission against which the \ttbar\ system recoils.}
set to 1.5\,\mtop~\cite{ATL-PHYS-PUB-2016-020}. The functional form of \muR and \muF is set
to the default scale $\sqrt{m_{\textrm{top}}^2 + \pt^2}$ where \pt is the transverse momentum of the top quark obtained using the underlying Born kinematics.  Top quarks are decayed at LO using \MADSPIN~\cite{Frixione:2007zp,Artoisenet:2012st}
to preserve all spin correlations. The events are interfaced with \PYTHIA[8.230] for the parton shower and hadronization,
using the A14 tune and the \NNPDF[2.3lo] PDF set.
The decays of bottom and charmed hadrons are simulated using \EVTGEN[1.6.0].
 
The NLO \ttbar\ inclusive production cross-section is corrected to the theory prediction at NNLO
in QCD including the resummation of next-to-next-to-leading logarithmic (NNLL) soft-gluon terms calculated using
\TOPpp[2.0]~\cite{Beneke:2011mq,Cacciari:2011hy,Baernreuther:2012ws,Czakon:2012zr,Czakon:2012pz,Czakon:2013goa,Czakon:2011xx}.
 
\POWHER[7.04] and \MGNLOPY[8] \ttbar samples are used to estimate the systematic uncertainty due to the choice
of MC model as explained in the following and the details of the configurations used are provided below.
 
\ttbar \POWHER[7.04]: The impact of using a different parton shower and hadronization model is evaluated
by comparing the nominal \ttbar sample with another event sample produced with the \POWHEGBOX[v2]
generator using the \NNPDF[3.0nlo] parton distribution function.
Events in the latter sample are interfaced with \HERWIG[7.04]~\cite{Bahr:2008pv,Bellm:2015jjp},
using the H7UE set of tuned parameters~\cite{Bellm:2015jjp} and the \MMHT[lo] PDF set~\cite{Harland-Lang:2014zoa}.
The decays of bottom and charmed hadrons are simulated using \EVTGEN[1.6.0]~\cite{Lange:2001uf}.
 
\ttbar \MGNLOPY[8]: The uncertainty in the matching of NLO matrix elements to the
parton shower is assessed by comparing the \POWHEG sample with events generated with \MGNLO[2.6.0]
interfaced with \PYTHIA[8.230]. The \MGNLO calculation used the \NNPDF[3.0nlo] set of PDFs
and \PYTHIA[8] used the A14 tune and the \NNPDF[2.3lo] set of
PDFs. The decays of bottom and charmed hadrons are simulated using \EVTGEN~{1.6.0}.
 
\subsubsection{$Wt$-channel single-top background samples}
 
Single-top $Wt$ associated production is modeled using the \POWHEGBOX[v2] generator which provides matrix elements
at NLO in the strong coupling constant \alphas\ in the five-flavor scheme with the \NNPDF[3.0nlo]
parton distribution function set. The functional form of \muR and \muF is set to the default scale
$\sqrt{m_{\textrm{top}}^2 + p_{\textrm T}^2}$. The diagram removal scheme~\cite{Frixione:2008yi} is employed to handle the interference
with \ttbar production~\cite{ATL-PHYS-PUB-2016-020}. Top quarks are decayed at LO using \MADSPIN
to preserve all spin correlations. The events are interfaced with \PYTHIA[8.230] using the A14 tune and
the \NNPDF[2.3lo] PDF set. The decays of bottom and charmed hadrons are simulated using \EVTGEN[1.6.0]. The inclusive cross-section
is corrected to the theory prediction calculated at NLO in QCD with NNLL soft gluon corrections~\cite{Aliev:2010zk,Kant:2014oha}.
 
\subsubsection{$t$-channel and $s$-channel single-top background samples}
 
Single-top $t$-channel ($s$-channel) production is modeled using the \POWHEGBOX[v2] generator at NLO in QCD using the
four-flavor (five-flavor) scheme and the corresponding \NNPDF[3.0nlo] set of PDFs. The events are interfaced with
\PYTHIA[8.230] using the A14 tune and the \NNPDF[2.3lo] set of PDFs.
 
The uncertainty due to initial-state radiation (ISR) is estimated by simultaneously varying the \hdamp parameter and \muR and \muF, and choosing the \textsc{Var3c} up and down variants of the A14 tune
as described in \Refn{\cite{ATL-PHYS-PUB-2017-007}}. The impact of final-state radiation (FSR) is evaluated by halving and doubling
the renormalization scale for emissions from the parton shower.
 
\subsubsection{$\ttbar+V$ background samples}
 
The production of $t\bar{t}V$ events, where $V$ denotes either \Wboson, \Zboson, or $\ell^+\ell^-$ produced through $Z/\gamma$ interference,
is modeled using the \MGNLO[2.3.3]~\cite{Alwall:2014hca} generator at NLO with the \NNPDF[3.0nlo] parton distribution
function. The events are interfaced with \PYTHIA[8.210] using the A14 tune and the
\NNPDF[2.3lo] PDF set. The uncertainty due to ISR is estimated by comparing the
nominal $t\bar{t}V$ sample with two additional samples, which have the same settings as the nominal one, but with the \textsc{Var3}
up or down variation of the A14 tune.
 
\subsubsection{Diboson background samples}
Samples of diboson final states (\(VV\)) are simulated with the
\SHERPA[2.2.1] or 2.2.2~\cite{Bothmann:2019yzt} generator depending on the process (see \Tab{\ref{tab:samples:mc}}),
including off-shell effects and Higgs boson contributions, where appropriate.
Fully leptonic final states and semileptonic final states, where one boson
decays leptonically and the other hadronically, are generated using
matrix elements at NLO accuracy in QCD for up to one additional parton
and at LO accuracy for up to three additional parton
emissions. Samples for the gluon-loop-induced processes \(gg \to VV\) are
generated using LO-accurate matrix elements for up to one
additional parton emission for both the cases of fully leptonic and
semileptonic final states. The matrix-element calculations are matched
and merged with the \SHERPA parton shower based on Catani--Seymour
dipole factorization  using the MEPS@NLO
prescription.
The virtual QCD corrections are provided by the
\OPENLOOPS library. The
\NNPDF[3.0nnlo] set of PDFs is used along with the
dedicated set of tuned parton-shower parameters developed by the
\SHERPA authors.
 
Matrix element to parton shower matching~\cite{Hoeche:2011fd} is employed for different jet
multiplicities, which are then merged into an inclusive sample using an improved CKKW matching
procedure which is extended to NLO accuracy using the MEPS@NLO prescription. These
simulations are NLO-accurate for up to one additional parton and LO-accurate for up to three additional
partons. The virtual QCD correction for matrix elements at NLO accuracy is provided by the \OPENLOOPS
library. The calculation is performed in the $G_\mu$ scheme~\cite{Denner:2003iy}, ensuring an optimal description of pure electroweak
interactions at the electroweak scale.


\FloatBarrier


\section{Object selection}
\label{sec:object_selection}
 
The selection and categorization of \WplusDmeson candidate events depend on the reconstruction and identification of electrons,
muons, tracks, and jets. Proton--proton interaction vertices are reconstructed from charged-particle tracks with $\pT > \SI{500}{\MeV}$
in the ID. The presence of at least one such vertex with a minimum of two associated tracks is required, and the vertex with the
largest sum of $\pT^2$ of associated tracks is chosen as the primary vertex (PV).
 
Three different categories of leptons are used in the analysis: \texttt{baseline}, \texttt{loose}, and \texttt{tight}. Here, \enquote{leptons}
include electrons and muons, but exclude $\tau$-leptons. \texttt{Baseline} leptons are required to have $\pt > \SI{20}{GeV}$, while
\texttt{loose} and \texttt{tight} leptons are required to have $\pt > \SI{30}{\GeV}$.  \texttt{Tight} leptons are required
to meet isolation requirements. Anti-\texttt{tight} leptons are required to pass the \texttt{loose}
requirements, but fail the \texttt{tight} requirements. They are used in the data-driven \MJ production estimation described in \Sect{\ref{sec:signal_and_bkg_modelling:qcd}}.
Full electron and muon selection criteria are given in the text below and summarized in \Tab{\ref{tab:lepton_selection}}.
 
Tracks used in the electron and muon reconstruction are required to be associated with the PV,
using constraints on the transverse impact parameter significance ($\dzerosigbl$) and on the
longitudinal impact parameter ($z_{0}^{\text{BL}}$).
The transverse impact parameter significance is calculated with respect to the measured beamline position and
must satisfy $\dzerosigbl < 3.0$ for muons and $\dzerosigbl < 5.0$ for electrons. The longitudinal
impact parameter of the track is the longitudinal distance along the beamline
between the point where \dzerosigbl is measured and the primary vertex. Tracks are required to have
$|z_{0}^{\text{BL}} \sin{\theta}| < \SI{0.5}{\mm}$, where $\theta$ is the polar angle of the track.
 
Electron candidates are reconstructed from an isolated energy deposit in the electromagnetic calorimeter
matched to a track in the ID and must pass the tight likelihood-based working point~\cite{EGAM-2018-01}.
Electrons must be in the fiducial pseudorapidity region of  $|\eta|<2.47$, excluding the transition region
$1.37 < \abseta < 1.52$ between the calorimeter barrel and endcaps.
The \texttt{tight} electrons are required to meet the \enquote{tight} isolation criteria~\cite{EGAM-2018-01},
based on a combination of the track-based and calorimeter-based isolation. The track-based isolation is
$\pt^{\mathrm{iso}}(\DeltaR_{\mathrm{var}} < 0.2)/\pt < 0.06$, with a variable cone size as defined in \Sect{\ref{sec:samples:data}}.
The tracks are required to have $\pt^{\text{trk}} > \SI{1}{\GeV}$ and are required to be associated with the
primary vertex. The calorimeter-based isolation is $E_{\mathrm{T}}^{\mathrm{cone20}}/\pT < 0.06$, where
$E_{\mathrm{T}}^{\mathrm{cone20}}$ is the sum of the transverse energy of positive-energy topological clusters whose
barycenter falls within a $\Delta R < 0.2$ cone centred around the electron, corrected for the energy leakage, \pileup,
and underlying event, as described in \Refn{\cite{EGAM-2018-01}}. Electron energy scale is calibrated following the procedure
given in \Refn{\cite{EGAM-2018-01}}.
 
Muon candidates are reconstructed in the region $\abseta<2.5$ by matching tracks in the MS with those in the ID. The global
re-fitting algorithm~\cite{PERF-2015-10} is used to combine the information from the ID and MS subdetectors. Muons are identified using
the \enquote{Tight} quality criteria~\cite{MUON-2018-03}, characterized by the numbers of hits in the ID and MS subsystems.
The \texttt{tight} muons are required to pass the \enquote{tight} isolation working point, based on a combination of the track-based
and particle-flow-based~\cite{PERF-2015-09} isolation. The requirement is
$(\pt^{\mathrm{iso}}(\DeltaR_{\mathrm{var}} < 0.3) + 0.4 \times E_{\mathrm{T}}^{\mathrm{neflow20}})/\pT < 0.045$, where the
track-based isolation uses a variable cone size as defined in \Sect{\ref{sec:samples:data}}, with a maximum size of $\Delta R = 0.3$.
The tracks are required to have $\pt^{\text{trk}} > \SI{500}{\MeV}$ and are required to be associated with the primary vertex.
The $E_{\mathrm{T}}^{\mathrm{neflow20}}$ is the sum of the transverse energy of neutral particle-flow objects in a cone of size
$\Delta R < 0.2$ around the muon~\cite{MUON-2018-03}. Muon momentum calibration is performed using the prescription in \Refn{\cite{PERF-2015-10}}.
 
Jets are reconstructed from particle-flow objects~\cite{PERF-2015-09} using the \antikt~\cite{Cacciari:2008gp,Fastjet} jet-reconstruction
algorithm with a distance parameter $R=0.4$. Candidate jets are required to have $\pT > \SI{20}{\GeV}$ and $|\eta| < 5.0$.
The jet energy scale (JES) calibration restores the jet energy to that of jets reconstructed at the particle level,
as described in \Refn{\cite{JETM-2018-05}}. The jets from \pileup interactions are suppressed using the Jet Vertex Tagger
algorithm (JVT)~\cite{ATLAS-CONF-2014-018}.

Jets with $\abseta < 2.5$ and $\pt > \SI{20}{\GeV}$ containing $b$-hadrons are identified by a deep neural network tagger, DL1r
\cite{FTAG-2018-01, ATL-PHYS-PUB-2017-013, FTAG-2019-07}, that uses displaced tracks, secondary vertices and decay topologies.
The chosen working point has 70\% efficiency for identifying \bjets in a simulated \ttbar sample and the measured rejection factor
(the inverse misidentification efficiency) for $c$-jets (light-jets) is about 11 (600)~\cite{FTAG-2019-07}. The \bjets are defined
according to the presence of $b$-hadrons with $\pT > \SI{5}{\GeV}$ within a cone of size $\Delta R = 0.3$ around the jet axis. If a $b$-hadron
is not found and a $c$-hadron is found, then the jet is labeled a $c$-jet. Light-jets are all the rest.
 
The missing transverse momentum (\MET) in the events is calculated as the
negative vector sum of the selected high-\pt calibrated objects (jets and \texttt{baseline} electrons and muons), plus a
\enquote{soft term} reconstructed from tracks not associated with any of the calibrated objects~\cite{PERF-2016-07,ATLAS-CONF-2018-023}.
 
To avoid cases where the detector response to a single physical object is reconstructed as two different final-state objects, e.g.,
an electron reconstructed as both an electron and a jet, an overlap removal strategy is used. If the two calorimeter energy clusters
from two electron candidates overlap, the electron with the highest \ET is retained. If a reconstructed electron and muon share the
same ID track, the muon is rejected if it is calorimeter-tagged, meaning the muon is identified as a reconstructed ID track that
extrapolates to the calorimeter energy deposit of a minimum-ionizing particle without an MS signal~\cite{MUON-2018-03}; otherwise the electron is rejected.
Next, jets within $\DeltaR = 0.2$ of electrons are removed. In the last step, electrons and muons within $\DeltaR = 0.4$ of any remaining
jet are removed. This overlap removal procedure is performed using the \texttt{baseline} leptons.

\begin{table}[htpb]
\caption{Lepton categories used in this analysis.}
\label{tab:lepton_selection}
\centering
\begin{tabular}{ l | c | c | c | c | c | c }
 
\toprule
 
& \multicolumn{3}{c|}{Electrons} & \multicolumn{3}{c}{Muons} \\
 
\midrule
 
Features & \texttt{baseline} & \texttt{loose} & \texttt{tight} & \texttt{baseline} & \texttt{loose} & \texttt{tight} \\
 
\midrule
 
\pT & $>\SI{20}{\GeV}$ & \multicolumn{2}{c|}{$>\SI{30}{\GeV}$} & $>\SI{20}{\GeV}$ & \multicolumn{2}{c}{$>\SI{30}{\GeV}$} \\
 
$\delzzerosinthetabl$ & \multicolumn{3}{c|}{$< \SI{0.5}{\mm}$} & \multicolumn{3}{c}{$< \SI{0.5}{\mm}$} \\
 
$\dzerosigbl$ & \multicolumn{3}{c|}{$< 5$} & \multicolumn{3}{c}{$< 3$} \\
 
Pseudorapidity & \multicolumn{3}{c|}{$(\abseta < 1.37) || (1.52 < \abseta < 2.47)$} & \multicolumn{3}{c}{$\abseta < 2.5$} \\
 
Identification & \multicolumn{3}{c|}{Tight} & \multicolumn{3}{c}{Tight} \\
 
Isolation & \multicolumn{2}{c|}{No} & \multicolumn{1}{c|}{Yes} & \multicolumn{2}{c|}{No} & \multicolumn{1}{c}{Yes} \\
 
\bottomrule
\end{tabular}
\end{table}


\FloatBarrier


\section{Charmed meson reconstruction}
\label{sec:meson_selection}
 
Events containing $c$-quarks are identified by explicitly reconstructing charmed mesons in charged, hadronic decay channels.
Two charmed hadron decay channels are used: $\DtoKpipi$ and $\DstarKpi$ (and charge conjugates). The invariant mass distribution
\mDplus (mass difference \mdiff) used in the fit for the \Dplus (\Dstarp) channel is described in \Sect{\ref{sec:likelihood_fit}}.
 
ID tracks satisfying
$|\eta| < 2.5$ and $|z_0\sin\theta| < \SI{5}{\mm}$ are used for \Dmeson meson reconstruction.
The $\text{Loose}$ track quality requirement is applied~\cite{ATL-PHYS-PUB-2015-051}.
The \Dplus (\Dzero) 
candidate is reconstructed using ID tracks with $\pt > \SI{800}{\MeV}$~($\SI{600}{\MeV})$.
A geometric separation of $\DeltaR < 0.6$ among the tracks is required.
Tracks corresponding to the \texttt{baseline} leptons used for
the \Wboson boson candidates are excluded.
The \Dplus candidates are required to have three tracks with  total charge $=\pm 1$.
The two tracks with the same charge are assigned the charged pion mass and the remaining track is assigned the kaon mass.
The \Dzero candidates are required to have two tracks with total charge $=0$.
One track is assigned the charged pion mass and the other is assigned the charged kaon mass.
Both possible choices for the mass assignment are retained until matching to the
prompt pion from the \Dstarp decay is performed.
Tracks from the \Dplus (\Dzero) candidate  are fitted to a common secondary vertex (SV), with a fit $\ChiSquared$ required to be
$\ChiSquared < 8.0\ (10.0)$. To reduce the contribution from \pileup and from $b$-hadron decays, the transverse impact parameter
of the \Dmeson candidate's flight path with respect to the PV is required to satisfy $\dzeroabs < \SI{1.0}{\mm}$ and the
candidate is required to have a 3D impact parameter significance $\ImpactSig < 4.0$,  where \ImpactSig is the distance of closest
approach of the candidate's flight path to the PV divided by the uncertainty in that distance. These selection criteria and those
described below were determined by optimizing the \OSminusSS signal significance, using MC predictions to estimate the signal, and
mass sidebands to estimate the background.
 
Several requirements are placed on  the \Dplus candidates to reduce combinatorial background.  The angle between the kaon track in the rest frame of the \Dplus candidate and the line of flight of the \Dplus candidate in the
center-of-mass frame is required to satisfy $\CombBkgReject > -0.8$.
The distance between the SV and the PV in the transverse plane is required to satisfy
$\Lxy > \SI{1.1}{\mm}$ for \Dplus candidates with $\pt < \SI{40}{\GeV}$ and $\Lxy > \SI{2.5}{\mm}$ for \Dplus candidates with
$\pt > \SI{40}{\GeV}$.
Kinematic requirements are applied to ensure orthogonality
to other \Dmeson decays with similar final states. The contamination from \DstarplustoKpipi, which has the same final-state content as
the \DtoKpipi channel, is reduced by requiring \DstarBkgReject. Background from the \DsubstoKKpi channel, with one of the kaons misidentified
as a pion, is removed by requiring the mass of each pair of oppositely charged particles, assuming the kaon mass hypothesis, to be \DsubsReject.
The world-average mass of the $\phi$ meson from the Particle Data Group (PDG) database~\cite{Workman:2022ynf}, $m_{\phi}=\SI{1019.455}{\MeV}$, is used.
Finally, a requirement is placed on the invariant mass of the \Dplus candidates, $\SI{1.7}{\GeV} < \mDplus < \SI{2.2}{\GeV}$.
 
The \Dstarp candidates are reconstructed by combining \Dzero candidates with prompt tracks that are assigned the charged pion mass.
Only combinations where the pion in the \Dzero candidate has the same charge as the prompt pion are considered.  The small mass difference between the \Dstarp and \Dzero mesons restricts the phase space of this associated prompt pion, which has low momentum
in the \Dzero rest frame and hence is referred to as the slow pion.
Slow pion tracks  are required to have $\pt > \SI{500}{\MeV}$ and
a transverse impact parameter of $\dzeroabs < \SI{1.0}{\mm}$ with respect to the primary vertex. An $\Lxy > \SI{0}{mm}$ requirement is applied to \Dzero candidates.
The mass of the \Dzero  candidate must be within \SI{40}{\MeV} of the PDG world-average
value of the \Dzero mass, $m_{\Dzero}=\SI{1864.83}{\MeV}$~\cite{Workman:2022ynf}.
Additionally, the angular separation between the slow pion and the \Dzero
meson must be small, $\SlowPiAngularSep < 0.3$, and the invariant mass cut
of $\SI{140}{\MeV} < \mdiff < \SI{180}{\MeV}$ is imposed.
 
Combinatorial background from light jets is reduced by requiring  \Dmeson candidates to be isolated. The transverse momenta of tracks
in a cone of size $\DeltaR = 0.4$ around the \Dmeson candidate are summed, and the sum
is required to be less than the \pt of the \Dmeson.  Background from semileptonic $B$ meson decays is
reduced by requiring $\QCDBkgReject > 0.3$. Finally, the \Dmeson candidates are required to have $\SI{8}{\GeV} < \pt < \SI{150}{\GeV}$
and $|\eta| < 2.2$. The $\eta$ cut is applied to avoid the edge of the ID, where the amount of the detector material increases rapidly
and thus reduces the reconstruction efficiency and degrades the resolution. The upper $\pT$ cut is applied to reject the
background from fake \Dmeson mesons at high momentum and, because the predicted fraction of \Dmeson mesons with $\pT(\Dmeson) > \SI{150}{\GeV}$
is small, it has no significant impact on the signal reconstruction efficiency. The full set of selection requirements for the \Dmeson
candidates is summarized in \Tab{\ref{tab:dmeson_selection}}.

\begin{table}[htpb]
\caption{\Dmeson object selection criteria. For \Dstarp candidates the cuts related to SV reconstruction are applied to the corresponding \Dzero candidate.}
\label{tab:dmeson_selection}
\centering
\setlength\tabcolsep{3.2pt}
\begin{tabular}{ l | r | r }
 
\toprule
 
\Dmeson cut & \Dplus cut value & \Dstarp cut value (\DzeroKpi) \\
 
\midrule
 
$N_{\mathrm{tracks}}$ at SV & $3$ & $2$ \\
 
SV charge & $\pm 1$ & $0$ \\
 
SV fit quality & $\ChiSquared < 8$ & $\ChiSquared < 10$ \\
 
Track $\pt$ & $\pt > \SI{800}{\MeV}$ & $\pt > \SI{600}{\MeV}$ \\
 
Track angular separation & $\DeltaR < 0.6$ & $\DeltaR < 0.6$ \\
 
\multirow{2}{*}{Flight length} & $\Lxy > \SI{1.1}{\mm}$ \big($\pt(\Dplus) < \SI{40}{\GeV}$\big)    & \multirow{2}{*}{$\Lxy > \SI{0}{\mm}$} \\
& $\Lxy > \SI{2.5}{\mm}$ \big($\pt(\Dplus) \geq \SI{40}{\GeV}$\big) & \\
 
SV impact parameter & $|d_0| < \SI{1}{\mm}$ &  $|d_0| < \SI{1}{\mm}$ \\
 
SV 3D impact significance & $\ImpactSig < 4.0 $ & $\ImpactSig < 4.0 $ \\
 
Combinatorial background rejection & $\CombBkgReject > -0.8$ & --- \\
 
Isolation & $\DplusIsolation < 1.0$ & $\DstarIsolation < 1.0$ \\
 
\midrule
 
\Dsubstophipi rejection & \DsubsReject & --- \\
 
\Dstarp background rejection & \DstarBkgReject & --- \\
 
\Dzero mass & --- & \DzeroMassCut \\
 
\midrule
 
\SlowPi $\pt$ & --- & $ \pt > \SI{500}{\MeV}$ \\
 
\SlowPi angular separation & --- & $ \SlowPiAngularSep < 0.3$ \\
 
\SlowPi $d_0$ & --- & $|d_0| < \SI{1}{\mm}$ \\
 
\midrule
 
QCD background rejection & $\QCDBkgRejectDplus > 0.3$ & $\QCDBkgRejectDstar > 0.3$ \\
 
\midrule
 
\Dmeson \pT & $\SI{8}{\GeV} < \pT(\Dplus) < \SI{150}{\GeV}$ & $\SI{8}{\GeV} < \pT(\Dstarp) < \SI{150}{\GeV}$ \\
 
\Dmeson $\eta$ & $|\eta(\Dplus)| < 2.2$ & $|\eta(\Dstarp)| < 2.2$ \\
 
Invariant mass & $\SI{1.7}{\GeV} < \mDplus < \SI{2.2}{\GeV}$ & $\SI{140}{\MeV} < \mdiff < \SI{180}{\MeV}$ \\
 
\bottomrule
\end{tabular}
\end{table}


\FloatBarrier


\section{Event selection}
\label{sec:event_selection}
 
Events for the analysis are selected through requirements on leptons, \MET, jets and \Dmeson mesons satisfying the criteria defined
in \Sects{\ref{sec:object_selection}}{\ref{sec:meson_selection}} and passing the single-lepton triggers as discussed
in \Sect{\ref{sec:samples}}. Reconstruction of \Wboson bosons is based on their leptonic decays to either an electron (\Wtoenu) or a muon (\Wtomunu).
The lepton is measured in the detector and the presence of a neutrino is inferred from \MET. Events are
required to have exactly one \texttt{tight} lepton with $\pt > \SI{30}{\GeV}$ and $\abseta < 2.5$. Events with additional \texttt{loose}
leptons are rejected. To reduce the \MJ background and enhance the \Wboson boson signal purity, additional requirements are imposed:
$\MET > \SI{30}{\GeV}$ and $\mT > \SI{60}{\GeV}$, where the \Wboson boson transverse mass (\mT) is defined as
$\sqrt{2 \pT(\text{lep}) \MET (1 - \cos(\Delta\phi))}$ and $\Delta\phi$ is the azimuthal separation between
the lepton and the missing transverse momentum. Candidate \Dmeson mesons are reconstructed using a secondary-vertex fit as described
in \Sect{\ref{sec:meson_selection}}. Any number of \Dmeson meson candidates satisfying these criteria are selected, which accounts
for the production of multiple mesons in a single event. Only events with one or more \Dmeson candidates are selected.
 
Events selected in this way are used to extract the \WplusDmeson observables with a profile likelihood fit defined in
\Sect{\ref{sec:likelihood_fit}}. Furthermore, the selected events are categorized according to the $b$-jet multiplicity to separate the
\WplusDmeson signal process from the \ttbar background with events containing $W \to cs$ decays. The ID tracks associated with the
reconstructed \Dmeson candidates are often also associated with a jet mis-tagged as a $b$-jet. To avoid categorizing these \WplusDmeson
signal events as events with one more more $b$-jets, the $b$-jets are required to be geometrically separated from reconstructed \Dmeson mesons
by satisfying $\DeltaR(\text{\textit{b}-jet}, \Dmeson) > 0.4$. Events with exactly zero such $b$-tagged jets are classified
as the \WplusDmeson signal region (SR) and events with one or more $b$-tagged jets comprise the Top control region (CR).
In this way about 80\% of the \ttbar background events are in the \TopCR
and about 99\% of \WplusDmeson signal events remain in the \SR, effectively reducing the amount of \ttbar background. Collectively,
the \SR and \TopCR are called the \enquote{fit regions}. These requirements are summarized in \Tab{\ref{tab:event_selection:reco_selection:sel_table:reco}}.
The measured signal and background yields in the \SR are given in \Tabs{\ref{tab:results:yieldtab:dplus}}{\ref{tab:results:yieldtab:dstar}}
in \Sect{\ref{sec:results}}. The yield of \WplusDmeson signal events is about \SI{5}{\percent} of the \ttbar background yield in the \TopCR.

\begin{table}[htpb]
\caption{Tables summarizing the event selection in the analysis:
\protect\subref{tab:event_selection:reco_selection:sel_table:reco} fit regions used in the statistical analysis,
\protect\subref{tab:event_selection:reco_selection:sel_table:truth} the \enquote{truth} fiducial selection.
The \WplusDmeson signal is defined by performing the \OSminusSS subtraction as described in the text.}
\centering
\subfloat[]{
\begin{tabular}{l c c}
\toprule
\multicolumn{3}{c}{Detector-level selection} \\
\midrule
Requirement              & \SR & \TopCR \\
\midrule
$N$($b$-jet)               & 0   & $\geq 1$ \\
\midrule
\MET                     & \multicolumn{2}{c}{\(>\SI{30}{GeV}\)} \\
\mT                      & \multicolumn{2}{c}{\(>\SI{60}{GeV}\)} \\
Lepton \pT               & \multicolumn{2}{c}{\(>\SI{30}{GeV}\)} \\
Lepton \(\abseta\)       & \multicolumn{2}{c}{\(< 2.5\)} \\
\midrule
$N$(\Dmeson)             & \multicolumn{2}{c}{$\geq 1$} \\
\Dmeson \pt              & \multicolumn{2}{c}{\(>\SI{8}{GeV}\) and \(<\SI{150}{GeV}\)} \\
\Dmeson \(\abseta\)      & \multicolumn{2}{c}{\(< 2.2\)} \\
\bottomrule
\end{tabular}
\label{tab:event_selection:reco_selection:sel_table:reco}
}
\qquad
\subfloat[]{
\begin{tabular}{l c}
\toprule
\multicolumn{2}{c}{Truth fiducial selection}   \\
\midrule
Requirement             & \WplusDmeson         \\
\midrule
$N$($b$-jet)              & ---                  \\
\midrule
\MET                    & ---                  \\
\mT                     & ---                  \\
Lepton \pT              & \(>\SI{30}{GeV}\)    \\
Lepton \(|\eta|\)       & \(< 2.5\)            \\
\midrule
$N$(\Dmeson)            & \(\geq 1\)           \\
\Dmeson \pt             & \(>\SI{8}{GeV}\)     \\
\Dmeson \(|\eta|\)      & \(< 2.2\)            \\
\bottomrule
\end{tabular}
\label{tab:event_selection:reco_selection:sel_table:truth}
}
\end{table}

The analysis exploits the charge correlation of the \Wboson boson and the charm quark
to enhance the signal  and reduce the backgrounds. The signal has a \Wboson boson and a \Dmeson meson of opposite charge,
while most backgrounds are symmetric in charge. Therefore, the signal is extracted by measuring the difference between the numbers of opposite-sign (OS)
and same-sign (SS) \WplusDmeson candidates, which is referred to as OS--SS. While the signal-to-background ratio is about unity in the OS
region, the OS--SS \WplusDmeson signal is an order of magnitude larger than the remaining background after the subtraction.
 
The \WplusDmeson measurement is unfolded to a \enquote{truth} fiducial region defined at MC particle level to have exactly one \enquote{truth} lepton with
$\pT(\ell) > \SI{30}{\GeV}$ and $|\eta(\ell)| < 2.5$. The lepton must originate from a \Wboson boson decay, with
$\tau$ decays excluded from the fiducial region. Lepton momenta are calculated using \enquote{dressed} leptons, where the four-momenta of photons radiated
from the final-state leptons within a cone of $\DeltaR = 0.1$ around the lepton are added to the four-momenta of leptons.
Truth \Dmeson mesons are selected by
requiring $\pT(\Dmeson) > \SI{8}{\GeV}$ and $|\eta(\Dmeson)| < 2.2$. The OS--SS subtraction is also applied to the truth fiducial events.
This removes any charge-symmetric processes, which are expected to originate mostly from gluon splitting in the final state.
The \MET and \mT requirements and $b$-jet veto are not applied in the fiducial selection. The truth fiducial selection is summarized in
\Tab{\ref{tab:event_selection:reco_selection:sel_table:truth}}. The fiducial efficiency is defined as the fraction of \WplusDmeson signal
events from the truth fiducial region that pass the detector-level reconstruction and requirements in \Tab{\ref{tab:event_selection:reco_selection:sel_table:reco}}.
In the unfolding, events where the reconstructed objects pass the event selection but the truth objects fail the
truth fiducial requirements are treated as fakes;  cases where the reconstructed objects fail the reconstruction fiducial selection
but the truth objects pass the truth selection are treated as inefficiencies.

\FloatBarrier


\section{Signal and background modeling}
\label{sec:signal_and_bkg_modelling}
 
MC samples are used to construct signal and background mass templates, except for the \MJ background, which is determined
using a data-driven method (\Sect{\ref{sec:signal_and_bkg_modelling:qcd}}). Generally, \SHERPA[2.2.11] MC samples are used to
model events containing a single $W$ boson and one or more reconstructed \Dmeson meson candidates because they provide
the highest precision when simulating QCD processes and the highest statistical power among the available samples. For specific
purposes, \MGLO and \aMGNLO MC samples are used in conjunction with \SHERPA to account for shortcomings in \SHERPA modeling
of \Dmeson meson decays as described in \Sects{\ref{sec:signal_and_bkg_modelling:signal_modelling}}{\ref{sec:signal_and_bkg_modelling:prompt_mc}}.
MC truth information is used to categorize the MC \WplusDmeson events according to the origin of the tracks used to reconstruct the \Dmeson meson candidate:
\begin{itemize}
\item{\WplusDmeson signal}: If all tracks originate from the signal charmed hadron species (\Dplus or \Dstar) and are assigned
in the reconstruction to the correct particle species ($K^{\mp}\pi^{\pm}\pi^{\pm}$), then that reconstructed \Dmeson candidate is
labeled as \WplusDmeson signal.
\item{\Wpluscmatch}: If all tracks originate either from a different charmed hadron species (\Dzero, \Dsubs, or $c$-baryon)
or from a different decay mode of a signal charmed meson (e.g.\ \DplustoKKpi), the reconstructed \Dmeson candidate is labeled as \Wpluscmatch.
\item{\Wpluscmismatch}: If at least one but not all tracks belong to a single charmed hadron, the reconstructed \Dmeson candidate
is labeled as \Wpluscmismatch.
\item{\Wjets}: if none of the tracks are matched to a particle originating from a charmed particle,
the \Dmeson candidate is labeled \Wjets. This is the combinatorial background from the underlying event
and \pileup.
\end{itemize}
 
Additional background categories modeled using MC simulation are:
\begin{itemize}
\item{Top}: Processes containing top quarks (\ttbar, single-$t$, $\ttbar X$) are jointly represented by the \enquote{Top} category,
which is dominated by the \ttbar process.
\item{Other}: Events from diboson and \Zjets processes are combined into the \enquote{Other} category.
\end{itemize}
 
The signal and background samples used in the \WplusD and \WplusDstar fits are given in
\Tab{\ref{tab:signal_bkg_templates}}. The rates at which $c$-quarks hadronize into different species of weakly decaying charmed
hadrons in the MC samples are reweighted to the world-average values~\cite{ATL-PHYS-PUB-2022-035}. The weights improve agreement
between data and MC simulation by modifying the signal and background normalizations and the shapes of the \WplusDmeson
background templates by changing the relative contribution of each species. The normalization of the background
templates changes by up to 3\%, depending on the \Dmeson species.
 
\begin{table}[htpb]
\caption{Single-$W$-boson MC samples employed to create mass templates used in the \WplusDmeson fits.
The \enquote{Normalization} and \enquote{Shape} columns indicate the source used to calculate the corresponding property.
\enquote{LIS} refers to the Loose Inclusive Selection explained in the text, and $m(\Dmeson)$ stands for \mDplus in the
\Dplus channel and \mdiff in the \Dstar channel. The MC configurations used to model
these backgrounds are described in \Sect{\ref{sec:samples:mc}}. Preferentially, \SHERPA samples are
used for signal and background modeling. There are some exceptions to account for the shortcomings as
explained in the text (e.g. incorrect \Dstarp decay with in \SHERPA).}
\label{tab:signal_bkg_templates}
\centering
\begin{tabular}{lcc}
\toprule
Category                     & Normalization                     & $m(\Dmeson)$ shape   \\
\midrule
\WplusDmeson (\Dplus channel) & \SHERPA[2.2.11]                 & \SHERPA[2.2.11]     \\
\WplusDmeson (\Dstar channel) & \SHERPA[2.2.11]                 & \aMGNLO               \\
\Wpluscmatch (\Dplus channel) & \MGLO                            & \MGLO                \\
\Wpluscmatch (\Dstar channel) & \SHERPA[2.2.11]                 & \SHERPA[2.2.11]     \\
\Wpluscmismatch               & \SHERPA[2.2.11]                 & LIS \SHERPA[2.2.11] \\
\Wjets (\Dplus channel)       & \SHERPA[2.2.11]                 & LIS \SHERPA[2.2.11] \\
\Wjets (\Dstar channel)       & \MGLO                            & LIS \MGLO            \\
\bottomrule
\end{tabular}
\end{table}

\subsection{Signal modeling}
\label{sec:signal_and_bkg_modelling:signal_modelling}
 
The \SHERPA[2.2.11] \WplusD signal sample with \EVTGEN decays is used for the modeling of the
mass template in the \Dplus channel. However, because the width of the \Dstarp meson is set incorrectly in
\SHERPA[2.2.11], the mass shape in the \Dstarp channel is taken from the \aMGNLO\ \WplusDstar
signal sample instead. In both channels the normalization is taken from \SHERPA[2.2.11] because
it provides the best available statistical power for calculating the fiducial efficiency.
 
\FloatBarrier


\subsection{Modeling backgrounds with a single $W$ boson}
\label{sec:signal_and_bkg_modelling:prompt_mc}
 
The \Wpluscmatch background in the \Dplus channel is modeled using \MGLO\ because the \EVTGEN decay tables and models used with
\MGLO\ provide a better description of the $D$ meson decay rates and kinematics than those implemented in \SHERPA[2.2.11].
Corrections to account for LO $\to$ NLO effects in \WplusDmeson production are applied by reweighting the \MGLO\ MC truth distribution
of $\pt(\Dplus)$ to the corresponding \SHERPA[2.2.11] distribution. \SHERPA[2.2.11] is also used in the \Dstarp channel.
 
The \Wpluscmismatch backgrounds are modeled using \SHERPA[2.2.11] in both the \Dplus and \Dstarp channels.
The \Wjets background is modeled using \SHERPA[2.2.11] in the \Dplus channel and \MGLO\ in the \Dstarp channel
because their descriptions of this background yield and invariant mass shape are closer to the data before the fit.
These background MC samples suffer from large
statistical uncertainties. A Loose Inclusive Selection (LIS) method was developed to reduce these uncertainties.
The LIS method is based on the observation that, for these backgrounds, the \Dmeson meson mass shapes are the same for both
\Wboson boson charges and do not depend on the \MET and \mT cuts. Therefore, the LIS can be used to construct mass templates
inclusively and without \MET and \mT cuts. These inclusive mass distributions are then used for both
\Wboson boson charges. In the \Dstarp channel, the LIS \Wjets background is fitted with a parametric
function. This parametric function is then used to generate the template histogram which is used in the \WplusDstar fit.


\subsection{Data-driven \MJ background estimation}
\label{sec:signal_and_bkg_modelling:qcd}
Multijet backgrounds arise if one or more constituents of a jet are misidentified as a prompt lepton.
In the electron channel, \MJ events pass the electron selection due to having misidentified hadrons,
converted photons or semileptonic heavy-flavor decays. In the muon channel, muons from heavy-flavor
hadron decays are the dominant source. Collectively, these backgrounds are called \enquote{fake and nonprompt leptons}.
MC-based predictions for the normalization and composition of these backgrounds suffer from large uncertainties.
The background rate is therefore determined using the data-driven Matrix Method~\cite{EGAM-2019-01}.
 
The Matrix Method takes advantage of the fact that fake and nonprompt leptons (F) are less well isolated than
real leptons (R). Leptons can be split independently in two ways: by origin, R and F, or by the \texttt{tight} (T)
and \texttt{loose} reconstruction criteria defined in \Tab{\ref{tab:lepton_selection}}. Leptons satisfying the \texttt{loose}
but not the \texttt{tight} criteria are labeled as anti-\texttt{tight} (!T). While the abundances of R and F leptons
($N_\text{R}$ and $N_\text{F}$) are not directly measurable in data, they can be related to the measurable numbers
of \texttt{tight} and anti-\texttt{tight} leptons ($N_\text{T}$ and $N_{!\text{T}}$) via the efficiency $r$ ($f$) for a \texttt{loose} real (fake) lepton
to also be \texttt{tight}:
 
$$
\left(\begin{array}{c}
N_{\text{T}} \\
N_{!\text{T}}
\end{array}\right)=\left(\begin{array}{lll}
r & f \\
1-r & 1-f
\end{array}\right)\left(\begin{array}{c}
N_{\text{R}}\\
N_{\text{F}}
\end{array}\right),
$$
 
This expression is inverted to give an expression for the number of fake and nonprompt leptons in the \SR,
dependent on measurable quantities:
 
$$
N^{\text{fake}}_{\text{T}} = \frac{f}{r-f} \left( (r-1) N_\text{T} + r N_{!\text{T}} \right).
$$
 
This Matrix Method relation is applied bin-by-bin to estimate the \MJ background yield in the variable of interest.
 
The real-lepton efficiency $r$ is determined from the data in auxiliary measurements~\cite{EGAM-2018-01,PERF-2015-10}
and extrapolated to the \WplusDmeson SR using MC samples. The real-lepton efficiency is estimated in 3 (4) bins in $\eta$ for
electrons (muons) and in \pT bins of \SI{6}{\GeV} width.
 
The fake-lepton efficiency $f$ is computed from the data in a dedicated region enriched in fake and nonprompt leptons, called the \FakeCR.
This region, orthogonal to the \SR, is selected by inverting the \MET and \mT requirements to $\MET < \SI{30}{\GeV}$ and $\mT < \SI{40}{\GeV}$. These
requirements reduce the contribution of real leptons originating from $\Wboson$ boson decays. To further increase the \FakeCR's purity in fake and nonprompt
leptons, processes with real leptons are estimated from MC simulation and subtracted from both \texttt{tight} and anti-\texttt{tight} subsets of \FakeCR.
The OS--SS subtraction is not performed for the calculation of the fake-lepton efficiencies because the \MJ background is largely symmetric in OS and SS events.
The number of \texttt{tight} leptons divided by the sum of \texttt{tight} and anti-\texttt{tight} gives the fake-lepton efficiency.
The efficiency is estimated in 3 (4) bins in $\eta$ for electrons (muons) and in \pT bins of \SI{5}{\GeV} to \SI{20}{\GeV} width,
depending on the available sample size. The fake-lepton efficiency, in the \FakeCR, is in the range 50\%--90\% or 10\%--70\% for
electrons and muons respectively.

Systematic uncertainties in the \MJ estimation arise from several sources. Statistical uncertainties in the determination
of the real- and fake-lepton efficiencies lead to systematic uncertainties of approximately 10\% to 20\% in the overall \MJ yield.
Uncertainties in the size of the real-lepton contamination in the \FakeCR  are assessed using two methods.
First, the change in rate due to varying the QCD renormalization and factorization scales in MC samples is obtained.
Second, the difference between the prompt rates determined using \MGLO\ or \SHERPA[2.2.11] \Wjets MC samples is evaluated.
These two variations together result in relative uncertainties on the \MJ yield of ${\sim}20\%$ for the \Dplus channel and ${\sim}30\%$ for the \Dstarp channel.
 
An additional systematic uncertainty is derived to account for the dependence of fake-lepton efficiencies on \MET, which may arise
from the different composition of fake background processes depending on the \MET (e.g. misidentified hadrons or semileptonic heavy-flavor decays),
the correlation between the lepton isolation variables and \MET, and the tendency of misidentified objects (e.g. jets misidentified as electrons) to give rise
to \MET due to incorrect assumption about the object type in their energy calibration.
To estimate this, the \FakeCR's \MET cut is inverted to require $\MET > \SI{30}{\GeV}$ while its \mT cut is retained to ensure orthogonality with the \SR. This
process provides an independent estimate of the \MJ background. Differences between the \MJ background yields in the \SR obtained with
these two choices of \FakeCR\ cuts are ${\sim}50\%$ for \Dplus and ${\sim}60\%$ for \Dstarp.  While this \MJ background estimate has
large systematic uncertainties, the \MJ yield in the \SR is only up to \SI{1}{\percent} of the signal yield in the electron channel and negligible
in the muon channel. Thus the \MJ background uncertainties are subdominant when estimating the overall background yield.
 
\begin{figure}[htbp]
\centering
\subfloat[]{
\includegraphics[width=0.50\textwidth]{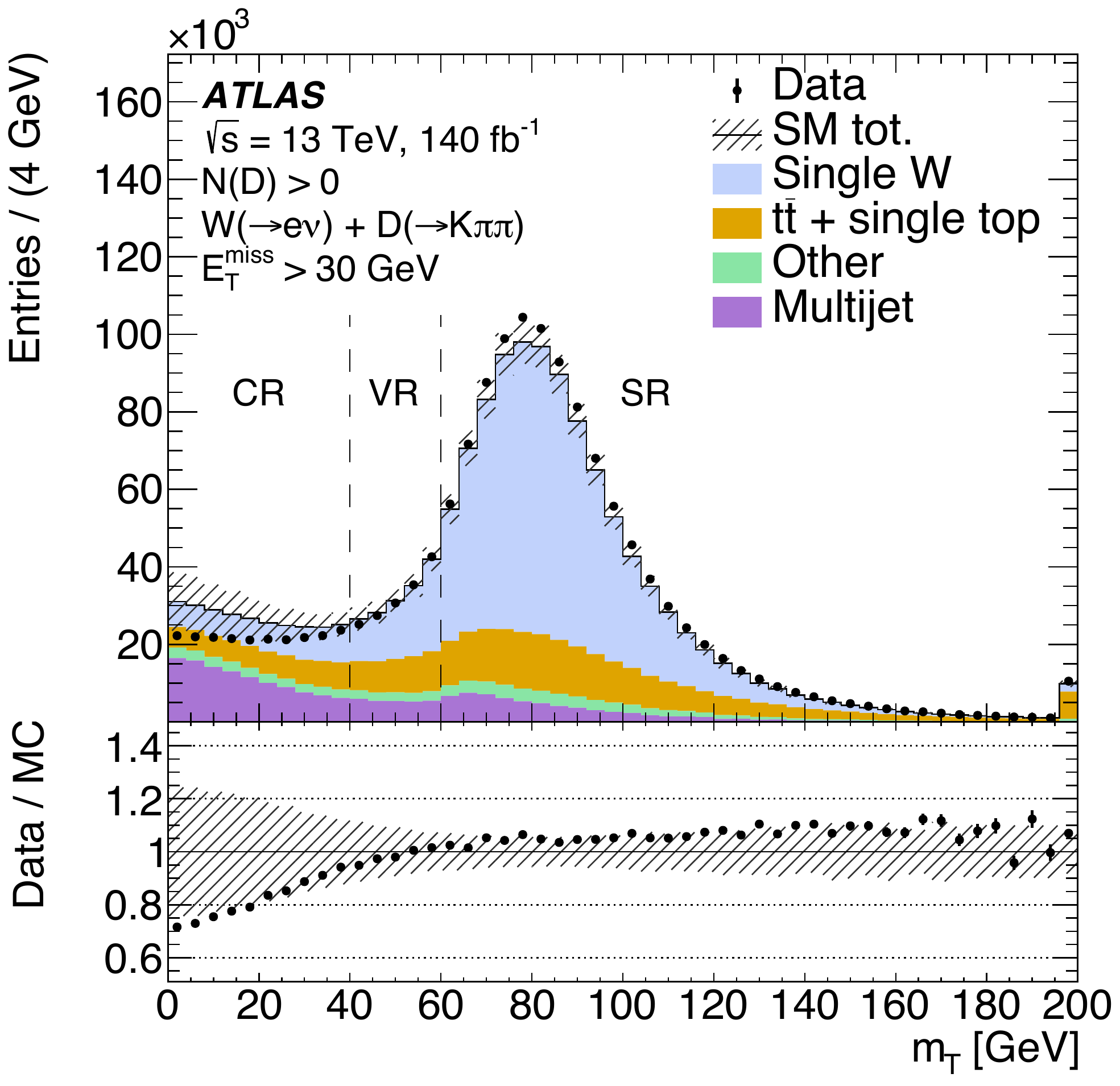}
\label{fig:matrix_method:dplus_el_mt}
}
\subfloat[]{
\includegraphics[width=0.50\textwidth]{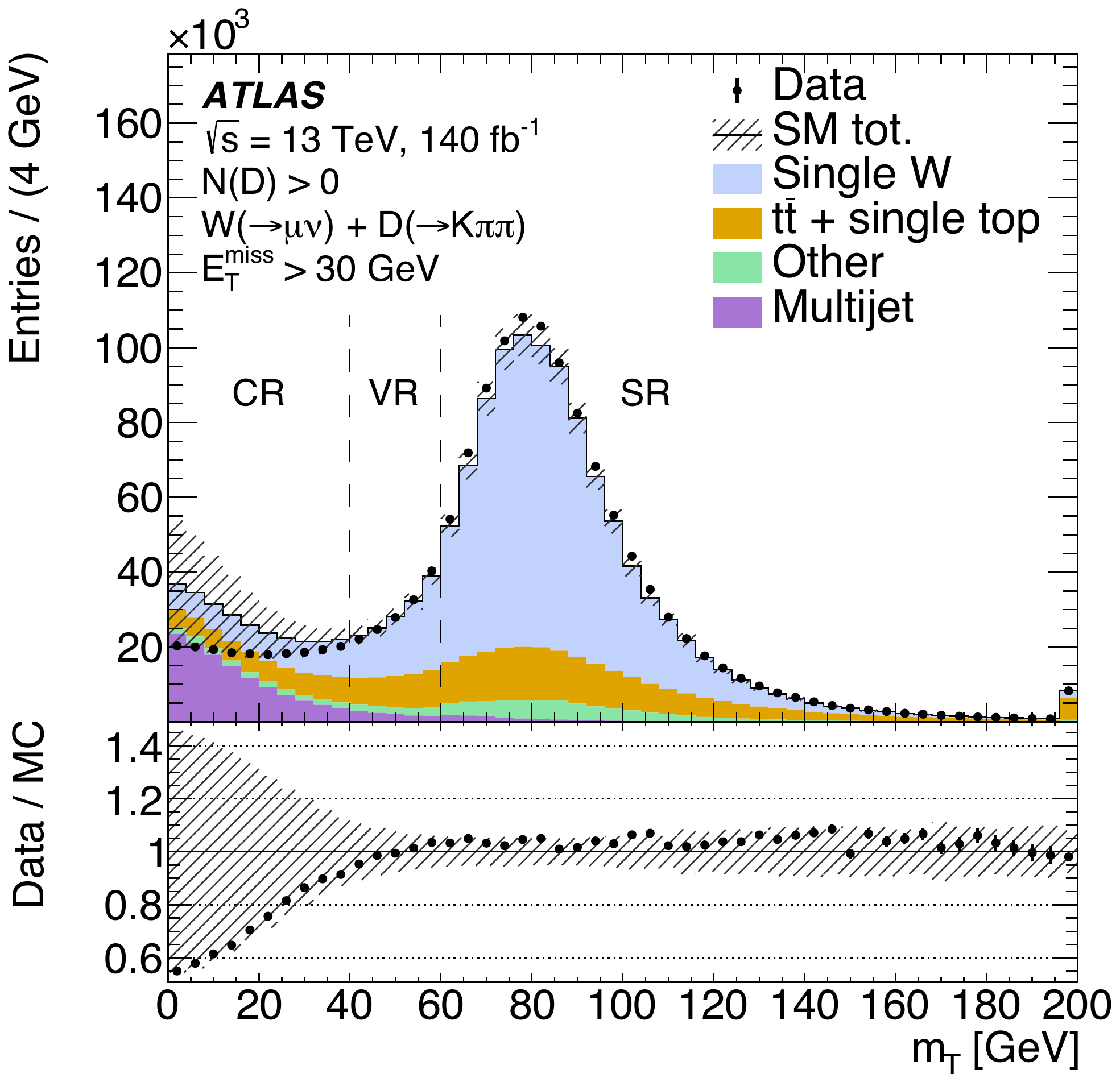}
\label{fig:matrix_method:dplus_mu_mt}
}
\\
\subfloat[]{
\includegraphics[width=0.50\textwidth]{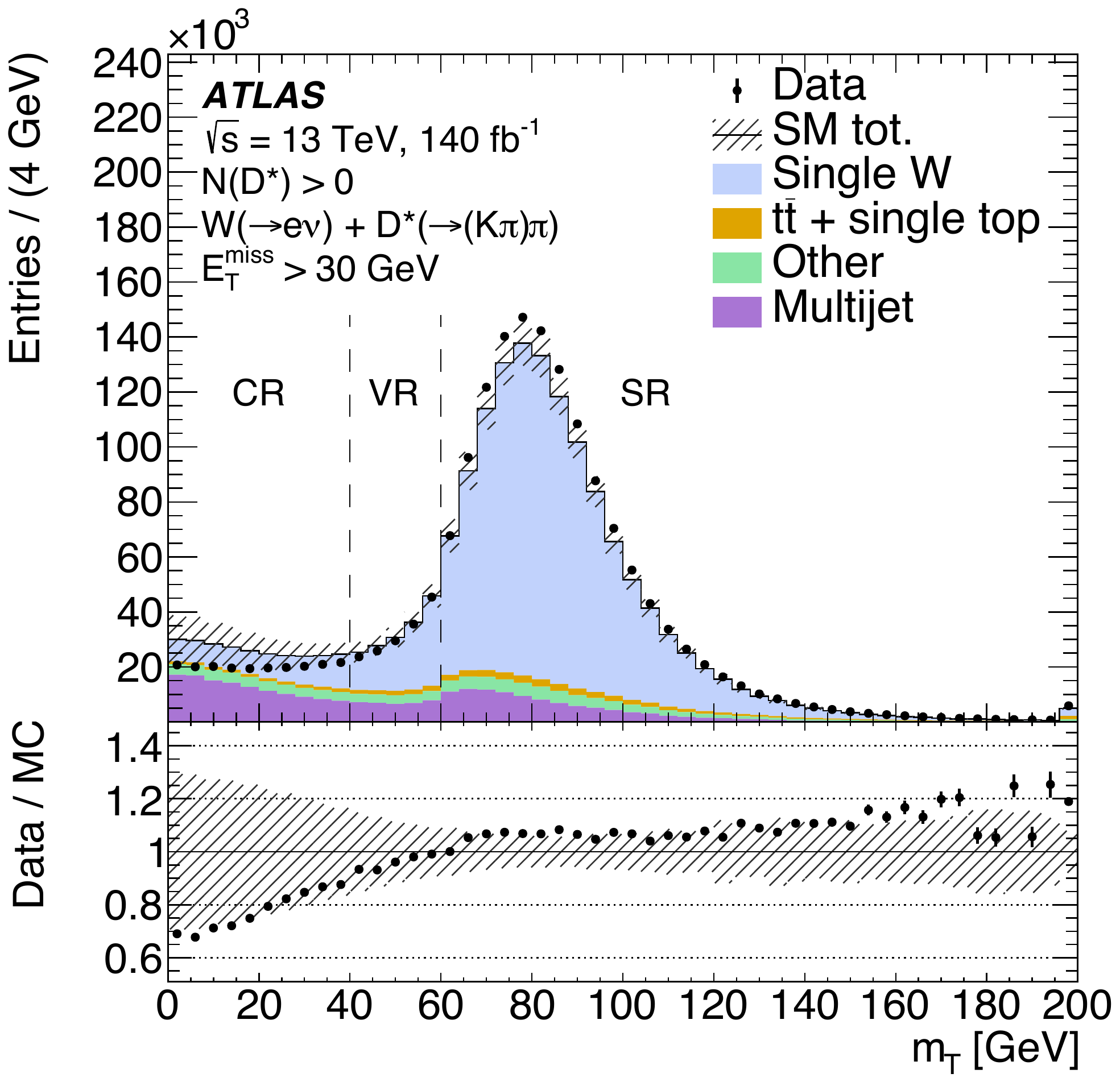}
\label{fig:matrix_method:dstar_el_mt}
}
\subfloat[]{
\includegraphics[width=0.50\textwidth]{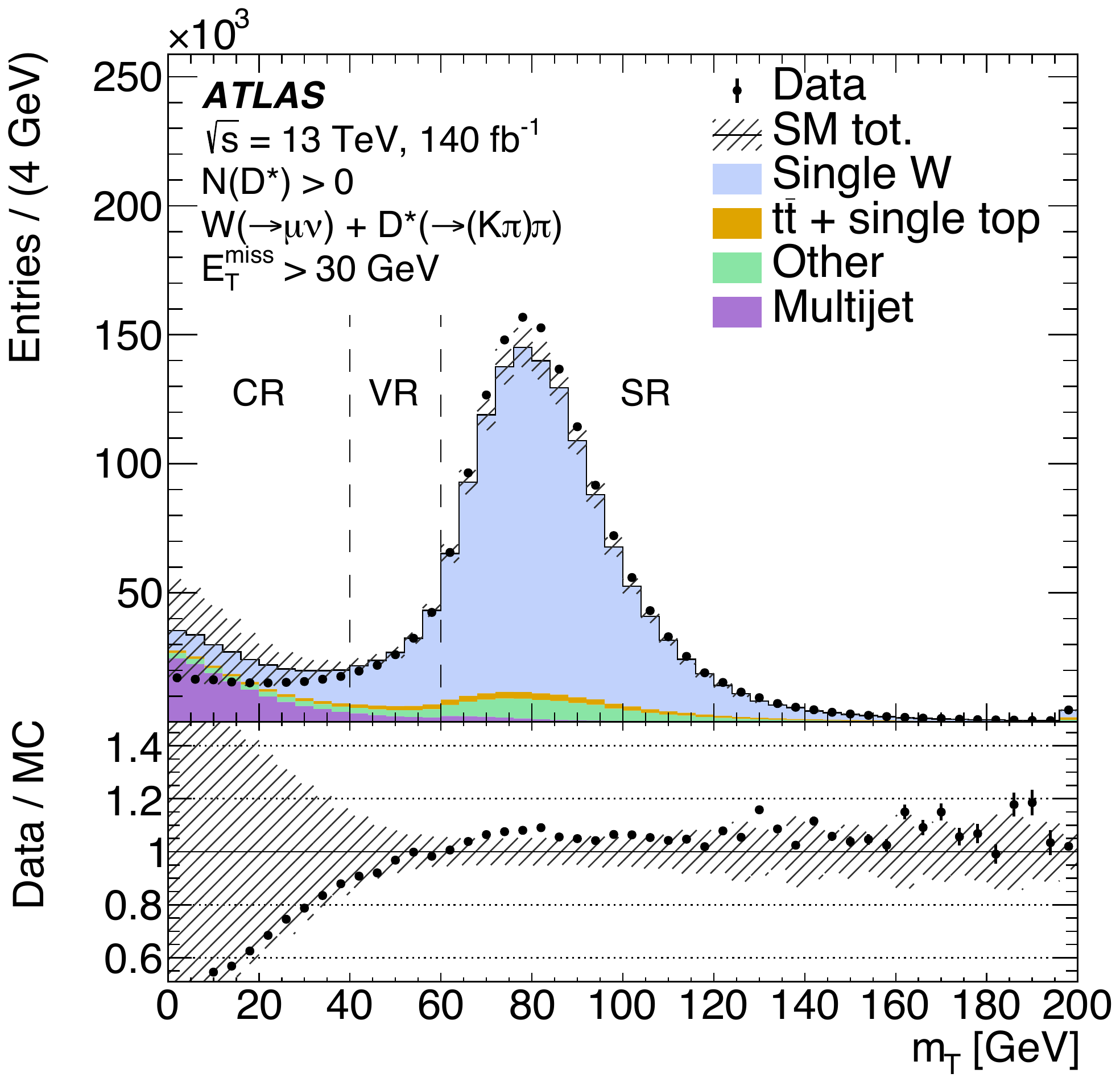}
\label{fig:matrix_method:dstar_mu_mt}
}
\caption{Modeling distributions of the \mT variable using the Matrix Method to estimate the \MJ background. The distributions are
\protect\subref{fig:matrix_method:dplus_el_mt} \mT in the \Dplus electron channel,
\protect\subref{fig:matrix_method:dplus_mu_mt} \mT in the \Dplus muon channel,
\protect\subref{fig:matrix_method:dstar_el_mt} \mT in the \Dstarp electron channel,
\protect\subref{fig:matrix_method:dstar_mu_mt} \mT in the \Dstarp muon channel.
The \enquote{SM Tot.} line represents the sum of all signal and background samples and the corresponding hatched
uncertainty band includes all Matrix Method systematic uncertainties, \MET systematic uncertainties, and QCD scale variations.
The \enquote{Single W} component includes all contributions from \Tab{\ref{tab:signal_bkg_templates}}. The $D$ and
$D^{*}$ stand for \Dplus and \Dstarp mesons respectively and pion and kaon charges are omitted for brevity. Dashed
vertical lines indicate the \mT values defining the control, validation, and signal regions (CR, VR, and SR) as explained
in the text. The last bin also includes the events with $\mT>\SI{200}{\GeV}$. The prompt processes estimated with MC
samples are normalized to the expected SM cross-sections given in \Sect{\ref{sec:samples:mc}}.
}
\label{fig:matrix_method:mt}
\end{figure}

\Fig{\ref{fig:matrix_method:mt}} demonstrates the extrapolation of the \MJ background from the \FakeCR to the \SR. Without the OS--SS subtraction,
most of the $D$ mesons in the Top background originate from $B$ meson decays. This background is larger in the \Dplus channel than in the \Dstar channel
because the slow pion in the \Dstar reconstruction chain is required to be associated with the PV and charmed mesons produced in $B$ meson
decays often fail this requirement due to the sizable average lifetime of the $B$ mesons. The central values
of the fake-lepton efficiencies are calculated in the $\mT < \SI{40}{\GeV}$ region, but with the \MET requirement inverted ($\MET < \SI{30}{\GeV}$).
The figure instead shows the events with the $\MET > \SI{30}{\GeV}$ requirement corresponding to the \SR selection. The prediction disagrees with
the data at low \mT due to an \MET dependence in the fake-lepton efficiencies that is not directly accounted for in the parameterization. A systematic
uncertainty is introduced, as described above, by calculating the fake-lepton efficiencies with the $\MET > \SI{30}{\GeV}$ requirement and taking the
full difference between the two \MJ predictions as the uncertainty. Since this is the largest systematic uncertainty in the \MJ background, the data
is almost exactly covered by the one-standard-deviation variation in this region. Furthermore, the \MJ prediction and the uncertainties are
extrapolated into the \SR with the $\mT > \SI{60}{\GeV}$ requirement. To validate the extrapolation, the prediction is evaluated in a validation region
(VR) with an \mT requirement of $\SI{40}{\GeV} < \mT < \SI{60}{\GeV}$. \Fig{\ref{fig:matrix_method:mt}} shows that the prediction in the VR is in agreement with
the data within the systematic uncertainties, indicating that the \MJ background is modeled well enough.
 
\section{Cross-section determination}
\label{sec:likelihood_fit}
 
A statistical fitting procedure based on the standard profile-likelihood formalism used in LHC
experiments~\cite{moneta2011roostats,RooFit} is used to extract the observables from the data with corresponding uncertainties:
\begin{itemize}
\item absolute fiducial cross-sections: \sigmaWmplusDmeson and \sigmaWpplusDmeson,
\item the cross-section ratio: $\Rc = \sigmaWpplusDmeson / \sigmaWmplusDmeson$,
\item differential cross-sections for OS-SS \WmplusDmeson and \WpplusDmeson.
\end{itemize}
 
The likelihood fit enables the estimation of background normalization and constraining of systematic
uncertainties in situ by extracting the information from the data in mass peak sidebands and control regions.
It is a crucial ingredient in achieving percent-level precision in the \WplusDmeson cross-section measurement.
The formalism of the profile likelihood fit is given in \Sect{\ref{sec:likelihood_fit:formalism}},
\Sect{\ref{sec:likelihood_fit:os-ss}} explains how the \enquote{OS--SS} subtraction is incorporated,
\Sect{\ref{sec:likelihood_fit:normalized}} introduces the measurement of normalized
differential cross-sections, and \Sect{\ref{sec:likelihood_fit:bins}} defines the bin edges of
the measured differential variables.
 
\subsection{The profile likelihood fit}
\label{sec:likelihood_fit:formalism}
 
A binned likelihood function, $\mathcal{L}(\vec{\sigma}, \vec{\theta})$, is constructed as the product of Poisson probability
terms for each bin of the input mass distributions, based on the number of data events and the expected signal and background yields.
The product over the mass bins is performed for each differential bin, in bins of either \diffpt or \diffeta.
The reconstructed invariant mass of the \Dplus meson, \mDplus, is used as input in the \Dplus channel and the mass difference \mdiff is
used in the \Dstar channel because it has better resolution than the \Dstar invariant mass. The invariant mass bins in the \SR
are narrower in the peak region (with about 8 bins) and wider in the tails, where the shape is more uniform
(up to 4 bins). Only a single bin is fitted in each \TopCR. The integrated \SR invariant mass distributions are shown in
\Fig{\ref{fig:fits:differential:OS-SS:0tag:postfit}} in \Sect{\ref{sec:results}}. The impact of systematic uncertainties is
included via nuisance parameters, $\vec{\theta}$. Separate likelihood fits are performed for the \Dplus and \Dstarp channels
and for \diffpt and \diffeta distributions. A likelihood equation describing this fitting procedure is given in
\Eqnrange{(\ref{eq:fits:differential})}{(\ref{eq:fits:differential:OS-SS})}:
 
\begin{linenomath}
\begin{equation}
\begin{aligned}
\mathcal{L}(\vec{\sigma}, \vec{\theta}) = \prod_{\alpha} \Bigg( \prod_i^{\Wboson^{-}\,\mathrm{OS}} \mathcal{L}(\vec{\sigma}, \vec{\theta})^{\alpha\,\mathrm{OS}}_{i} \times \prod_i^{\Wboson^{-}\,\mathrm{SS}} \mathcal{L}(\vec{\theta})^{\alpha\,\mathrm{SS}}_{i} \times \prod_i^{\Wboson^{+}\,\mathrm{OS}} \mathcal{L}(\vec{\sigma}, \vec{\theta})^{\alpha\,\mathrm{OS}}_{i} \times \prod_i^{\Wboson^{+}\,\mathrm{SS}} \mathcal{L}(\vec{\theta})^{\alpha\,\mathrm{SS}}_{i} \Bigg) \times \mathcal{L}^{\mathrm{constr.}},
\end{aligned}
\label{eq:fits:differential}
\end{equation}
\end{linenomath}
 
\begin{linenomath}
\begin{equation}
\begin{aligned}
\mathcal{L}(\vec{\sigma}, \vec{\theta})^{\alpha\,\mathrm{OS}}_{i} = f\Bigg(N_i^{\alpha}|\gamma_i^{\alpha} \cdot \bigg( \sum_{\beta} \big[ \sigma_{\mathrm{fid}}^{\beta} \cdot r^{\alpha \beta}(\vec{\theta}) \cdot \mathscr{P}^{\alpha \beta}_i(\vec{\theta}) \big] \cdot \mathscr{L}(\theta_{\mathrm{lumi}}) \cdot B_{\Dmeson} + \mathscr{B}_i^{\alpha}(\vec{\theta},\mu_{\mathrm{Top}}) \bigg) + \small{\mathscr{C}}_i^{\alpha} \Bigg),
\end{aligned}
\label{eq:fits:differential:OS}
\end{equation}
\end{linenomath}
 
\begin{linenomath}
\begin{equation}
\begin{aligned}
\mathcal{L}(\vec{\theta})^{\alpha\,\mathrm{SS}}_{i} = f\big(N_i^{\alpha}|\gamma_i^{\alpha} \cdot \mathscr{B}_i^{\alpha}(\vec{\theta},\mu_{\mathrm{Top}}) + \small{\mathscr{C}}_i^{\alpha} \big),
\end{aligned}
\label{eq:fits:differential:SS}
\end{equation}
\end{linenomath}
 
\begin{figure}[htbp]
\centering
\subfloat[]{
\includegraphics[width=0.50\textwidth]{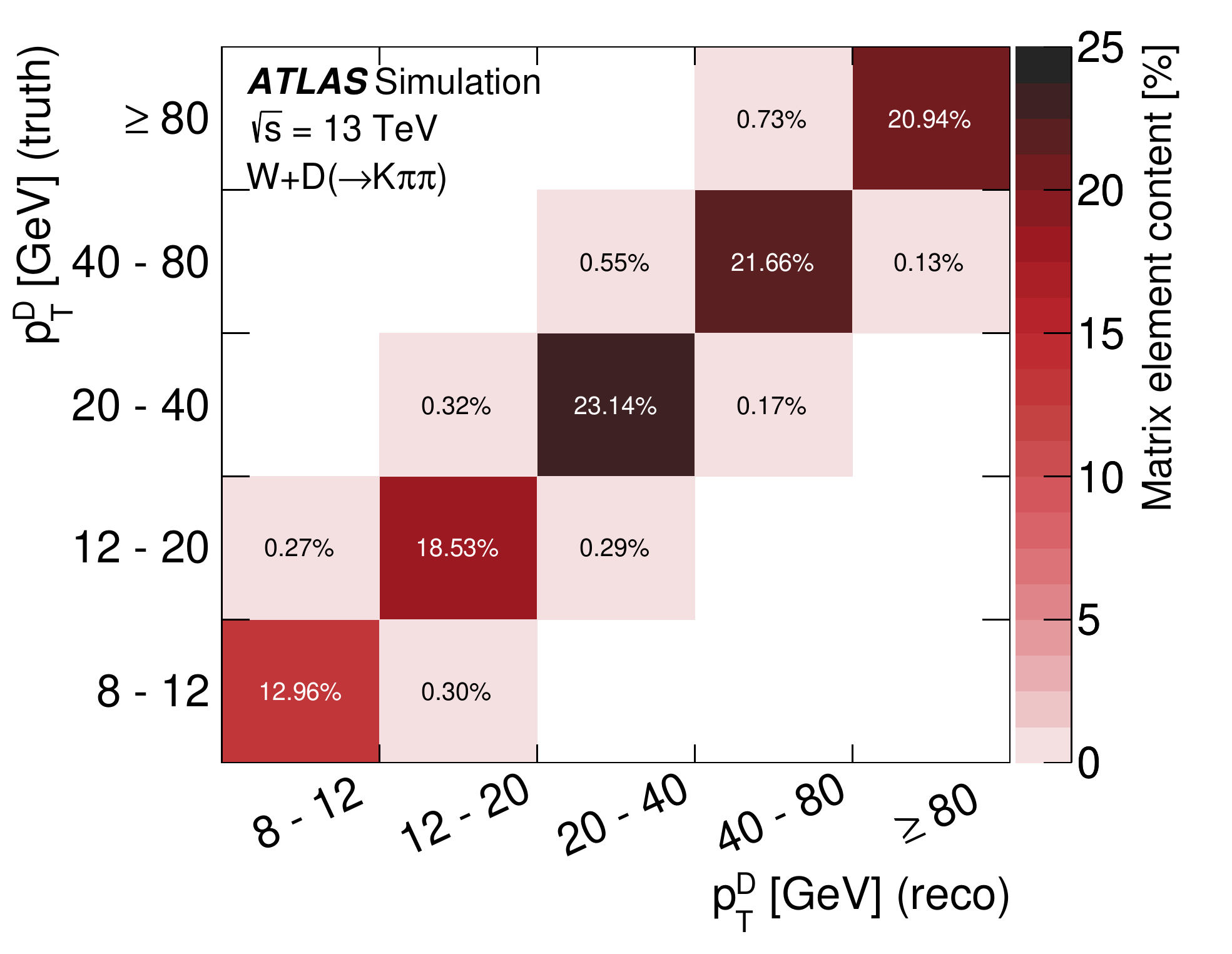}
\label{fig:fit:inputs:pt:OS-SS_Dplus_Kpipi_fid_eff_per_bin_norm}
}
\subfloat[]{
\includegraphics[width=0.50\textwidth]{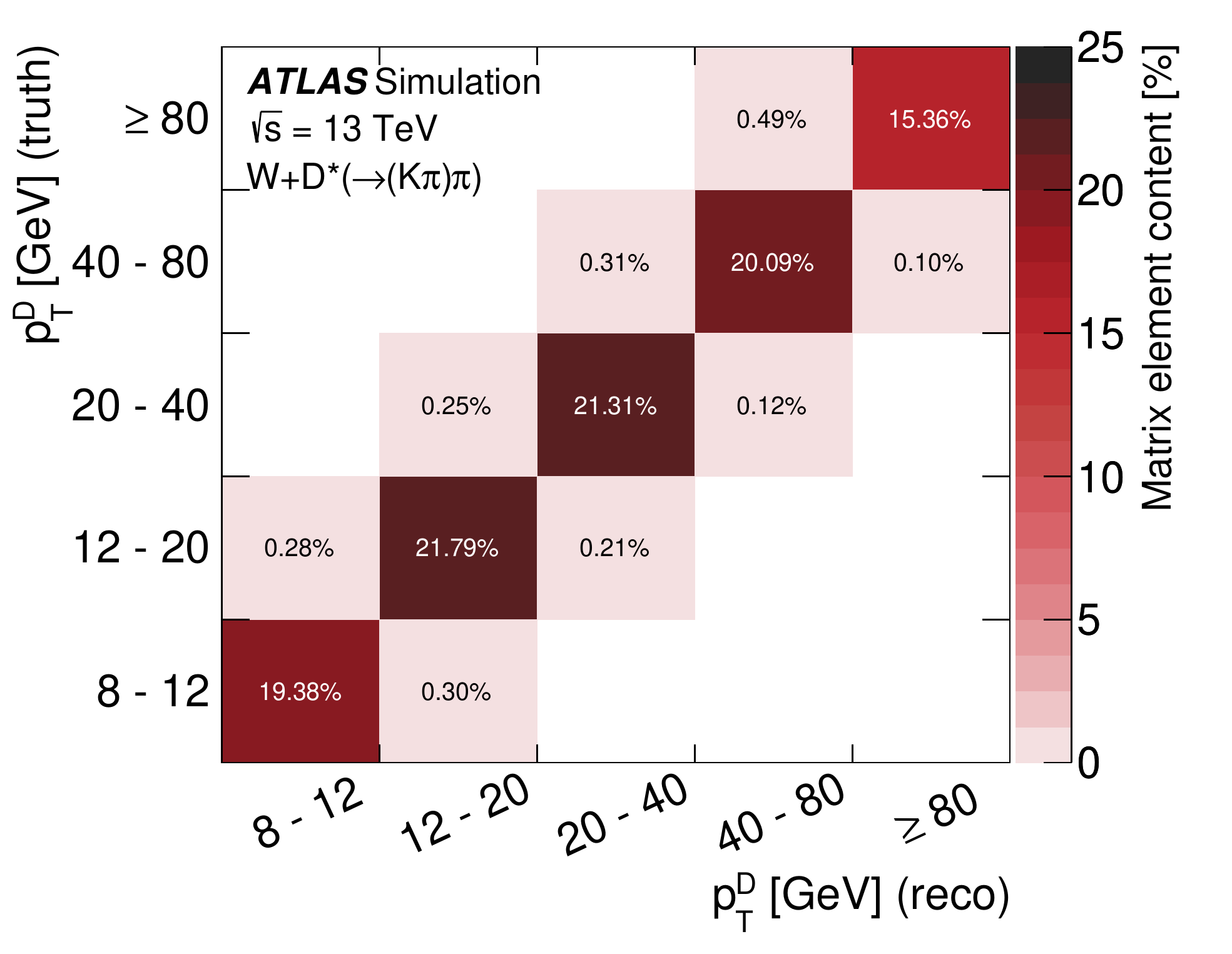}
\label{fig:fit:inputs:pt:OS-SS_Dstar_Kpipi_fid_eff_per_bin_norm}
}
\\
\subfloat[]{
\includegraphics[width=0.50\textwidth]{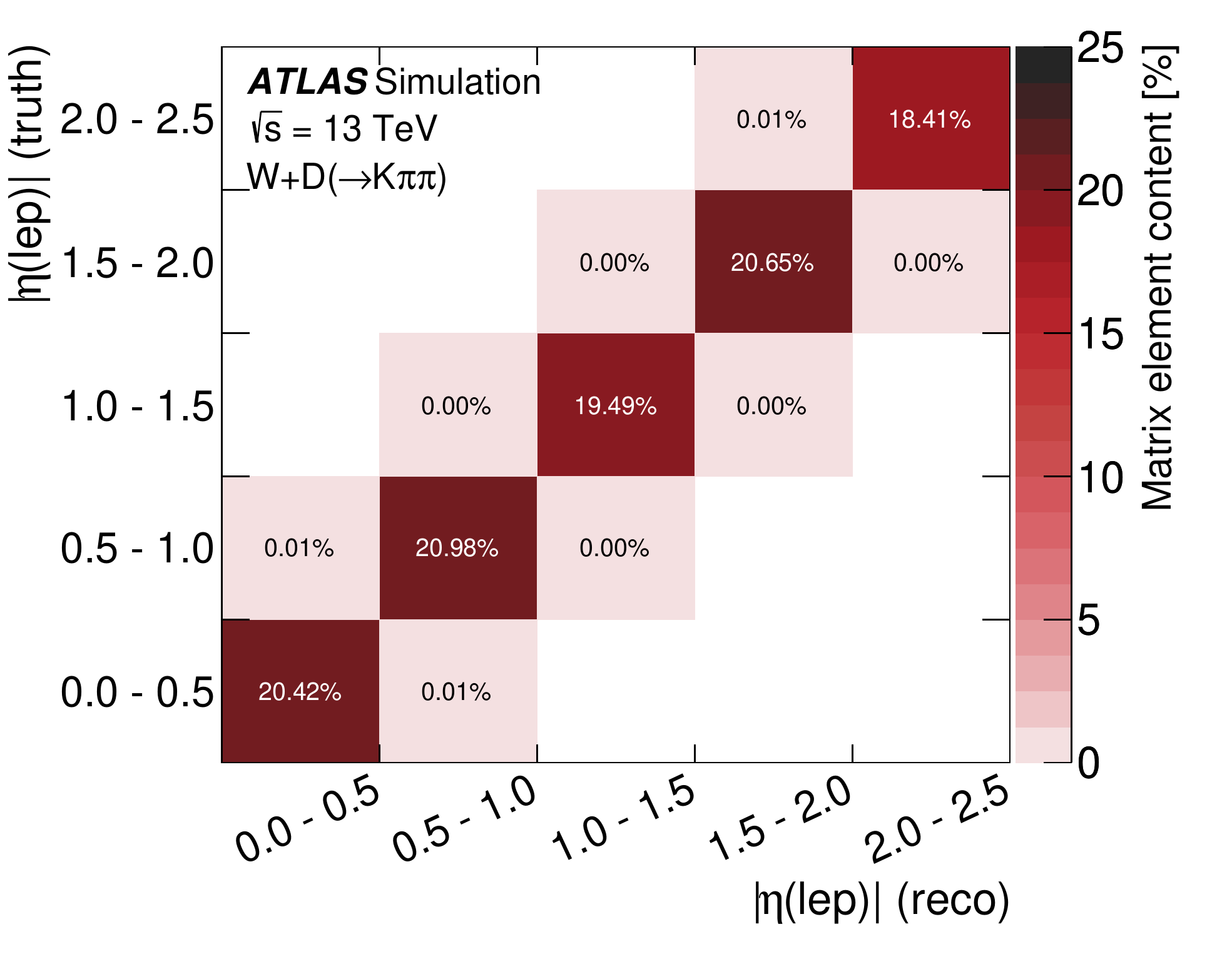}
\label{fig:fit:inputs:eta:OS-SS_Dplus_Kpipi_fid_eff_per_bin_norm}
}
\subfloat[]{
\includegraphics[width=0.50\textwidth]{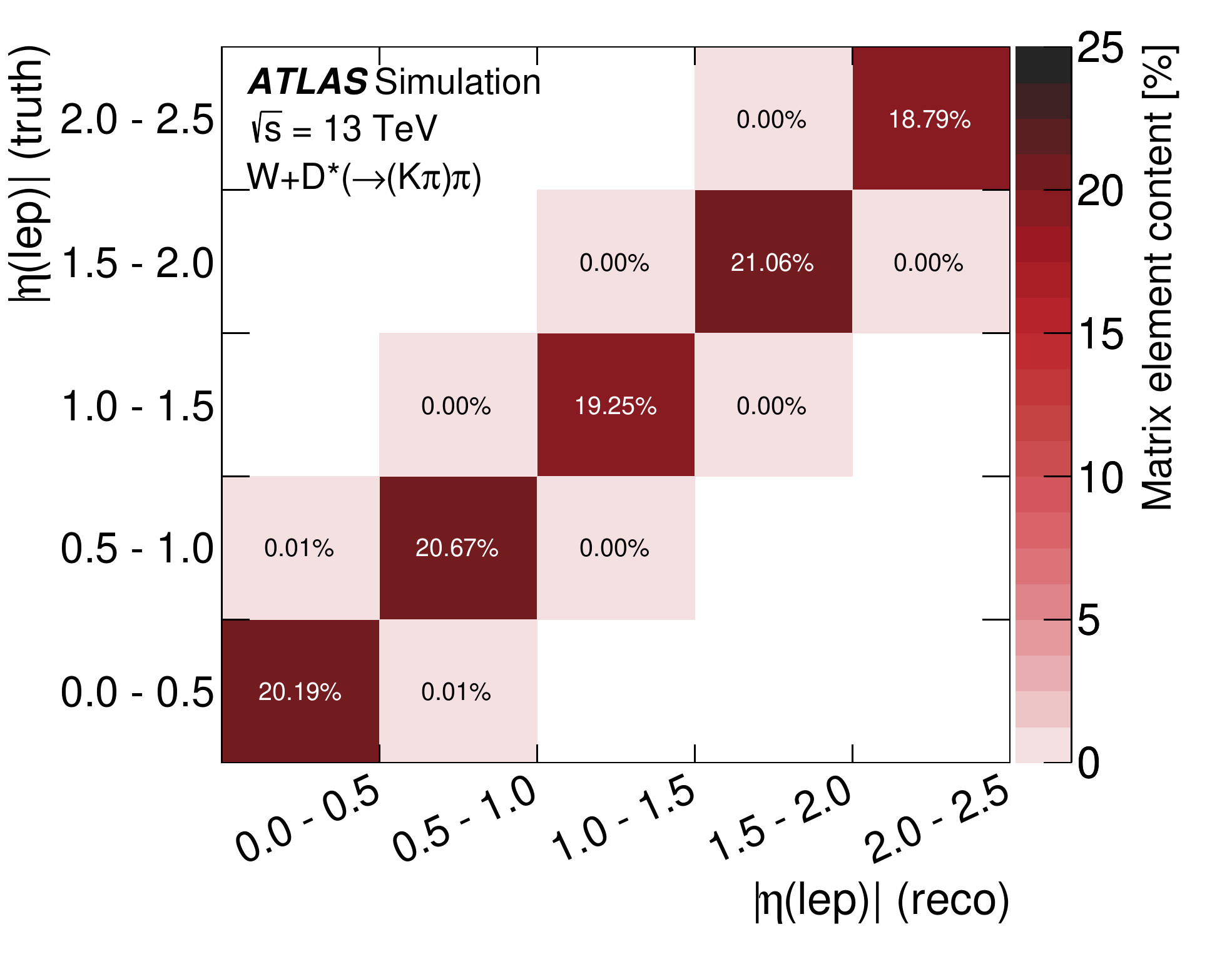}
\label{fig:fit:inputs:eta:OS-SS_Dstar_Kpipi_fid_eff_per_bin_norm}
}
\caption{
The \WplusDmeson detector response matrix in differential \diffpt bins:
\protect\subref{fig:fit:inputs:pt:OS-SS_Dplus_Kpipi_fid_eff_per_bin_norm} \WplusD,
\protect\subref{fig:fit:inputs:pt:OS-SS_Dstar_Kpipi_fid_eff_per_bin_norm} \WplusDstar, and
in differential \diffeta bins:
\protect\subref{fig:fit:inputs:eta:OS-SS_Dplus_Kpipi_fid_eff_per_bin_norm} \WplusD,
\protect\subref{fig:fit:inputs:eta:OS-SS_Dstar_Kpipi_fid_eff_per_bin_norm} \WplusDstar.
The detector response matrix is calculated with \SHERPA[2.2.11] \WplusDmeson samples.
The detector response matrices are normalized to unity such that the sum of all elements is \SI{100}{\percent}.
The last \diffpt bin has an upper cut of \SI{150}{\GeV} at the detector level, while there is no upper cut at the truth level.
}
\label{fig:fit:inputs:fid_eff_per_bin_norm}
\end{figure}

\begin{linenomath}
\begin{equation}
\begin{aligned}
\mathcal{L}^{\mathrm{constr.}} = \prod_t g(\theta_t) \times \prod_{\alpha,\,i} f(\gamma_i^{\alpha}),
\end{aligned}
\label{eq:fits:differential:OS-SS}
\end{equation}
\end{linenomath}
 
where the index $i$ represents the bins of the \Dmeson mass distribution (either OS or SS)
both in the $\nbjets = 0$ \SR, as well as the single bin used in the $\nbjets > 0$ \TopCR. Indices $\alpha$ and $\beta$ represent
the detector-level and truth differential bins respectively, and the index $t$ represents the nuisance parameters $\vec{\theta}$.
The expression $f(k | \lambda) = \lambda^{k} e^{-\lambda} / k!$ is the Poisson probability density function. Furthermore,
\begin{itemize}
\item $N_i^{\alpha}$ is the number of observed events in mass bin $i$ and reconstructed differential bin $\alpha$,
\item $\sigma_{\mathrm{fid}}^{\beta}$ is the fiducial cross-section in differential bin $\beta$ (one parameter per differential bin and \Wboson boson charge),
\item $r^{\alpha \beta}(\vec{\theta})$ is the detector response matrix, defined as the fraction of \WplusDmeson events produced in truth fiducial bin $\beta$
that also satisfy the \SR reconstruction criteria in bin $\alpha$,
\item $\mathscr{P}^{\alpha \beta}_i(\vec{\theta})$ is the $i$-th bin of the mass shape distribution of the signal sample corresponding to truth differential bin $\beta$ in reconstructed differential bin $\alpha$ (a separate invariant mass distribution for every non-zero bin in \Fig{\ref{fig:fit:inputs:fid_eff_per_bin_norm}})
\item $\mathscr{L}(\theta_{\mathrm{lumi}})$ is the integrated luminosity,
\item $B_{\Dmeson}$ is the branching ratio of either the \Dplus or \Dstarp decaying into $K\pi\pi$ (\Refn{\cite{Workman:2022ynf}}),
\item $\mathscr{B}^{\alpha}_i(\vec{\theta},\mu_{\mathrm{Top}})$ is the total number of background events in mass bin $i$ and reconstructed differential bin $\alpha$,
including the \WplusDmeson signal events failing the truth fiducial selection (\Tab{\ref{tab:event_selection:reco_selection:sel_table:truth}}),
\item $\mu_{\mathrm{Top}}$ is the normalization factor for the top quark background,
\item $\small{\mathscr{C}}_i^{\alpha}$ is the ``\commonfloat'' in mass bin $i$ and reconstructed differential bin $\alpha$ (mathematical construct to enable likelihood minimization in OS--SS, described further in \Sect{\ref{sec:likelihood_fit:os-ss}}),
\item $\vec{\theta}$ represents all nuisance parameters that are profiled in the likelihood fit,
\item $\gamma_i^{\alpha}$ parameters are the Poisson-constrained parameters accounting for the MC statistical
uncertainties in the combined signal-plus-background mass templates, following the simplified Beeston--Barlow technique~\cite{Barlow:1993dm}.
\end{itemize}

The nuisance parameters $\vec{\theta}$ have Gaussian constraints $g(\theta)$ in the likelihood with a mean of $0$ and a
standard deviation that corresponds to the one-standard-deviation variations of the associated systematic uncertainties,
determined from auxiliary measurements (e.g.\ lepton calibration described in \Sect{\ref{sec:object_selection}}). The $\gamma_i^{\alpha}$ parameters are centered
around $1$ and may deviate from unity within the corresponding Poisson constraints reflecting the combined
signal-plus-background statistical uncertainty in the invariant mass templates.
 
Response matrices for the \Dplus and \Dstarp channels are shown in \Fig{\ref{fig:fit:inputs:fid_eff_per_bin_norm}} for differential
\diffpt and \diffeta bins for nominal values of the nuisance parameters. Differential cross-sections extracted in this way correspond
to unfolding with matrix inversion. No regularization techniques were used because the detector response matrices are nearly
diagonal and because the statistical uncertainties are sufficiently low.
 
\subsection{The OS--SS subtraction}
\label{sec:likelihood_fit:os-ss}
A fitting procedure exploiting the charge correlation between the \Wboson boson and the \Dmeson meson was developed to
perform the OS--SS subtraction within the likelihood fit. Instead of using OS--SS distributions in the fit, both the OS and SS
regions enter the likelihood function and a \commonfloat is added in both regions. The additional component has one free parameter per invariant mass bin, and this parameter is correlated between the corresponding OS and SS regions. The \commonfloat is configured to absorb all charge-symmetric processes,
which effectively translates the maximization of separate OS and SS likelihoods into a maximization of the OS--SS
likelihood. This is done because the OS--SS event yields do not follow the Poisson distributions, which is a
requirement for the data yields in the profile likelihood fit. Furthermore, this fitting procedure ensures that the yields
of the individual signal and background components remain positive in the fit even though their OS--SS difference
could be negative.
 
\begin{figure}[htbp]
\centering
\subfloat[]{
\includegraphics[width=0.38\textwidth]{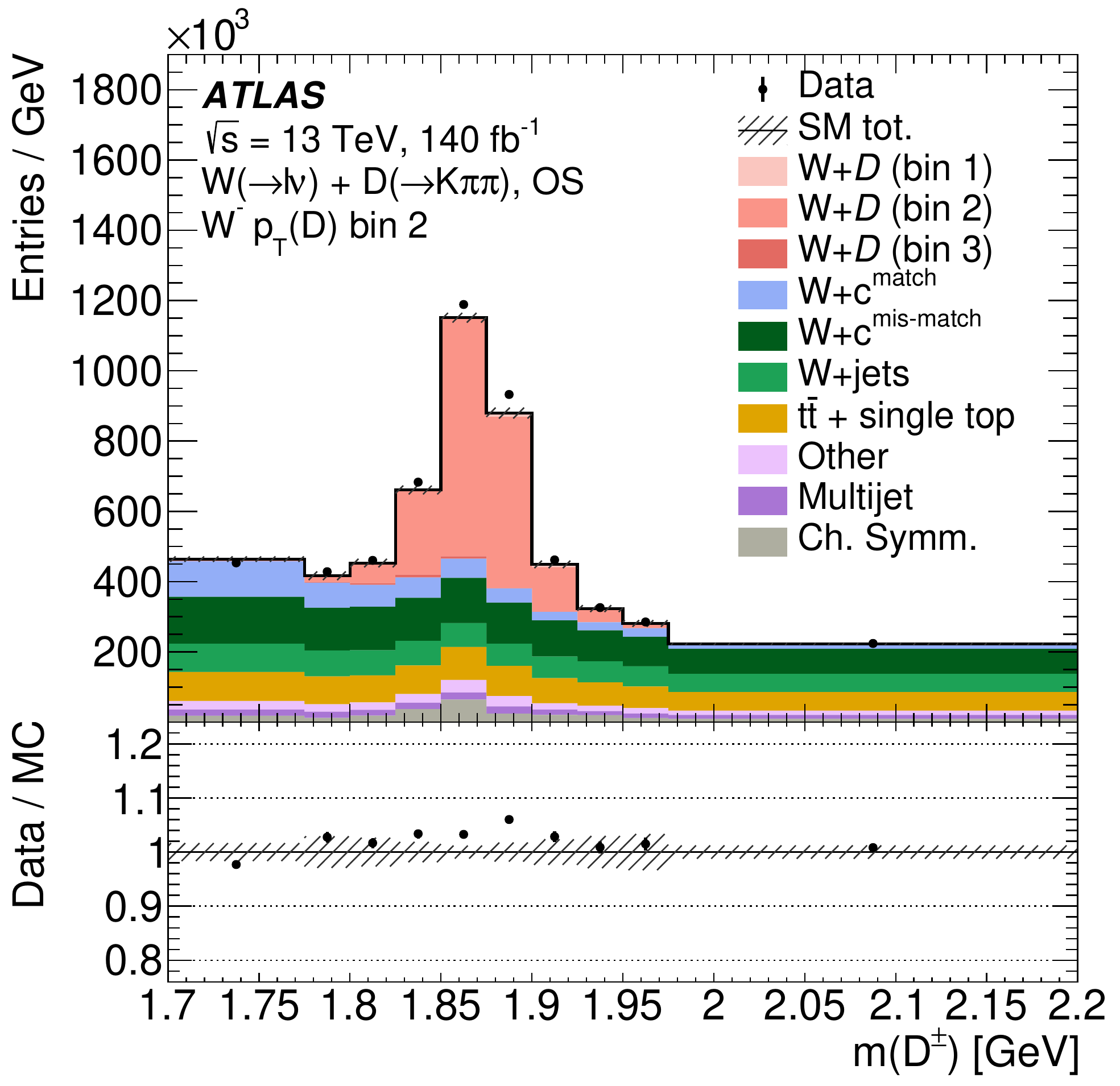}
\label{fig:fits:charge_symm:dplus:OS_lep_minus_0tag_Dplus_pt_bin2_Dmeson_m_fit}
}
\subfloat[]{
\includegraphics[width=0.38\textwidth]{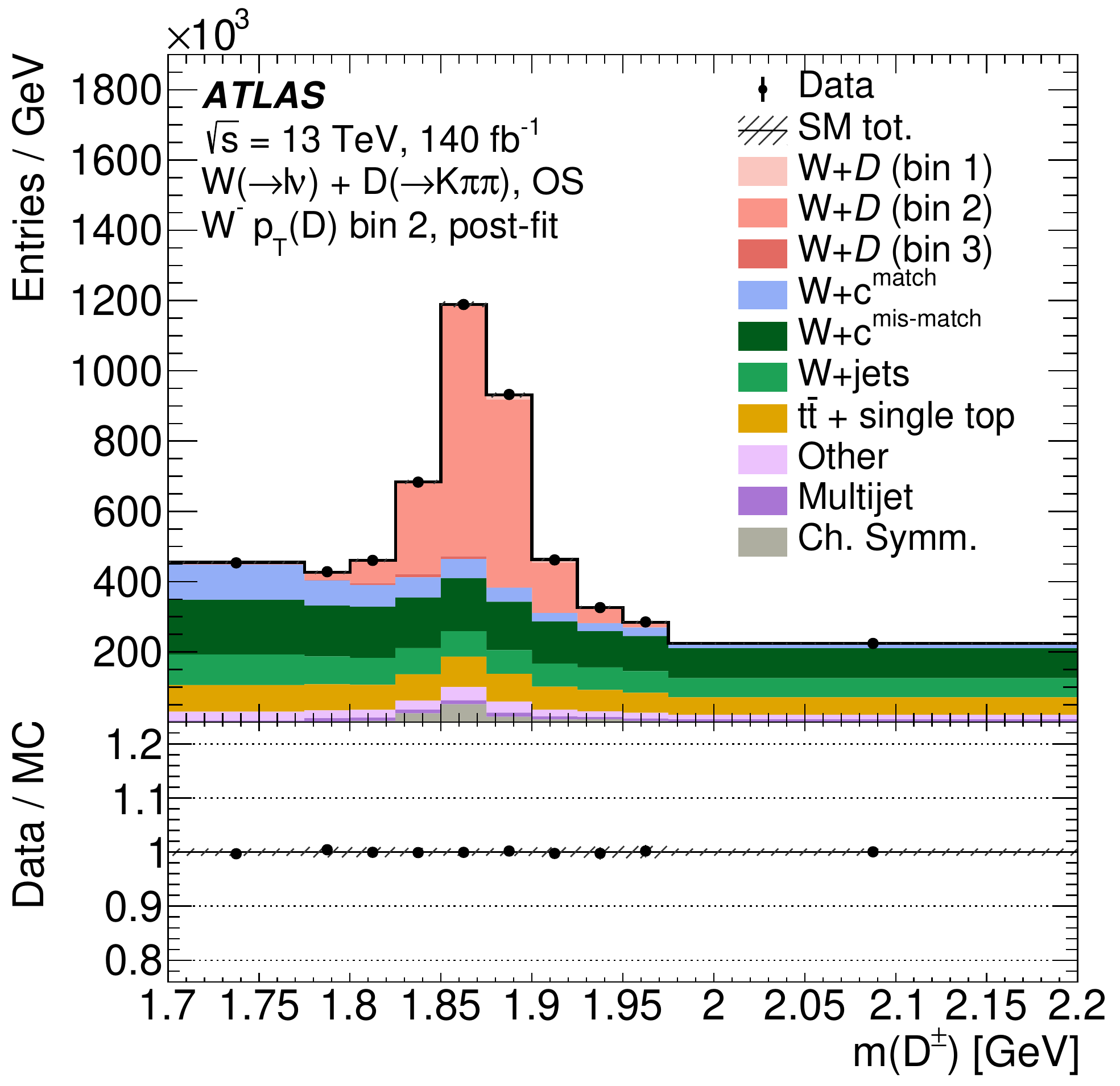}
\label{fig:fits:charge_symm:dplus:post_fit:OS_lep_minus_0tag_Dplus_pt_bin2_Dmeson_m_fit}
}
\\
\subfloat[]{
\includegraphics[width=0.38\textwidth]{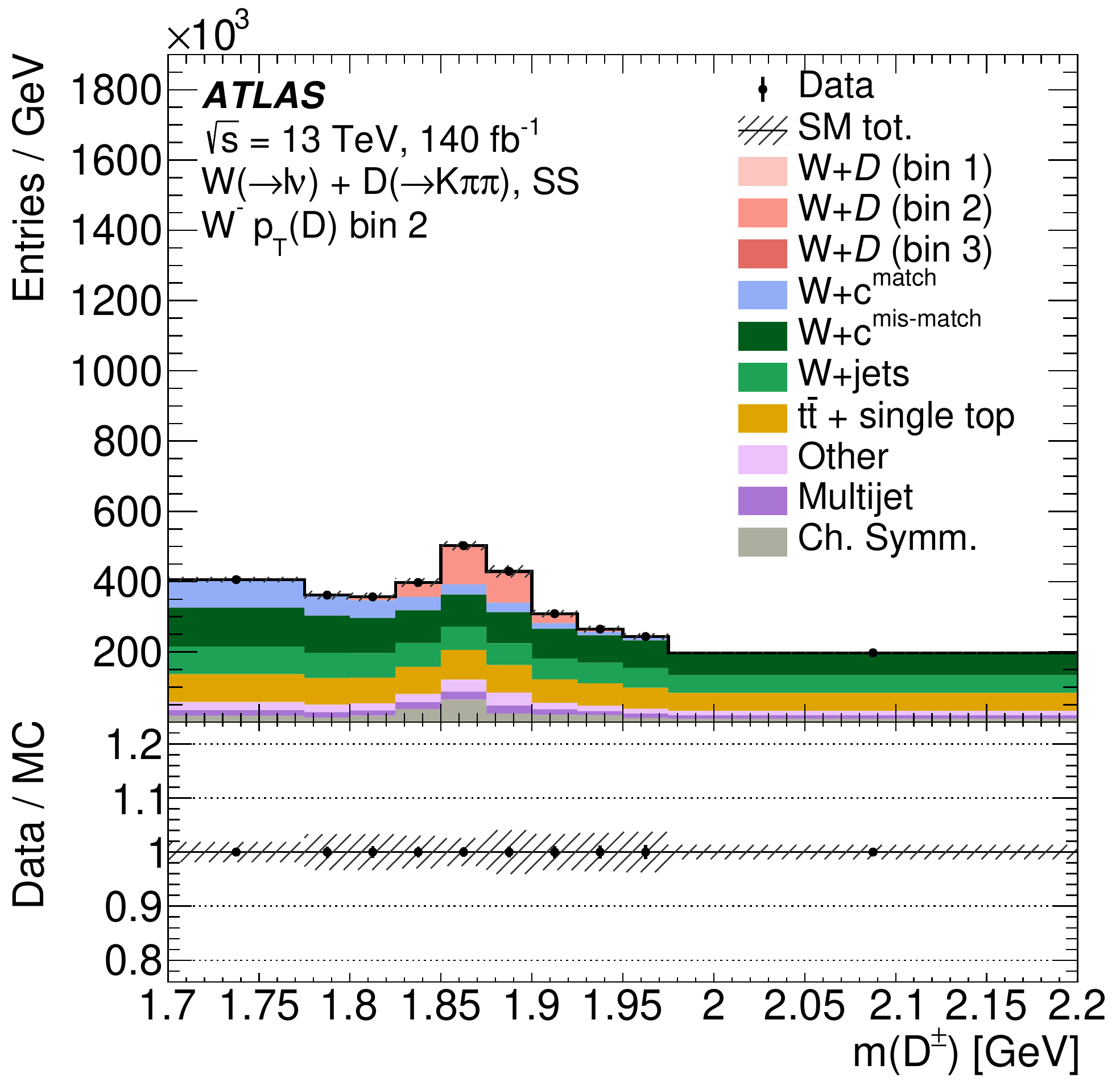}
\label{fig:fits:charge_symm:dplus:SS_lep_minus_0tag_Dplus_pt_bin2_Dmeson_m_fit}
}
\subfloat[]{
\includegraphics[width=0.38\textwidth]{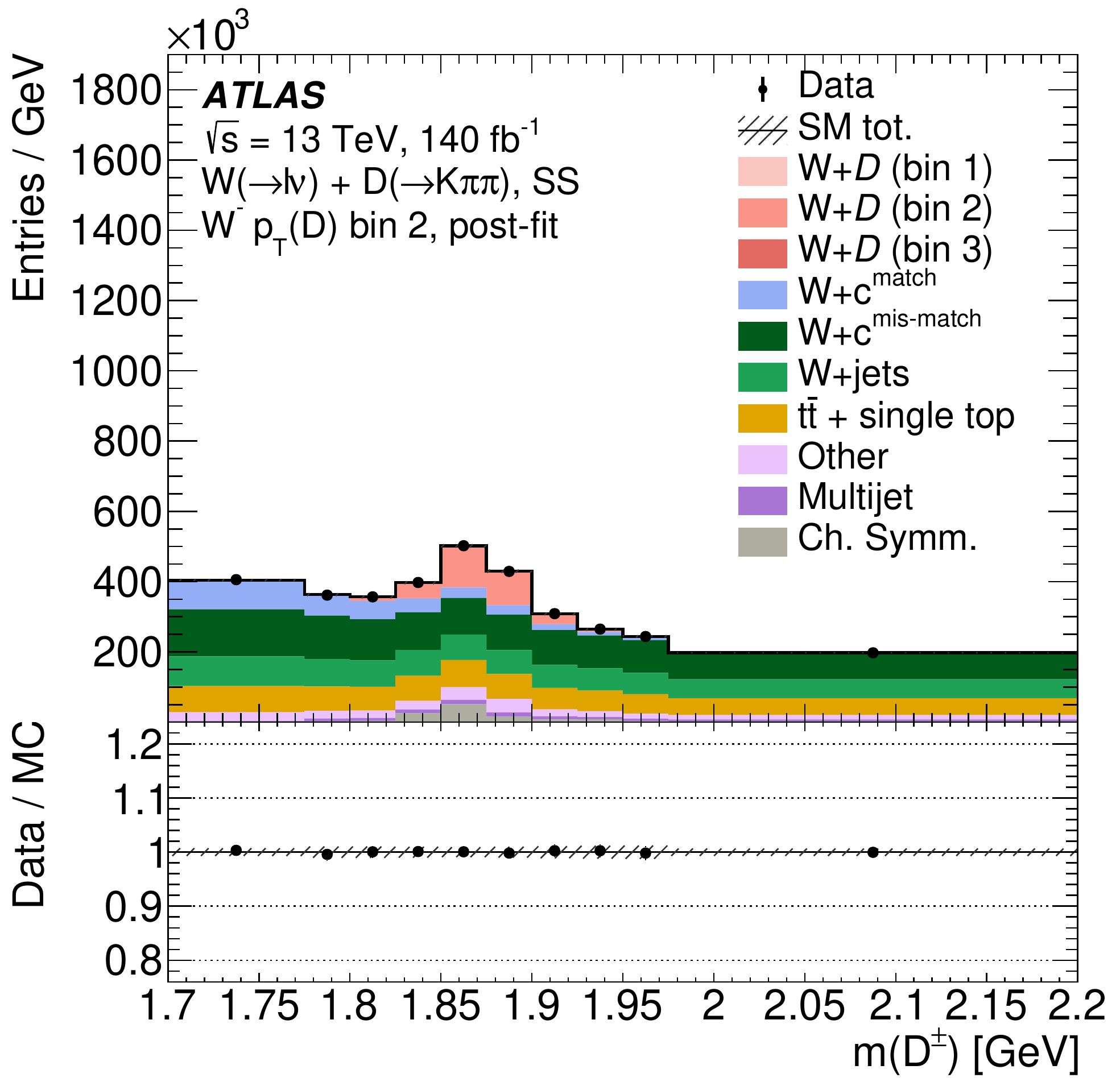}
\label{fig:fits:charge_symm:dplus:post_fit:SS_lep_minus_0tag_Dplus_pt_bin2_Dmeson_m_fit}
}
\\
\subfloat[]{
\includegraphics[width=0.38\textwidth]{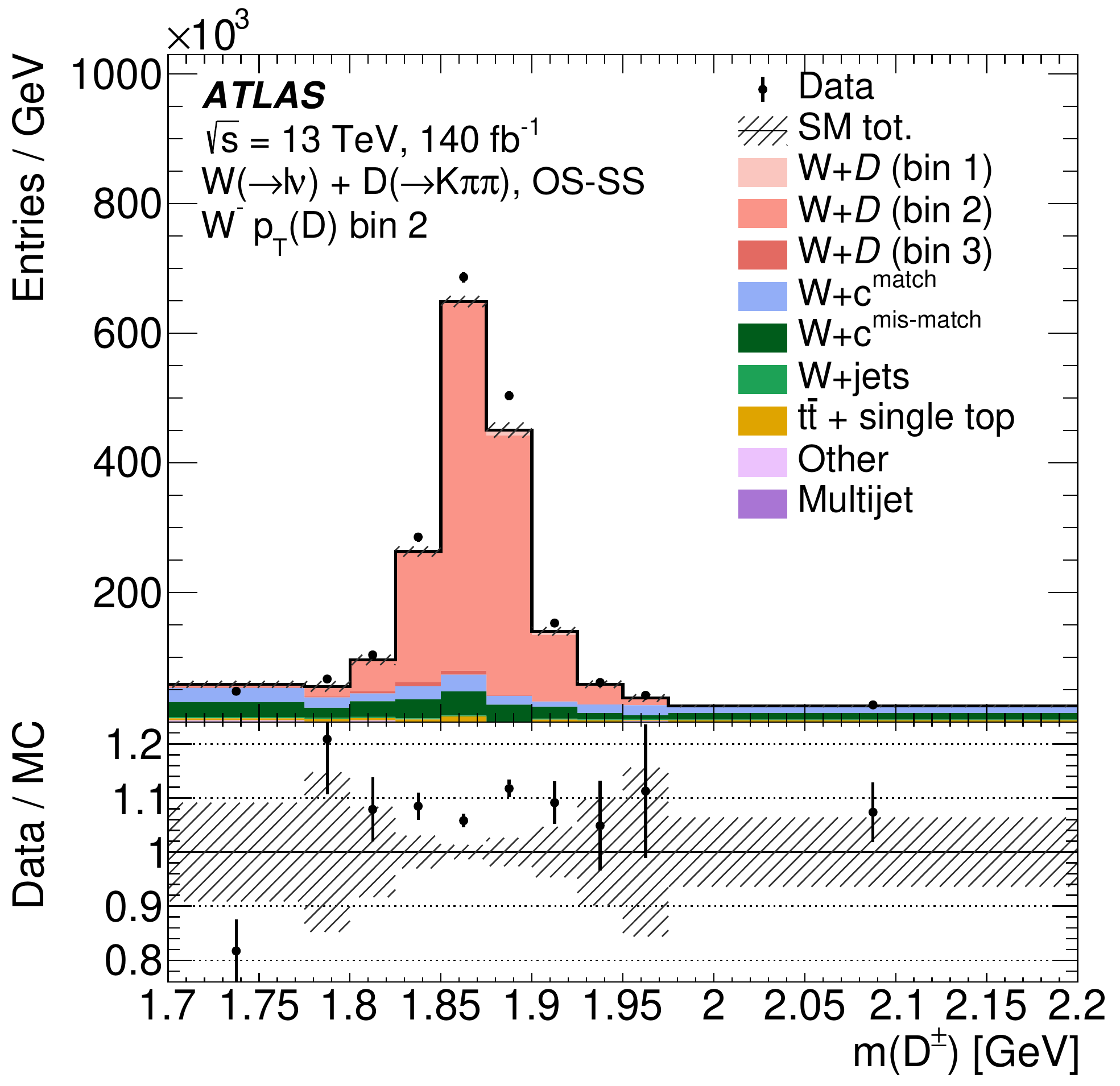}
\label{fig:fits:charge_symm:dplus:OS-SS_lep_minus_0tag_Dplus_pt_bin2_Dmeson_m_fit}
}
\subfloat[]{
\includegraphics[width=0.38\textwidth]{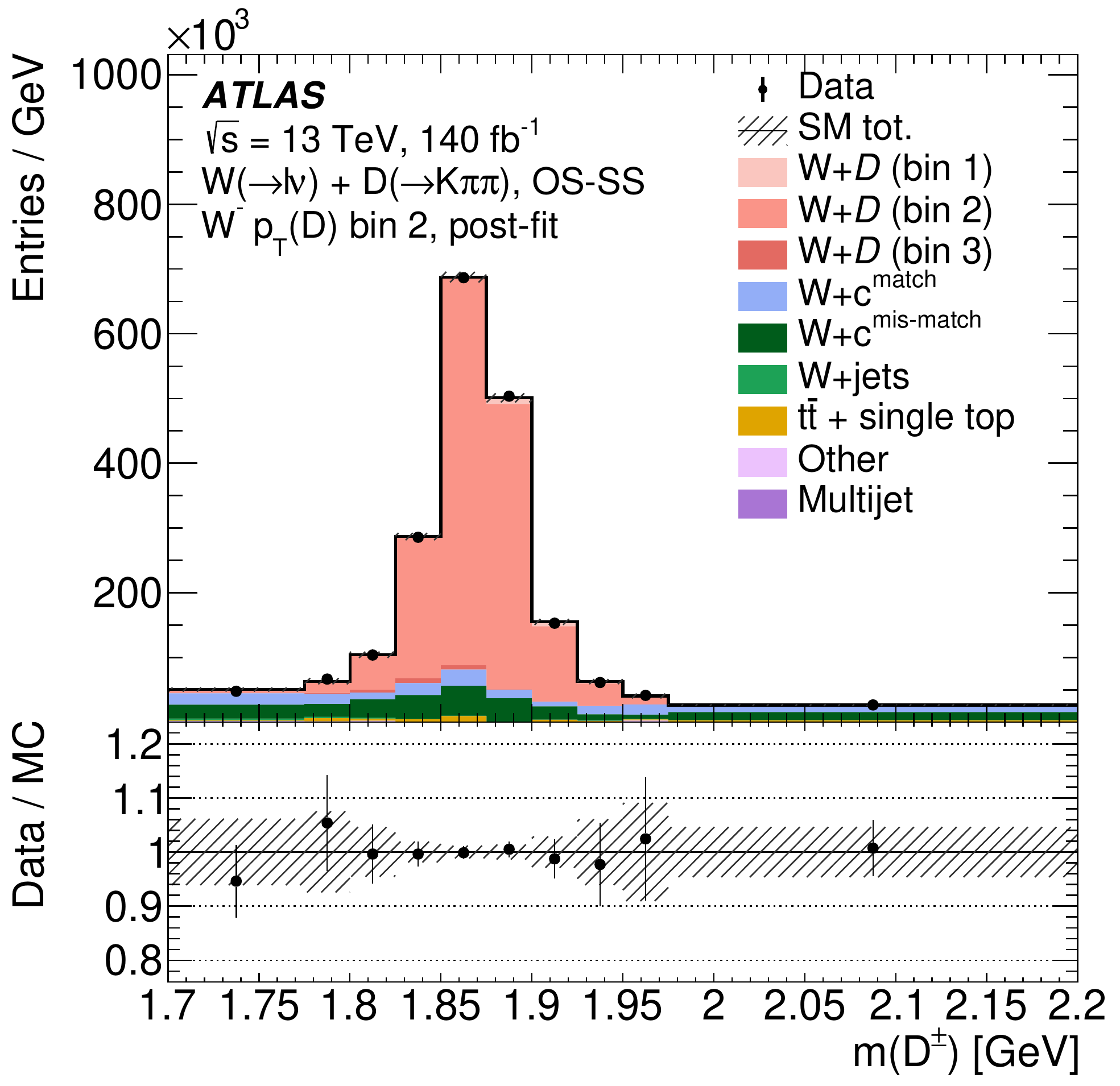}
\label{fig:fits:charge_symm:dplus:post_fit:OS_lep_minus_0tag_Dplus_pt_bin2_SS_lep_minus_0tag_Dplus_pt_bin2_Dmeson_m_fit}
}
\caption{
A demonstration of the OS--SS \WplusDmeson cross-section fit. Pre-fit $m(\Dplus)$ distributions
for the \WmplusD $\pT(\Dplus)$ bin 2:
\protect\subref{fig:fits:charge_symm:dplus:OS_lep_minus_0tag_Dplus_pt_bin2_Dmeson_m_fit} OS,
\protect\subref{fig:fits:charge_symm:dplus:SS_lep_minus_0tag_Dplus_pt_bin2_Dmeson_m_fit} SS, and
\protect\subref{fig:fits:charge_symm:dplus:OS-SS_lep_minus_0tag_Dplus_pt_bin2_Dmeson_m_fit} OS--SS.
The corresponding post-fit distributions:
\protect\subref{fig:fits:charge_symm:dplus:post_fit:OS_lep_minus_0tag_Dplus_pt_bin2_Dmeson_m_fit} OS,
\protect\subref{fig:fits:charge_symm:dplus:post_fit:SS_lep_minus_0tag_Dplus_pt_bin2_Dmeson_m_fit} SS, and
\protect\subref{fig:fits:charge_symm:dplus:post_fit:OS_lep_minus_0tag_Dplus_pt_bin2_SS_lep_minus_0tag_Dplus_pt_bin2_Dmeson_m_fit} OS--SS.
The \enquote{SM Tot.} line represents the sum of all signal and background samples.
The corresponding pre-fit uncertainty bands include MC statistical uncertainties only and
the post-fit uncertainty bands include the total uncertainty extracted from the fit. The gray histograms
represent the charge-symmetric common floating component and the three histograms associated with the signal
samples are the truth bins of the \diffpt differential distribution.
}
\label{fig:fits:charge_symm:OS-SS}
\end{figure}

The method used to extract the OS--SS \sigmaWplusD cross-section from a simultaneous fit to OS and SS regions with
the \commonfloat is demonstrated in \Fig{\ref{fig:fits:charge_symm:OS-SS}} for the second bin of the \diffpt distribution
in the \Dplus channel. The pre-fit OS, SS, and OS--SS distributions are shown at the left-hand side of \Fig{\ref{fig:fits:charge_symm:OS-SS}}
and the corresponding post-fit distributions are at the right-hand side. The \WplusD signal sample is split into three components
(labelled bin 1, bin 2 and bin 3), which corresponding to the diagonal and two off-diagonal elements immediately above and beneath the
diagonal in \Fig{\ref{fig:fit:inputs:pt:OS-SS_Dplus_Kpipi_fid_eff_per_bin_norm}}.
Since all other nondiagonal elements are zero, signal samples corresponding to truth fiducial bins 4 and 5 are not included.
The \commonfloat is shown with the gray histograms named \enquote{Ch.\ Symm.} in the legend.
The initial pre-fit values of the \commonfloat are arbitrary because every bin has a corresponding free parameter in the fit.
This component is merely a mathematical construct to translate the minimization of separate OS and SS negative log likelihoods into a minimization in OS--SS.
The initial values in both the OS and SS regions are set to the difference between the data and the MC prediction
in the SS region (different results were not observed with other initial values). This ensures that the initial signal-plus-background
predictions are positive and not too far away from the minimum. The plots illustrate the effectiveness of the OS--SS subtraction; the
backgrounds are almost symmetric in OS and SS regions, so the resulting OS--SS distributions are largely dominated by the \WplusDmeson signal.
 
\subsection{Normalized differential cross-section}
\label{sec:likelihood_fit:normalized}
 
Normalized differential cross-sections are generally more powerful than absolute differential cross-sections
in distinguishing between the observed data and the theory predictions since overall systematic uncertainties
such as those in the integrated luminosity and branching ratio cancel out in the normalized differential cross-sections.
To extract the normalized differential cross-sections and the corresponding uncertainties, the fit is performed to extract the
four normalized cross-sections and the total fiducial cross-section, $\sigma^{\mathrm{tot.}}_{\mathrm{fid}}$, instead of extracting
the five absolute differential cross-sections. By default, a substitution of the free parameters in the likelihood
fit is made as shown in \Eqn{(\ref{eq:fits:differential:normalized})}:
 
\begin{linenomath}
\begin{equation}
\begin{split}
\sigma^1_{\mathrm{fid}} & \to \sigma^{\mathrm{tot.}}_{\mathrm{fid}} \times \sigma^1_{\mathrm{rel}}, \\
\sigma^2_{\mathrm{fid}} & \to \sigma^{\mathrm{tot.}}_{\mathrm{fid}} \times \sigma^2_{\mathrm{rel}}, \\
\cdots & \\
\sigma^{N}_{\mathrm{fid}} & \to \sigma^{\mathrm{tot.}}_{\mathrm{fid}} \times \left[
1 - \sum_{i = 1}^{N-1} \sigma^i_{\mathrm{rel}}
\right], \\
\end{split}
\label{eq:fits:differential:normalized}
\end{equation}
\end{linenomath}
 
where $\sigma_{\mathrm{fid}}^{i}$ is the absolute fiducial cross-section in truth differential bin $i$
and $\sigma^{i}_{\mathrm{rel}}$ is the corresponding normalized differential cross-section. The value of $N$
is five in all cases. By definition, the
sum of all normalized differential cross-sections is one. This substitution is performed separately for each charge,
\WpplusDmeson and \WmplusDmeson. Furthermore, a similar substitution is made for the \Rc parameter.
The normalization factor for the \sigmaWpplusDmeson total fiducial cross-section is replaced by the expression
shown in \Eqn{(\ref{eq:fits:differential:rc})}:
 
\begin{linenomath}
\begin{equation}
\sigma^{\mathrm{tot.}}_{\mathrm{fid}}(\WpplusDmeson) \to \Rc \times \sigma^{\mathrm{tot.}}_{\mathrm{fid}}(\WmplusDmeson),
\label{eq:fits:differential:rc}
\end{equation}
\end{linenomath}
 
The free parameters in the fit after these substitutions are $\sigma^{1}_{\mathrm{rel}}$, $\dotsc$,
$\sigma^{N-1}_{\mathrm{rel}}$ for each charge (8 parameters in total),
\Rc, and $\sigma^{\mathrm{tot.}}_{\mathrm{fid}}(\WmplusDmeson)$. The central values of all additional
observables can be deduced from these free parameters; however, systematic uncertainties can only
be calculated for the parameters directly included in the fit (with a likelihood scan, explained in
\Sect{\ref{sec:sys:eval}}). To achieve this, several fits with different substitutions of parameters
are performed:
\begin{enumerate}
\item $\sigma^{1}$(\WmplusDmeson), $\dotsc$, $\sigma^{N}(\WmplusDmeson)$, $\sigma^{1}(\WpplusDmeson)$, $\dotsc$, $\sigma^{N}(\WpplusDmeson)$,
\item $\sigma^{1}_{\mathrm{rel}}(\WmplusDmeson)$, $\dotsc$, $\sigma^{N-1}_{\mathrm{rel}}(\WmplusDmeson)$, $\sigma^{1}_{\mathrm{rel}}(\WpplusDmeson)$, $\dotsc$, $\sigma^{N-1}_{\mathrm{rel}}(\WpplusDmeson)$, \Rc, $\sigma^{\mathrm{tot.}}_{\mathrm{fid}}(\WmplusDmeson)$,
\item $\sigma^{2}_{\mathrm{rel}}(\WmplusDmeson)$, $\dotsc$, $\sigma^{N}_{\mathrm{rel}}(\WmplusDmeson)$, $\sigma^{2}_{\mathrm{rel}}(\WpplusDmeson)$, $\dotsc$, $\sigma^{N}_{\mathrm{rel}}(\WpplusDmeson)$, $\sigma^{\mathrm{tot.}}_{\mathrm{fid}}(\WpplusDmeson)$, \Rc.
\end{enumerate}
 
These three fits allow a precise determination of the central values and systematic uncertainties
of all observables, including absolute and normalized differential cross-sections. In all cases the
number of free parameters is the same and the minimization procedure reaches the same minimum, yielding
identical results.
 
\subsection{Differential cross-section bins}
\label{sec:likelihood_fit:bins}
The bin edges of the five differential \diffpt
bins are given in \Tab{\ref{tab:fit_inputs:diff_bins}}. The last bin starts at \SI{80}{\GeV} and has no upper limit.
The number of bins and the bin edges were chosen such that the expected data statistical uncertainty is about
1\%--2\% in the first four bins. The available MC sample sizes also play an important role
in determining the bin size; up to a \SI{1}{\percent} statistical uncertainty is present in the diagonal
elements of the detector response matrix. Similarly to the \diffpt fits, five bins are chosen in \diffeta
to provide percent-level precision. Furthermore, the absolute value of the pseudorapidity is used to
further reduce the statistical uncertainty because there is no additional discriminating power in measuring
the sign of the pseudorapidity. The \diffeta bin edges are also given in \Tab{\ref{tab:fit_inputs:diff_bins}}.
 
With five differential bins per \Wboson boson charge there are 10 differential cross-sections in
total represented with the free parameters $\vec{\sigma}$ in the likelihood fit. Regions with
both charges of the \Wboson boson are included in the fit at the same time in order to extract the cross-section
ratio $\Rc$. SRs in the $\nbjets = 0$ category are split between the two \Wboson charges, into OS and SS events
and into the five differential bins: $[\Wboson^-,\,\Wboson^+] \times [\mathrm{OS},\,\mathrm{SS}] \times \mathrm{5} = 20$ regions.
The $\nbjets > 0$ CRs are split in the same way with the exception of differential bins since the normalization of
the backgrounds from top-quark production is extracted from the data only inclusively. The relative contribution
of the top-quark background in the \SR is small (about \SI{5}{\percent} of the signal yield for \Dplus and negligible for \Dstar),
so the modeling of the differential spectrum in the top-quark background simulation has a negligible impact on the result.
The regions used in the fit are summarized in \Tab{\ref{tab:fits:differential:regions}}.
 
\begin{table}[htpb]
\caption{The differential \diffpt and \diffeta bins used in the measurement.
The last \diffpt bin has no upper limit.}
\label{tab:fit_inputs:diff_bins}
\renewcommand{\arraystretch}{1.3}
\centering
\setlength\tabcolsep{5.2pt}
\begin{tabular}{
l | r r r r r
}
\toprule
Bin number & 1  & 2  & 3  & 4  & 5 \\
\midrule
\diffpt bin edges [\GeV]  & [8, 12]  & [12, 20] & [20, 40] & [40, 80] & [80, $\infty$)  \\
\diffeta bin edges & [0.0, 0.5] & [0.5, 1.0] & [1.0, 1.5] & [1.5, 2.0] & [2.0, 2.5]  \\
\bottomrule
\end{tabular}
\end{table}
 
\begin{table}[htpb]
\caption{A schematic of the signal and control regions (SR and CR) used in the fit.
The bin numbers correspond to either the \diffpt or \diffeta differential
bins listed in \Tab{\ref{tab:fit_inputs:diff_bins}}. The table indicates that the
invariant mass distribution is fitted in each \SR, with $m(\Dmeson)$ standing for \mDplus in the
\Dplus channel and \mdiff in the \Dstar channel, while only a single bin is fitted in the \TopCR.}
\label{tab:fits:differential:regions}
\renewcommand{\arraystretch}{1.3}
\setlength\tabcolsep{5.2pt}
\centering
\begin{tabular}{
l | *{7}{>{\centering\arraybackslash}p{2em}|} p{2em}
}
\toprule
& \multicolumn{4}{c|}{\SR ($\nbjets = 0$)} & \multicolumn{4}{c}{\TopCR ($\nbjets > 0$)} \\
\midrule
\Wboson charge & \multicolumn{2}{c|}{$\Wboson^-$} & \multicolumn{2}{c|}{$\Wboson^+$} & \multicolumn{2}{c|}{$\Wboson^-$} & \multicolumn{2}{c}{$\Wboson^+$} \\
\midrule
\Dmeson charge & OS & SS                          & OS & SS                          & OS & SS                          & OS & SS                         \\
\hline
Bin 1          & \multicolumn{4}{c|}{\multirow{5}{*}{Fit the $m(\Dmeson)$ distribution}} & \multicolumn{4}{c}{\multirow{5}{*}{Fit total yield}} \\
\cline{1-1}
Bin 2          & \multicolumn{4}{c|}{} & \multicolumn{4}{c}{} \\
\cline{1-1}
Bin 3          & \multicolumn{4}{c|}{} & \multicolumn{4}{c}{} \\
\cline{1-1}
Bin 4          & \multicolumn{4}{c|}{} & \multicolumn{4}{c}{} \\
\cline{1-1}
Bin 5          & \multicolumn{4}{c|}{} & \multicolumn{4}{c}{} \\
\bottomrule
\end{tabular}
\end{table}


\section{Systematic uncertainties}
\label{sec:sys}
 
The measurements in this analysis are affected by several sources of systematic uncertainty.
The first category, related to detector-interaction and reconstruction processes, includes uncertainties
in lepton and jet reconstruction, energy resolution, and energy scale, in lepton identification, isolation, and
trigger efficiencies, in $b$-jet tagging efficiencies, and in the total integrated luminosity
and pileup reweighting. These uncertainties affect the \WplusDmeson signal efficiency by altering
the detector response matrix, yields of the background processes estimated with MC simulation, and
the signal and background invariant mass templates used in the profile likelihood. These uncertainties are
correlated between all samples and regions in the likelihood fit and are generally derived from
auxiliary measurements:
 
\textbf{Charged leptons:} Electron and muon reconstruction, isolation, identification, and trigger efficiencies,
and the energy/momentum scale and resolution are derived from data using large samples of $\jpsi \to \ell \ell$
and $Z \to \ell \ell$ events~\cite{EGAM-2018-01,PERF-2015-10}. Systematic variations of the MC efficiency
corrections and energy/momentum calibrations applied to MC samples are used to estimate the signal selection uncertainties.
 
\textbf{Jets and missing transverse momentum:} Jet energy scale and energy resolution uncertainties affect the signal efficiency
and background yields indirectly by altering the reconstructed \MET in the event and hence
the selection efficiency of the \MET and \mT cuts. Systematic variations of the jet energy calibration
are applied to MC samples to estimate signal section uncertainties using the methodology described in \Refn{\cite{JETM-2018-05}}.
In total, there are 20 independent jet energy scale variations and 8 independent jet energy
resolution variations. None of the single variations have an impact of more than \SI{1}{\percent} on
the signal selection efficiency. Similarly, variations in \MET reconstruction are derived specifically
for the soft-term estimation following the methodology in \Refn{\cite{ATLAS-CONF-2018-023}}. Furthermore,
a single nuisance parameter is included to model the uncertainty in the JVT selection efficiency.
 
\textbf{Flavor tagging:} The uncertainty in the calibration of the $b$-tagging efficiencies and mis-tag rates
is derived from data using samples of dileptonic \ttbar events for $b$-jets and $c$-jets~\cite{FTAG-2018-01,FTAG-2020-08}
and a data sample enriched in light-flavor jets for light-jets~\cite{ATLAS:2023lwk}. Since
the majority (${>}\SI{99}{\percent}$) of \WplusDmeson signal events have no additional $b$-tagged jets,
these variations have a negligible impact on the signal efficiency. Nevertheless, the variations
in $b$-tagging efficiency have an impact of up to \SI{10}{\percent} on the relative yields of the
top quark backgrounds in the \SR and \TopCR.
 
\textbf{\Pileup and luminosity:} The uncertainty in the integrated luminosity is \SI{0.83}{\percent}~\cite{DAPR-2021-01},
which is obtained using the LUCID-2 detector~\cite{LUCID2} for the primary luminosity measurements.
MC samples are reweighted to have the number of \pileup vertices match the \pileup distribution measured
in the \RunTwo data. To account for the uncertainty in the \pileup estimation, variations of the reweighting
are applied to the MC samples. In addition to affecting the background yields, it also has a small impact on
the resolution of the reconstructed \Dplus meson mass peak and the \mdiff mass difference.
 
\textbf{SV reconstruction:}
Uncertainties in the secondary-vertex reconstruction efficiency arise from potential mismodeling of the amount and location of ID material,
from the modeling of hadronic interactions in \GEANT and from possible differences between the impact parameter
resolutions in data and MC events.  These uncertainties are evaluated by generating large single-particle samples of \Dplus and \Dstarp
decays with the same \pT and $\eta$ distributions as the baseline \WplusDmeson MC samples. These \enquote{single-particle gun} (SPG) samples are
simulated multiple times with different simulation parameters, mirroring the procedure in \Refn{\cite{ATL-PHYS-PUB-2015-051}}:
passive material in the whole ID scaled up by \SI{5}{\percent}, passive material in the IBL scaled up by \SI{10}{\percent},
and passive material in the Pixel detector services scaled by \SI{25}{\percent}. In addition to the variations in the amount of detector material,
a SPG sample where the physics model in the \GEANT toolkit was changed to QGSP\_BIC from FTFP\_BERT~\cite{Agostinelli:2002hh}
was generated.
 
The impact of the uncertainty in the ID material distribution is evaluated by comparing the efficiency obtained using the
baseline simulation and that obtained using altered material distributions. For each variation the relative change in the
\Dmeson reconstruction efficiency is parameterized as a function of
$\pT(\Dmeson)$ and $\eta(\Dmeson)$ separately for positive and negative charges of the mesons and separately for \Dplus and \Dstarp mesons.
The impact of changing the physics model was found to be negligible. The relative change in the reconstruction
efficiency due to the increased amount of the ID material was found to vary by 1\%--4\%. The
uncertainty is largest for low $\pT(\Dmeson)$ and high $\eta(\Dmeson)$. Because \Dmeson candidates in the signal and in the \ttbar
background do not necessarily have the same $\pT(\Dmeson)$ spectrum, their tracking efficiency NPs are treated as separate parameters
to minimize the correlation between them. The \ttbar background has large yields in the \TopCR and could affect the shape of the
$\pT(\Dmeson)$ signal distribution via pulls in the tracking efficiency uncertainties. The measured cross-sections would change by
up to \SI{1.0}{\percent} if the parameters were correlated, but this difference is covered by the associated systematic uncertainties.
 
Furthermore, the effect of the ID
material variations on the shape of the \Dplus invariant mass peak and the \mdiff mass difference is evaluated
by fitting the mass distributions with a double-sided Crystal Ball function, with yield modeling decoupled from the peak position.
The width and position of the peak are characterized with the width and mean of the central Gaussian distribution respectively.
The shift in the position of the \Dplus (\Dstarp) peak was found to be up to \SI{0.2}{\MeV} (\SI{0.05}{\MeV}).
The impact on the resolution of the peak was evaluated from the difference between the squares of the nominal width and the width obtained from
each variation. The resolution was found to be smeared by up to \SI{4.0}{\MeV} (\SI{0.2}{\MeV}) for the \Dplus (\Dstarp)
peak. The variations in the peak position and resolution are implemented in the likelihood fit as shape uncertainties
with no impact on the signal yield, but this additional freedom in the fit is necessary to achieve good agreement
between the data and the fit model.
 
An additional systematic uncertainty is applied to cover ID track impact-parameter resolution differences
between simulation and data after the ID alignment is performed~\cite{IDTR-2019-05}. The difference is evaluated
using minimum-bias data and the resulting uncertainty is extrapolated to higher \pT with muon tracks from \Zboson boson
decays~\cite{ATL-PHYS-PUB-2015-051}. The uncertainty is propagated to the \WplusDmeson
measurement by generating \Dplus and \Dstarp SPG samples where the impact parameters of the ID tracks are smeared before
performing the SV fit for the \Dmeson reconstruction. The relative change in the \Dmeson reconstruction efficiency was found to be
up to \SI{5}{\percent} for high-\pT \Dmeson mesons and about \SI{1.5}{\percent} at low \pT (i.e.\ $\pT < \SI{40}{\GeV}$).
The systematic uncertainties in the \Dmeson meson reconstruction efficiency related to ID track impact-parameter
resolution and ID material variations are among the largest systematic uncertainties in the analysis.
 
\textbf{Signal modeling:} The signal modeling uncertainty is derived by comparing the fiducial region efficiencies
for the signal \SHERPA[2.2.11], \MGFX, and \aMGNLO\ \WplusDmeson simulations. In each differential
bin, the maximum difference between the nominal MC simulation (\SHERPA) and either of the \MGNLO simulations
is taken and a symmetric systematic uncertainty is applied in the two directions. The uncertainty is correlated
between the differential bins and \Wboson boson charges. It accounts for the fact that the choice of MC
simulation for unfolding affects the measured values of the observables because of differences in the ME calculation,
PS simulation, and heavy-flavor quark fragmentation and hadronization. The uncertainty ranges from 1\% to 4\%,
depending on the bin, and is generally one of the largest uncertainties in the analysis. The relatively large difference
in fiducial efficiency between \SHERPA and \MGNLO simulations arises from the modeling of the correlation
between \Wboson boson and \Dmeson meson kinematics when the \MET and \mT cuts are applied at the detector level.
Including the same \MET and \mT cuts in the truth fiducial definition would reduce the uncertainty; however, it would
give rise to a large background from signal \WplusDmeson events that fail the truth \MET and \mT
selection, but pass the detector-level selection due to the poor \MET resolution, ultimately
increasing the total uncertainty.
 
Additional uncertainties are considered by varying the QCD scales,
the PDFs, \alphas, and the virtual EW corrections in \SHERPA[2.2.11]. The PDF variations, \alphas uncertainty, and EW corrections
were found to have a negligible effect on the fiducial efficiency. The effect of QCD scale uncertainties is defined by
the envelope of variations resulting from changing the renormalization and factorization
scales by factors of two with an additional constraint of $0.5 \leq \muR/\muF \leq 2$. In most differential
bins the effect was found to be smaller than the corresponding difference between \SHERPA and \MGNLO.
Lastly, the uncertainties in the \DtoKpipi and \DstartoKpipi branching ratios~\cite{Workman:2022ynf} are applied as
uncertainties of \SI{1.7}{\percent} and \SI{1.1}{\percent}, respectively, in the signal yield in the likelihood fit.
 
\textbf{Background MC modeling:} The implementation of the background modeling uncertainties varies
between the backgrounds. For \Wpluscmatch, \Wpluscmismatch, and \Wjets backgrounds, \SHERPA[2.2.11]
QCD scale, PDF, and \alphas variations are used. Among the three, the QCD scale uncertainty
generally has the largest effect and leads to a 10\%--30\% uncertainty in the
yield of the corresponding background process, depending on the differential bin. The uncertainty
is constrained in the likelihood fit by the small statistical uncertainties in the tails of
the invariant mass distributions in the \Dplus and \Dstarp channels, reducing its impact on the
observables. As in the case of the signal process, these uncertainties are correlated between
the differential bins. An additional modeling uncertainty is included by taking the
full difference between \SHERPA and \MGNLO predicted background yields.
To be conservative, this uncertainty is taken to be uncorrelated between the differential bins. This
avoids the assumption that either of the simulations have an a priori perfect description of the
shape of the differential variable (i.e.\ \diffpt or \diffeta), and provides more flexibility
in the likelihood fit.
 
Internal event weight variations in the \MGNLO[2.3.3] \ttbar
simulation are used to determine the effect of the PDF uncertainty on the top quark background.
The uncertainty due to initial-state radiation is estimated by simultaneously varying the \hdamp
parameter and the \muR and \muF scales, and choosing the \textsc{Var3c} up and down variants of the A14 tune
as described in \Refn{\cite{ATL-PHYS-PUB-2017-007}}. The impact of final-state radiation is evaluated
by halving and doubling the renormalization scale for emissions from the parton shower.
Uncertainties in the \ttbar ME calculation and PS are estimated by replacing the nominal \ttbar
prediction with two alternative simulations: \POWHER[7.04] and \MGNLOPY[8] and
taking the full difference as a systematic uncertainty. For other small backgrounds (\Zjets and
diboson events) a conservative \SI{20}{\percent} uncertainty in their yields is used.
Due to the high purity of the \WplusDmeson signal process in the \SR selection, background modeling
uncertainties are subdominant in the statistical analysis.
 
\textbf{Charm hadronization:} The \Wpluscmatch and \Wpluscmismatch backgrounds in the \Dplus
channel have large contributions from weakly decaying charmed mesons incorrectly reconstructed
as \DtoKpipi (e.g.\ \DsubstoKKpi reconstructed as \DtoKpipi). Two sources of associated systematic uncertainty are included:
uncertainties in the charmed hadron production fractions and uncertainties in the charmed hadron branching ratios.
Charmed hadron production fractions in the MC samples are reweighted to the world-average
values as described in \Sect{\ref{sec:signal_and_bkg_modelling:prompt_mc}}. Following the
procedure in \Refn{\cite{ATL-PHYS-PUB-2022-035}}, three eigenvector variations of the event
weights are derived to describe the correlated experimental systematic uncertainties associated
with the measurements of the charmed hadron production fractions. The uncertainty affects the
relative background yield by up to \SI{3}{\percent} and also the shape of the
background invariant mass distribution because the different charmed hadron species
populate different ranges of the reconstructed \Dplus invariant mass. The impact of the uncertainties
in the charmed hadron branching ratios is estimated in a conservative way by generating SPG \Dplus
samples with all branching ratios shifted simultaneously in a correlated manner to cover the systematic
uncertainties in charmed hadron decays reported in \Refn{\cite{Workman:2022ynf}}. The relative change
in the background yield and shape of the \Wpluscmatch background with respect to the nominal SPG
configuration is propagated to the \SHERPA MC sample and implemented in the statistical
analysis. The size of the uncertainty is up to \SI{5}{\percent}.
Both sources of charmed hadronization uncertainty related to background processes were found to have
a negligible impact on all observables.
 
\textbf{Multijet estimation:} The \MJ background and its uncertainties are estimated in the \FakeCR, as
described in \Sect{\ref{sec:signal_and_bkg_modelling:qcd}} and the corresponding systematic uncertainties
are implemented as nuisance parameters in the likelihood fit. Due to the difficulty of estimating
the \MJ background in the \SR selection, the relative uncertainties are large (${>}\SI{50}{\percent}$).
However, the \MJ background is largely symmetric between OS and SS regions and its relative
size is reduced in the OS--SS subtraction. Despite the large relative uncertainty in
the \MJ yield, the impact on the measured observables is therefore negligible.
 
\textbf{Finite size of MC samples:} MC statistical uncertainties affect the measurement
in several ways. The binomial uncertainties in the \WplusDmeson fiducial efficiencies calculated
with the \SHERPA MC samples are propagated into the likelihood fit via nuisance parameters
affecting the yield of the signal sample. There is one parameter per nonzero element of the
detector response matrix. The statistical uncertainty in the diagonal elements is less than \SI{1}{\percent},
while the uncertainty in the off-diagonal elements exceeds \SI{10}{\percent}. However, because
the off-diagonal elements have small values compared to diagonal ones, the corresponding
statistical uncertainty has a negligible impact on the results. Furthermore, statistical uncertainties
associated with the bins of the invariant mass distributions are implemented as constrained \enquote{$\gamma$}
parameters in the likelihood fit as explained in \Sect{\ref{sec:likelihood_fit}}. There is one
such parameter per invariant mass bin and their impact on the observables is of the order of \SI{1}{\percent}.

\subsection{Evaluation of the overall systematic uncertainty}
\label{sec:sys:eval}
 
The impact of each individual systematic uncertainty on the observables is calculated by
performing two likelihood fits with the corresponding nuisance parameter ($\theta$)
fixed to its post-fit one-standard-deviation bounds. The changes in the values of the
normalization factors associated with the observables, relative to the unconditional
likelihood fit, are then taken as the impact of the given systematic uncertainty on the observables.
Several nuisance parameters are grouped together by summing their impact on the observables in quadrature.
A summary of the dominant systematic uncertainties is given in \Tab{\ref{tab:fits:systematics:total:rc}}
for inclusive cross-sections and the cross-section ratio \Rc. The table demonstrates that most of the
systematic uncertainties are correlated between the positive and negative charge channels and therefore
cancel out in the \Rc calculation. The dominant uncertainties in \Rc are the data and MC statistical
uncertainties. Uncertainties in differential bins are summarized in \App{\ref{appendix:sys}}.
Similarly to the \Rc calculation, uncertainties with no dependence on the differential variable cancel out in the normalized cross-section.
For example, the SV reconstruction efficiency uncertainties almost completely cancel out in normalized \diffeta cross-sections
because the \Dmeson SV reconstruction has no dependence on the lepton pseudorapidity. However, the same
uncertainties do not cancel out in the $\pT(\Dmeson)$ measurement because there is a strong dependence
on $\pT(\Dmeson)$.
 
\begin{table}[htpb]
\caption{Summary of the main systematic uncertainties as percentages of the measured observable
for \sigmaWmplusDmeson, \sigmaWpplusDmeson, and \Rc in the \Dplus and \Dstarp channels. The individual
groups of uncertainties are defined in the text.}
\label{tab:fits:systematics:total:rc}
\renewcommand{\arraystretch}{1.0}
\setlength\tabcolsep{3.2pt}
\centering
\resizebox*{1.0\textwidth}{!}{
\begin{tabular}{l |
S[table-format=1.1, round-precision=1, round-mode=places]
S[table-format=1.1, round-precision=1, round-mode=places]
S[table-format=1.1, round-precision=1, round-mode=places] |
S[table-format=1.1, round-precision=1, round-mode=places]
S[table-format=1.1, round-precision=1, round-mode=places]
S[table-format=1.1, round-precision=1, round-mode=places]
}
\toprule
& \multicolumn{3}{c|}{\Dplus channel}
& \multicolumn{3}{c}{\Dstarp channel} \\
\midrule
Uncertainty [\%]               & \multicolumn{1}{c}{\sigmaWmplusD}
& \multicolumn{1}{c}{\sigmaWpplusD}
& \multicolumn{1}{c|}{$\Rc(\Dplus)$}
& \multicolumn{1}{c}{\sigmaWmplusDstar}
& \multicolumn{1}{c}{\sigmaWpplusDstar}
& \multicolumn{1}{c}{$\Rc(\Dstarp)$} \\
\midrule
SV reconstruction             & 3.0270 & 2.8655 & 0.4588 & 2.3115 & 2.3100 & 0.4151 \\
Jets and \MET        & 1.6929 & 1.8735 & 0.2299 & 1.4646 & 1.4900 & 0.3597 \\
Luminosity                    & 0.7992 & 0.8325 & 0.0043 & 0.8040 & 0.8017 & 0.0019 \\
Muon reconstruction           & 0.6369 & 0.6944 & 0.2827 & 0.6590 & 0.6565 & 0.2601 \\
Electron reconstruction       & 0.2111 & 0.2252 & 0.0156 & 0.1837 & 0.1796 & 0.0043 \\
Multijet background           & 0.2081 & 0.1560 & 0.1196 & 0.0814 & 0.0601 & 0.1081 \\
\midrule
Signal modeling               & 2.0659 & 2.0526 & 0.0979 & 1.2050 & 1.1962 & 0.0102 \\
Signal branching ratio        & 1.6070 & 1.5895 & 0.0100 & 1.0625 & 1.0527 & 0.0095 \\
Background modeling           & 1.1394 & 1.2029 & 0.2731 & 1.3248 & 1.3090 & 0.4553 \\
\midrule
Finite size of MC samples     & 1.1976 & 1.2103 & 1.1001 & 1.3517 & 1.3559 & 1.2977 \\
Data statistical uncertainty  & 0.4831 & 0.5124 & 0.6986 & 0.6937 & 0.7332 & 1.0125 \\
\midrule
Total                         & 4.6362 & 4.5897 & 1.3870 & 3.7196 & 3.7291 & 1.6919 \\
\bottomrule
\end{tabular}
}
\end{table}
 
\FloatBarrier


\section{Results and comparison with theoretical predictions}
\label{sec:results}
 
\begin{figure}[htbp]
\centering
\subfloat[]{
\includegraphics[width=0.45\textwidth]{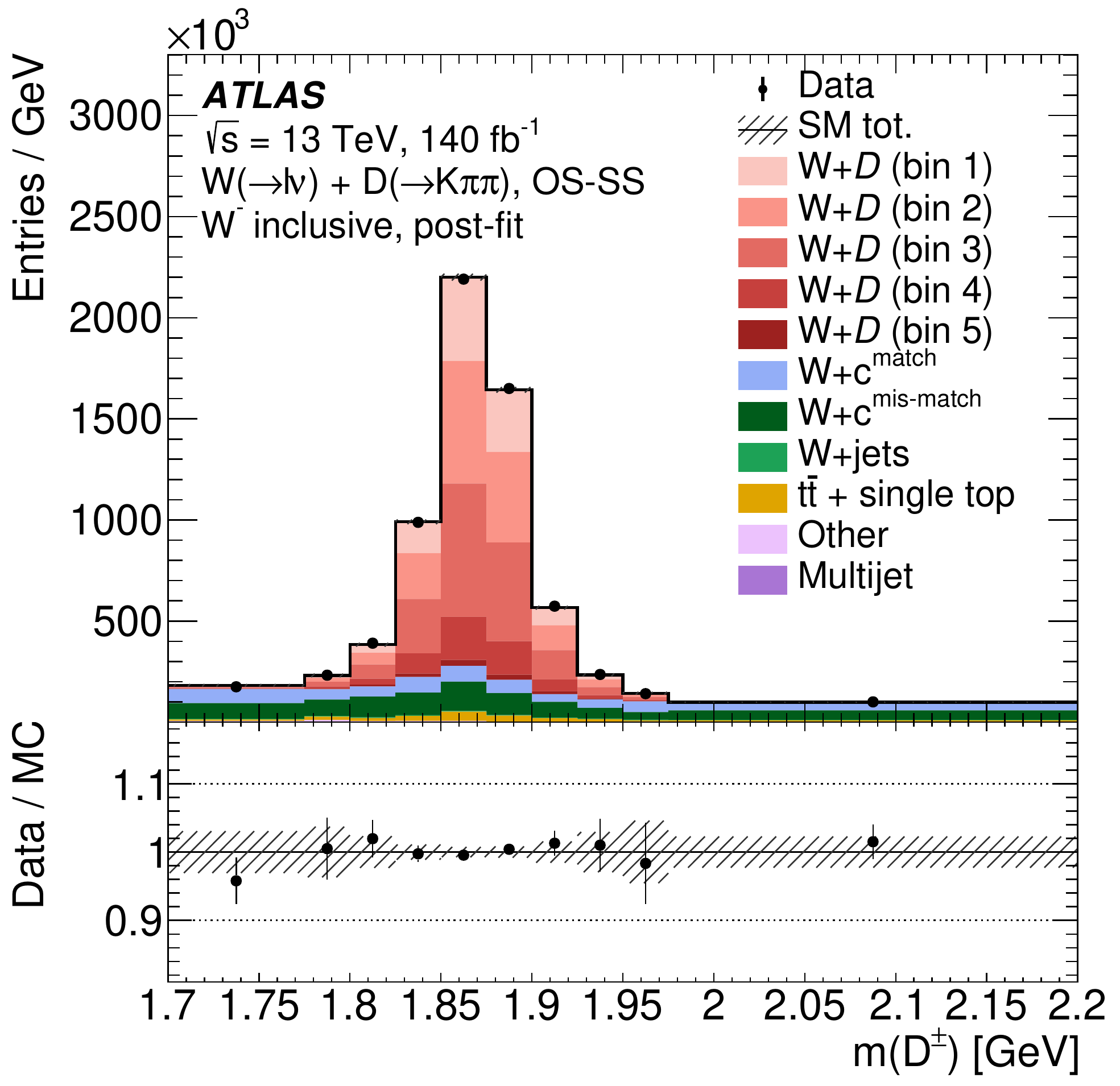}
\label{fig:fits:differential:dplus:lep_minus:0tag_inclusive_minus_Dmeson_m_fit}
}
\subfloat[]{
\includegraphics[width=0.45\textwidth]{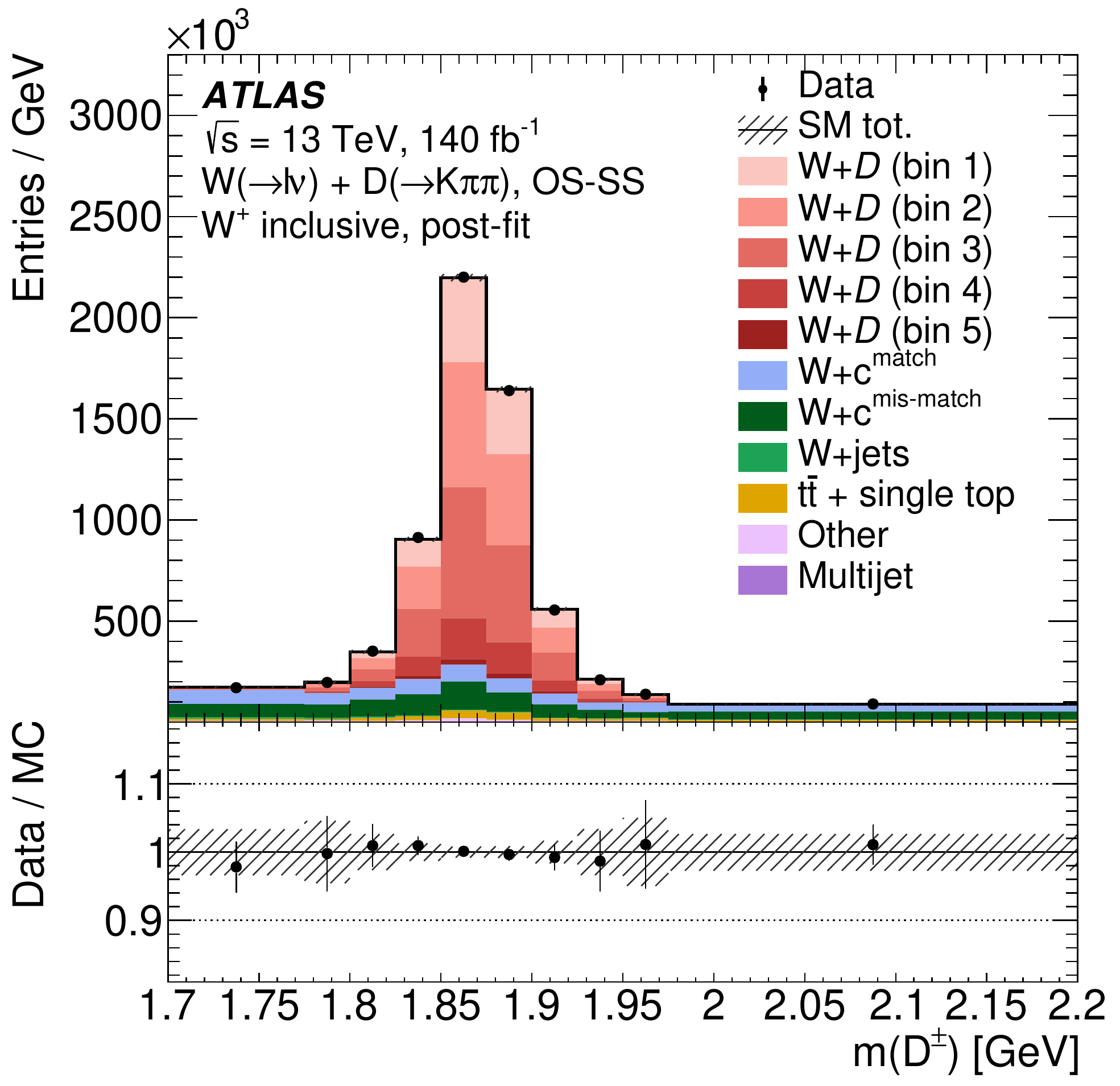}
\label{fig:fits:differential:dplus:lep_plus:0tag_inclusive_plus_Dmeson_m_fit}
}
\\
\subfloat[]{
\includegraphics[width=0.45\textwidth]{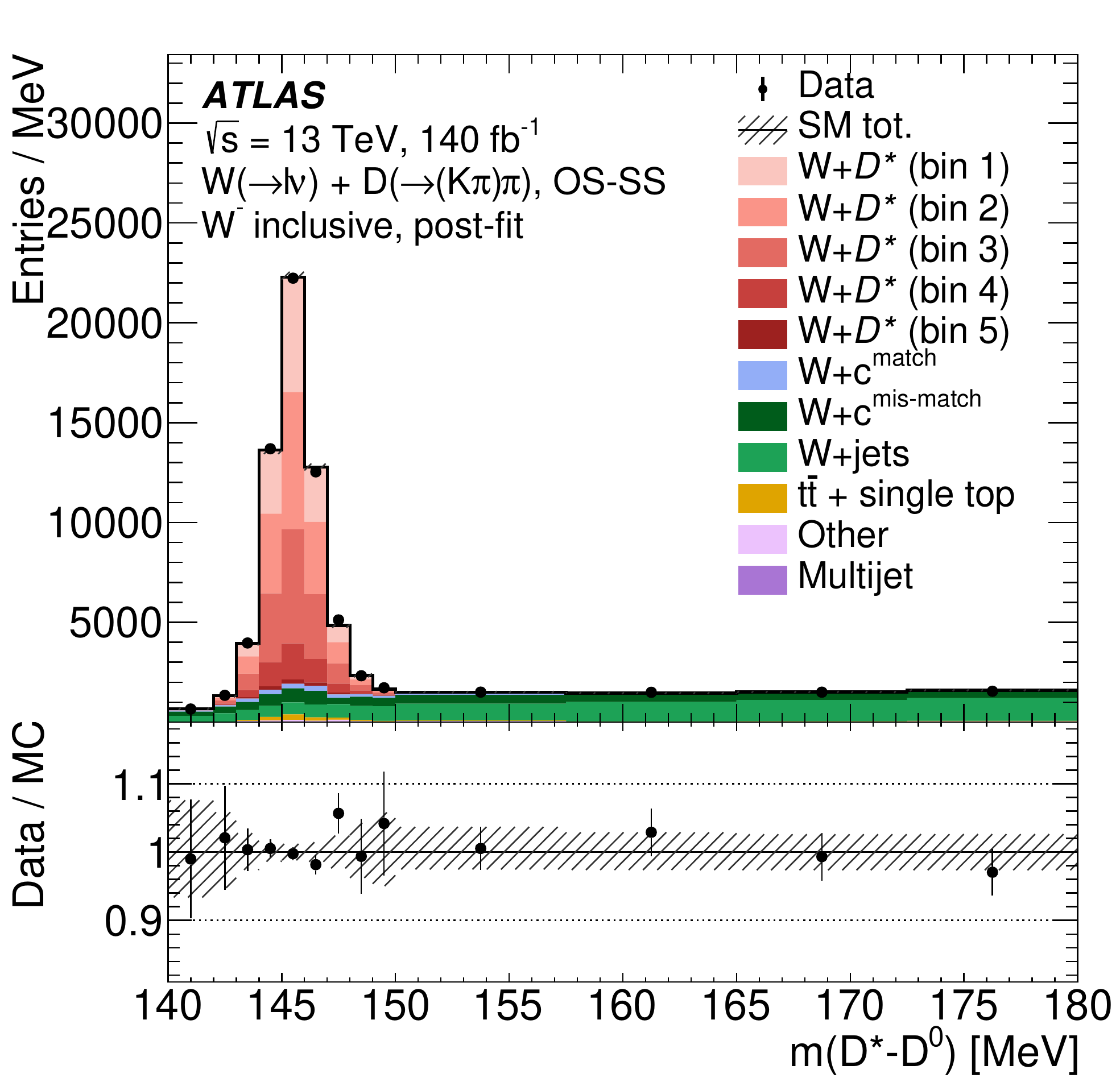}
\label{fig:fits:differential:dstar:lep_minus:0tag_inclusive_minus_Dmeson_mdiff_fit}
}
\subfloat[]{
\includegraphics[width=0.45\textwidth]{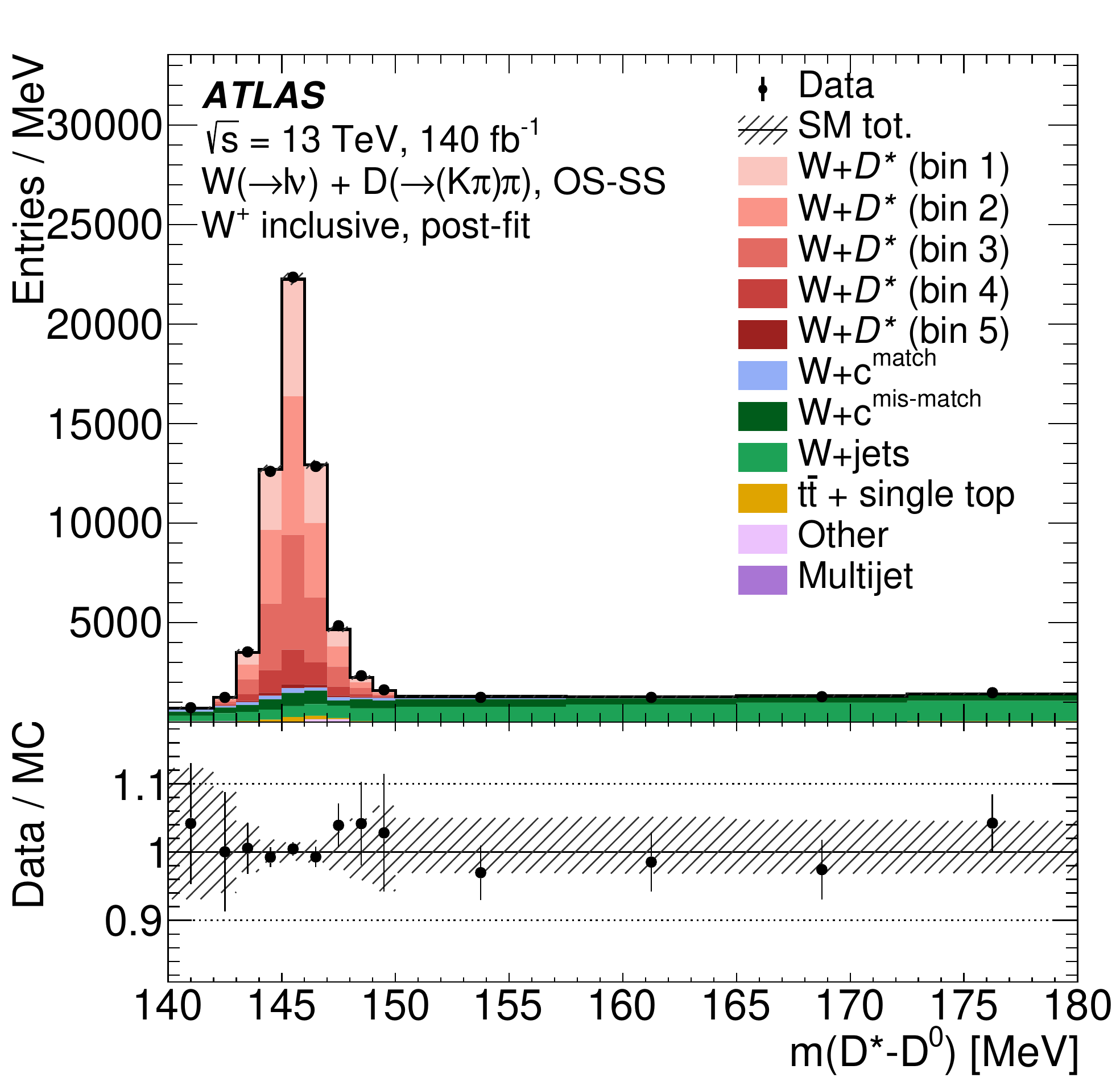}
\label{fig:fits:differential:dstar:lep_plus:0tag_inclusive_plus_Dmeson_mdiff_fit}
}
\caption{
Post-fit OS--SS \WplusDmeson signal and background predictions compared with data:
\protect\subref{fig:fits:differential:dplus:lep_minus:0tag_inclusive_minus_Dmeson_m_fit} \WmplusD channel,
\protect\subref{fig:fits:differential:dplus:lep_plus:0tag_inclusive_plus_Dmeson_m_fit} \WpplusD channel,
\protect\subref{fig:fits:differential:dstar:lep_minus:0tag_inclusive_minus_Dmeson_mdiff_fit} \WmplusDstar channel, and
\protect\subref{fig:fits:differential:dstar:lep_plus:0tag_inclusive_plus_Dmeson_mdiff_fit} \WpplusDstar channel.
The \enquote{SM Tot.} line represents the sum of all signal and background samples and the corresponding hatched
band shows the full post-fit systematic uncertainty. The five bins associated with the signal
samples are the truth bins of the \diffpt differential distribution.
}
\label{fig:fits:differential:OS-SS:0tag:postfit}
\end{figure}

Post-fit comparisons between the data and MC distributions for the \Dplus and \Dstarp channels are shown in
\Fig{\ref{fig:fits:differential:OS-SS:0tag:postfit}} separately for the $\Wboson^{-}$ and $\Wboson^{+}$
channels. Most of the data points are within the resulting $1\sigma$ systematic uncertainty band.
The \SR post-fit yields obtained with the likelihood fit are given in \Tabs{{\ref{tab:results:yieldtab:dplus}}}{\ref{tab:results:yieldtab:dstar}}. Yields are shown for both the \diffpt and \diffeta fits.
Background yields and the integrated signal yields are consistent between
the two fits in both the \Dplus and \Dstarp channels.
The systematic uncertainties in the integrated yields are slightly lower in the \diffpt fits than in the
\diffeta fits because the dominant systematic uncertainties depend more strongly on \diffpt
and are therefore more constrained in the fit.
 
The resulting cross-sections $\sigmaWplusD \times B(W\rightarrow \ell \nu)$ and \Rc are presented in
\Tab{\ref{tab:results:sigmaWD}}.
The results presented here are obtained using the \diffpt fit;
results from the differential \diffeta fit are compatible.
Ratios of cross-sections obtained in the \Dplus\ and \Dstarp\
channels are consistent with predictions obtained using the world-average production fractions,
$\sigma(\WplusDstar)/\sigma(\WplusD) = 1.01\pm 0.034$, where the 3.4\%\ uncertainty is obtained using
the (correlated) uncertainties in the \Dstar\ and \Dplus\ production fractions~\cite{Lisovyi:2015uqa}.
The measured differential cross-sections in bins of \diffpt and \diffeta are
given in \App{\ref{appendix:results}}. The statistical uncertainty is larger in the \Dstarp
channel because the branching ratio for that mode is smaller than the one for \Dplus; the relative sizes
of the systematic uncertainties are similar because they are largely independent of the decay mode.
A combined value of $\Rc(\Dmeson)$ is derived from the individual measurements of $\Rc(\Dplus)$ and
$\Rc(\Dstarp)$.  Systematic uncertainties are largely uncorrelated between the channels.
As shown in \Tab{\ref{tab:fits:systematics:total:rc}}, they are  dominated by the uncorrelated MC
statistical uncertainties. After correcting for differences between the chosen fiducial regions, these
measurements are consistent with, but more precise than, the CMS \WplusDstar results presented in \Refn{\cite{CMS-SMP-17-014}},
performed with \SI{35.7}{\ifb} of data.
 
\begin{table}[htpb]
\caption{Post-fit yields in the OS--SS \WplusD SR from the $\pT(\Dplus)$ differential fit.
The data statistical uncertainty is calculated as $\sqrt{N_{\mathrm{OS}} + N_{\mathrm{SS}}}$.
Uncertainties in individual SM components are the full post-fit systematic uncertainties.}
\label{tab:results:yieldtab:dplus}
\renewcommand{\arraystretch}{1.0}
\setlength\tabcolsep{5.2pt}
\centering
\begin{tabular}{l |
S[table-format=6.0, round-precision=0, round-mode=places]@{$\,\pm\,$}
S[table-format=4.0, round-precision=0, round-mode=places] |
S[table-format=6.0, round-precision=0, round-mode=places]@{$\,\pm\,$}
S[table-format=4.0, round-precision=0, round-mode=places] |
S[table-format=6.0, round-precision=0, round-mode=places]@{$\,\pm\,$}
S[table-format=4.0, round-precision=0, round-mode=places] |
S[table-format=6.0, round-precision=0, round-mode=places]@{$\,\pm\,$}
S[table-format=4.0, round-precision=0, round-mode=places]
}
\toprule
& \multicolumn{4}{c|}{OS--SS \WplusD SR ($\pT(\Dplus)$ fit)} & \multicolumn{4}{c}{OS--SS \WplusD SR (\diffeta fit)} \\
\midrule
Sample                   & \multicolumn{2}{c|}{\WmplusD} & \multicolumn{2}{c|}{\WpplusD} & \multicolumn{2}{c|}{\WmplusD} & \multicolumn{2}{c}{\WpplusD} \\
\hline
\WpmplusD (bin 1)        &      26430 &    510 &      26180 &    550 &      31530 &    530 &      30920 &    560 \\
\WpmplusD (bin 2)        &      39090 &    660 &      38610 &    660 &      30560 &    650 &      30790 &    620 \\
\WpmplusD (bin 3)        &      43520 &    660 &      41510 &    670 &      25640 &    470 &      24940 &    450 \\
\WpmplusD (bin 4)        &      15330 &    350 &      14520 &    350 &      23890 &    450 &      22380 &    500 \\
\WpmplusD (bin 5)        &       2740 &    120 &       2346 &     93 &      15860 &    480 &      14630 &    470 \\
\Wpluscmatch             &      24800 &   2400 &      24300 &   2400 &      23500 &   2600 &      22800 &   2700 \\
\Wpluscmismatch          &      34300 &   2500 &      29700 &   2400 &      33900 &   2500 &      29200 &   2500 \\
\Wjets                   &       1300 &   1400 &       1900 &   1500 &       2200 &   1500 &       2500 &   1800 \\
$t\bar{t}$ + single top  &       6500 &    550 &       6220 &    590 &       6520 &    540 &       6160 &    590 \\
Other                    &       1030 &    430 &       1830 &    460 &       1060 &    450 &       1940 &    470 \\
Multijet                 &        730 &    410 &       1070 &    450 &       1180 &    640 &       1600 &    690 \\ \cline{1-9}
Total SM                 &     195800 &   1200 &     188200 &   1300 &     195800 &   1300 &     187900 &   1400 \\ \cline{1-9}
Data                     &     195800 &   1100 &     188200 &   1100 &     195800 &   1100 &     188200 &   1100 \\
\bottomrule
\end{tabular}
\end{table}
 
\begin{table}[htpb]
\caption{Post-fit yields in the OS--SS \WplusDstar SR from the $\pT(\Dstarp)$ differential fit.
The data statistical uncertainty is calculated as $\sqrt{N_{\mathrm{OS}} + N_{\mathrm{SS}}}$.
Uncertainties in individual SM components are the full post-fit systematic uncertainties.}
\label{tab:results:yieldtab:dstar}
\renewcommand{\arraystretch}{1.0}
\setlength\tabcolsep{5.2pt}
\centering
\begin{tabular}{l |
S[table-format=6.0, round-precision=0, round-mode=places]@{$\,\pm\,$}
S[table-format=4.0, round-precision=0, round-mode=places] |
S[table-format=6.0, round-precision=0, round-mode=places]@{$\,\pm\,$}
S[table-format=4.0, round-precision=0, round-mode=places] |
S[table-format=6.0, round-precision=0, round-mode=places]@{$\,\pm\,$}
S[table-format=4.0, round-precision=0, round-mode=places] |
S[table-format=6.0, round-precision=0, round-mode=places]@{$\,\pm\,$}
S[table-format=4.0, round-precision=0, round-mode=places]
}
\toprule
& \multicolumn{4}{c|}{OS--SS \WplusDstar SR ($\pT(\Dstarp)$ fit)} & \multicolumn{4}{c}{OS--SS \WplusDstar SR (\diffeta fit)} \\
\midrule
Sample                   & \multicolumn{2}{c|}{\WmplusDstar} & \multicolumn{2}{c|}{\WpplusDstar} & \multicolumn{2}{c|}{\WmplusDstar} & \multicolumn{2}{c}{\WpplusDstar} \\
\hline
\WpmplusDstar (bin 1)    &      13670 &    280 &      13880 &    260 &      12640 &    260 &      12980 &    230 \\
\WpmplusDstar (bin 2)    &      17210 &    250 &      16950 &    280 &      12470 &    260 &      12910 &    280 \\
\WpmplusDstar (bin 3)    &      15000 &    200 &      14890 &    200 &      10370 &    220 &      10250 &    200 \\
\WpmplusDstar (bin 4)    &       5402 &     89 &       5139 &     95 &       9500 &    230 &       9120 &    240 \\
\WpmplusDstar (bin 5)    &        822 &     45 &        744 &     41 &       6900 &    290 &       6390 &    290 \\
\Wpluscmatch             &       2800 &    530 &       2730 &    530 &       3060 &    450 &       2690 &    480 \\
\Wpluscmismatch          &      15900 &   1700 &      14000 &   1600 &      16400 &   1400 &      14200 &   1400 \\
\Wjets                   &      35600 &   1800 &      32000 &   1700 &      35600 &   1800 &      31900 &   1700 \\
$t\bar{t}$ + single top  &       1580 &    200 &       1320 &    180 &       1480 &    180 &       1350 &    160 \\
Other                    &       1710 &    540 &        650 &    480 &       1480 &    480 &        510 &    420 \\
Multijet                 &        -90 &    190 &        -20 &    200 &       -160 &    220 &       -120 &    240 \\ \cline{1-9}
Total SM                 &     109600 &   1100 &     102200 &   1500 &     109700 &   1000 &     102200 &   1000 \\ \cline{1-9}
Data                     &     109690 &    900 &     102320 &    970 &     109690 &    900 &     102320 &    970 \\
\bottomrule
\end{tabular}
\end{table}
 
\begin{table}[htpb]
\caption{
Measured fiducial cross-sections times the single-lepton-flavor $\Wboson$ boson branching ratio and
the cross-section ratios. $\Rc(\Dmeson)$ is obtained by combining the individual measurements of
$\Rc(\Dplus)$ and $\Rc(\Dstarp)$ as explained in the text.
}
\label{tab:results:sigmaWD}
\renewcommand{\arraystretch}{1.3}
\centering
\begin{tabular}{l r @{$\,\pm\,$} l}
\toprule
Channel        & \multicolumn{2}{c}{\sigmaWplusD $\times B(W\rightarrow \ell \nu)$ [pb]} \\
\midrule
\WmplusD       & $50.2$  & $0.2$ (stat.) $^{+2.4}_{-2.3}$ (syst.) \\
\WpplusD       & $48.5$  & $0.2$ (stat.) $^{+2.3}_{-2.2}$ (syst.) \\
\WmplusDstar   & $51.1$  & $0.4$ (stat.) $^{+1.9}_{-1.8}$ (syst.) \\
\WpplusDstar   & $50.0$  & $0.4$ (stat.) $^{+1.9}_{-1.8}$ (syst.) \\
\midrule
& \multicolumn{2}{c}{$\Rc = \sigmaWpplusDmeson / \sigmaWmplusDmeson$} \\
\midrule
$\Rc(\Dplus)$  & $0.965$ & $0.007$ (stat.) $\pm0.012$ (syst.) \\
$\Rc(\Dstarp)$ & $0.980$ & $0.010$ (stat.) $\pm0.013$ (syst.) \\
\midrule
$\Rc(\Dmeson)$ & $0.971$ & $0.006$ (stat.) $\pm0.011$ (syst.) \\
\bottomrule
\end{tabular}
\end{table}
 
\begin{figure}[htbp]
\centering
\subfloat[]{
\includegraphics[width=0.5\textwidth]{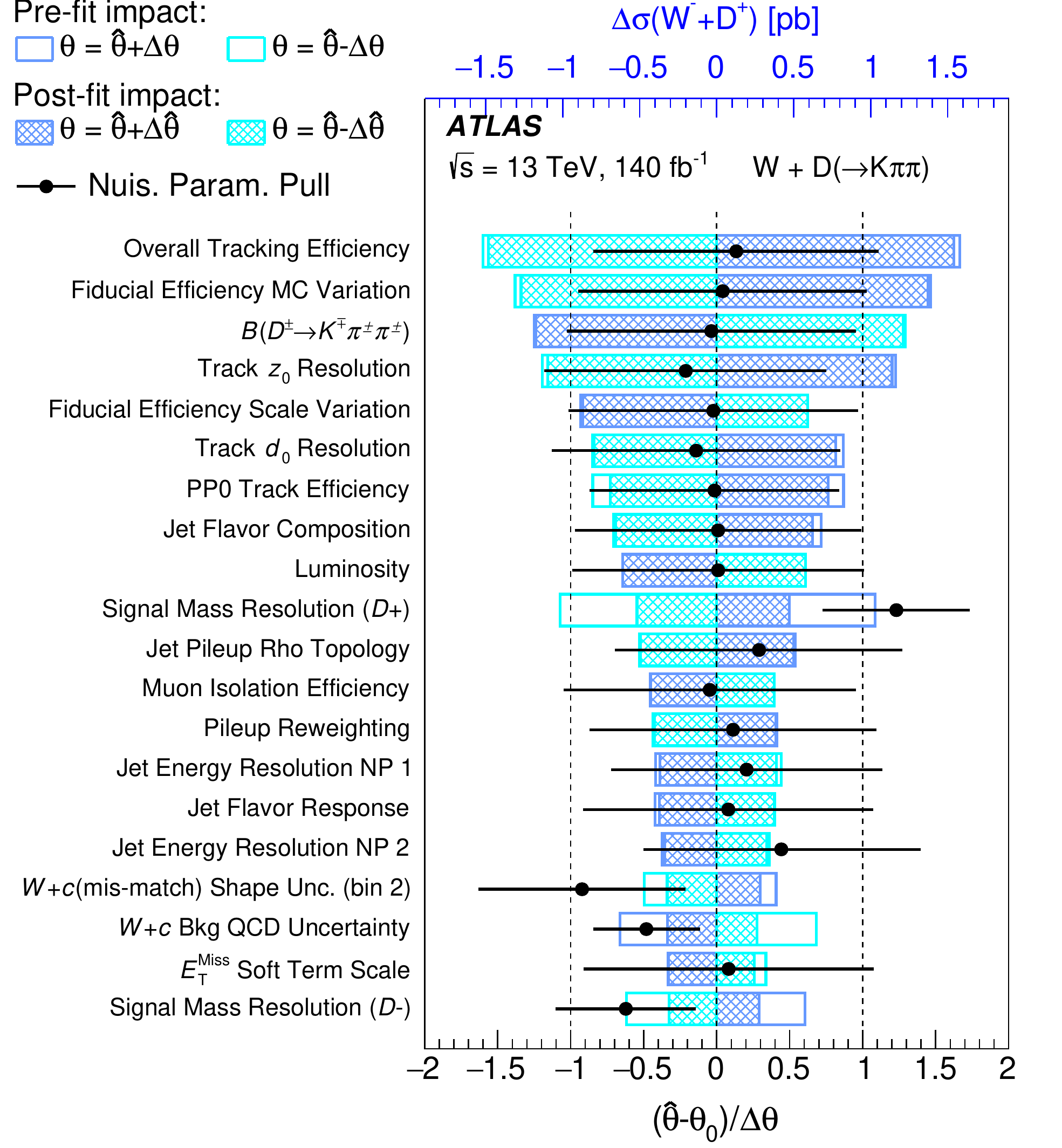}
\label{fig:fits:differential:results:dplus:pt:Ranking_mu_Wminus_tot}
}
\subfloat[]{
\includegraphics[width=0.5\textwidth]{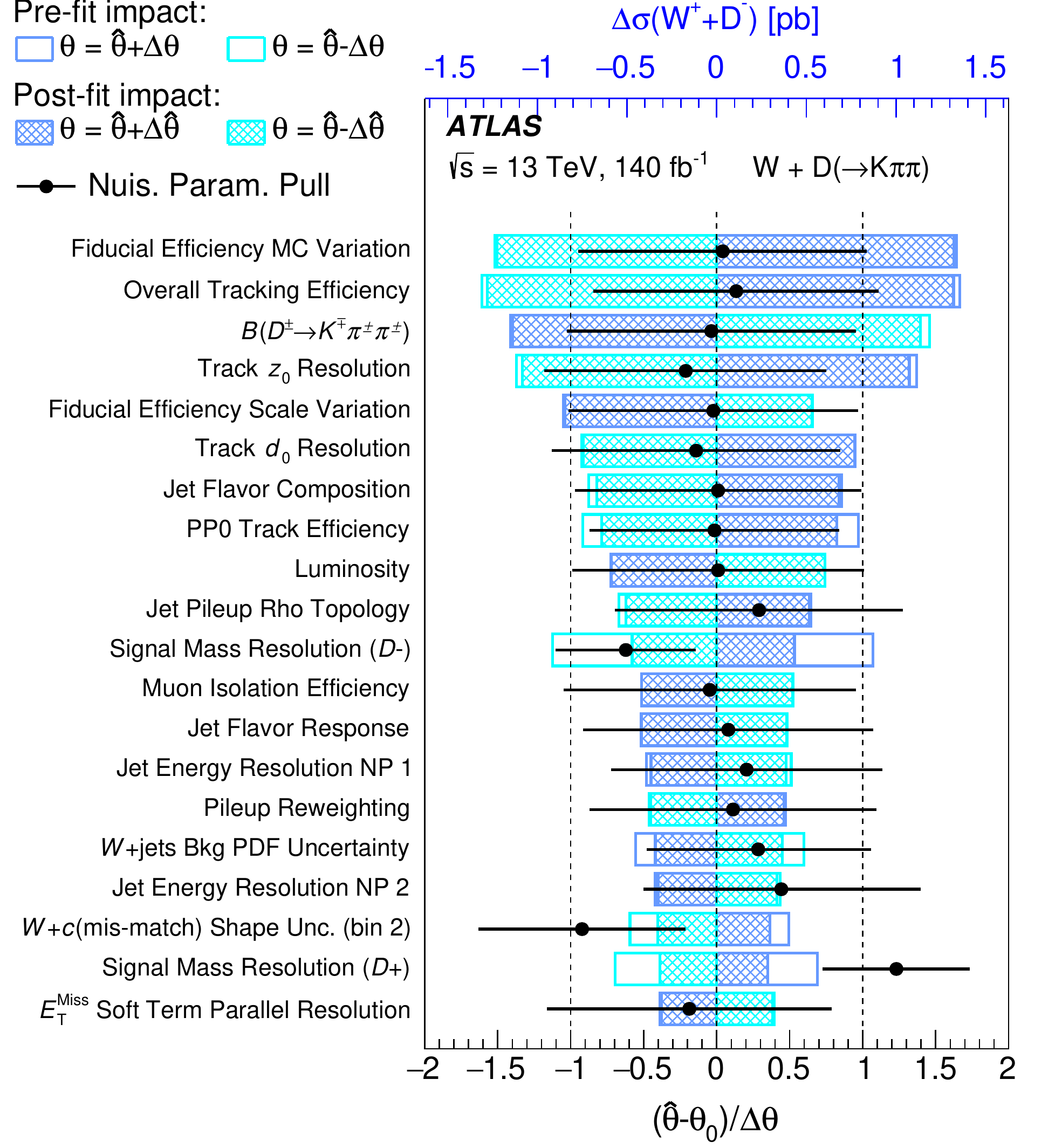}
\label{fig:fits:differential:results:dplus:pt:Ranking_mu_Wplus_tot}
}
\\
\subfloat[]{
\includegraphics[width=0.5\textwidth]{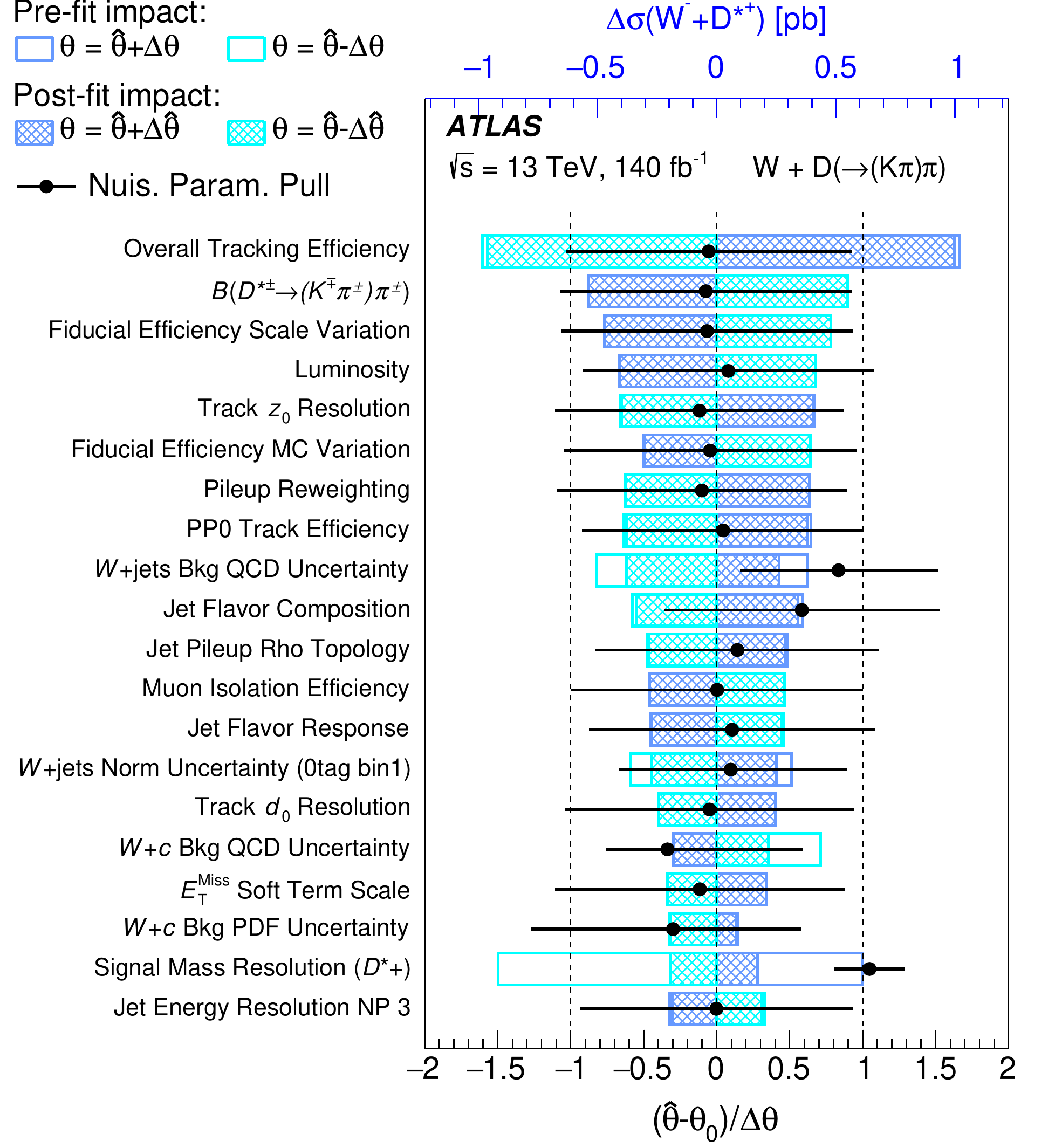}
\label{fig:fits:differential:results:dstar:pt:Ranking_mu_Wminus_tot}
}
\subfloat[]{
\includegraphics[width=0.5\textwidth]{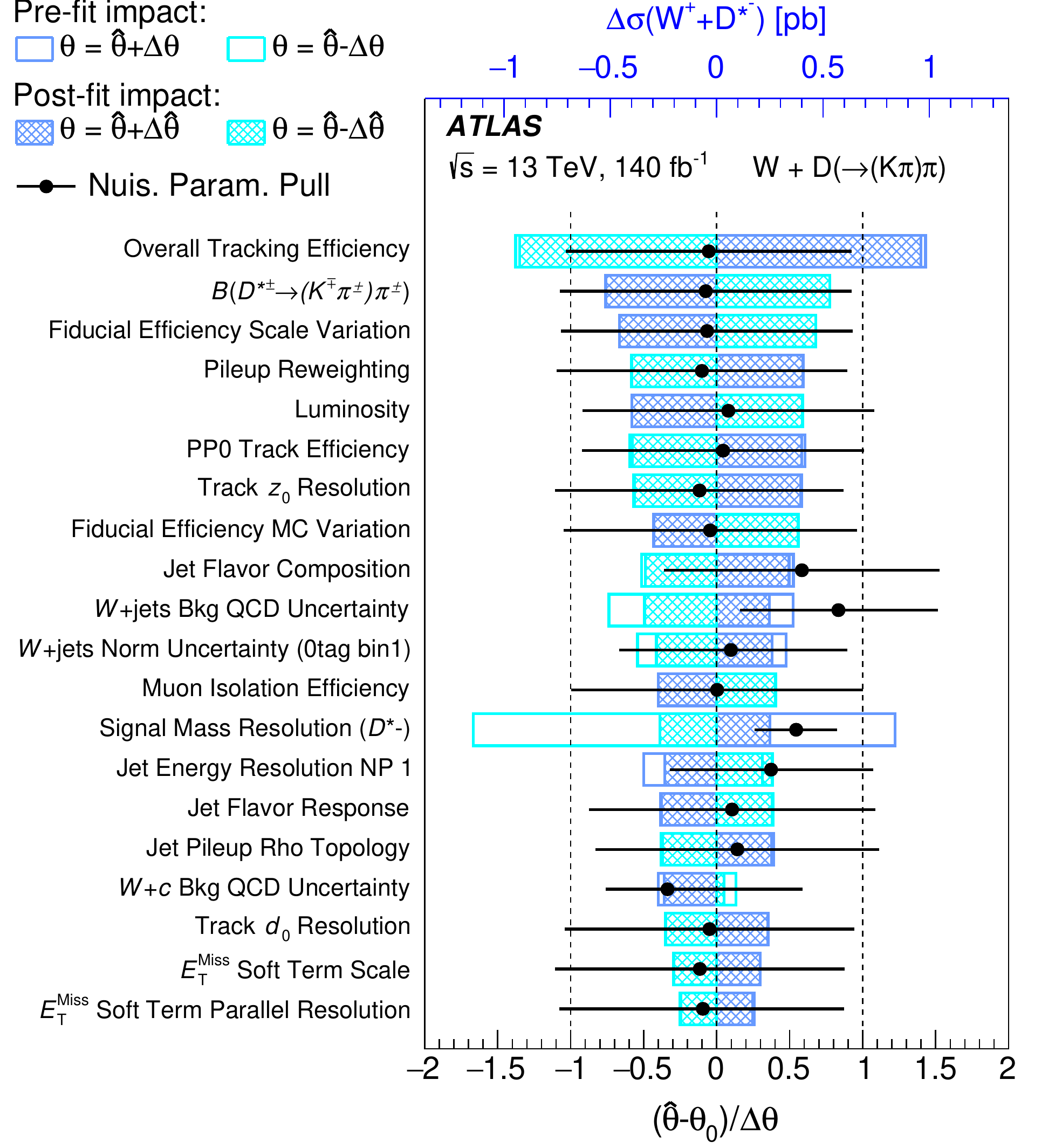}
\label{fig:fits:differential:results:dstar:pt:Ranking_mu_Wplus_tot}
}
\caption{
Impact of systematic uncertainties, for the 20 largest contributions, on the fitted cross-section
from the \diffpt fits, sorted in decreasing order. Impact on:
\protect\subref{fig:fits:differential:results:dplus:pt:Ranking_mu_Wminus_tot} \sigmaWmplusD,
\protect\subref{fig:fits:differential:results:dplus:pt:Ranking_mu_Wplus_tot} \sigmaWpplusD,
\protect\subref{fig:fits:differential:results:dstar:pt:Ranking_mu_Wminus_tot} \sigmaWmplusDstar, and
\protect\subref{fig:fits:differential:results:dstar:pt:Ranking_mu_Wplus_tot} \sigmaWpplusDstar.
The impact of pre-fit (post-fit) nuisance parameters $\vec{\theta}$ on the signal strength are shown with
empty (colored) boxes. The post-fit central value ($\hat{\theta}$) and uncertainty are
shown for each parameter with black dots.
}
\label{fig:fits:differential:results:rankings:total}
\end{figure}

The impact of the nuisance parameters on the fitted values of the absolute fiducial cross-section in the differential \diffpt fits
is shown as a \enquote{ranking plot} in \Fig{\ref{fig:fits:differential:results:rankings:total}}. The 20 nuisance parameters with the largest contribution
are ordered by decreasing impact on the corresponding observable. The post-fit central values and uncertainties of
the corresponding parameters are given in the same plots.
The ranking plots demonstrate that most nuisance parameters with large impact on the integrated fiducial
cross-section do not deviate significantly from the initial values in the likelihood fit. The
parameters associated with the signal mass-peak shape uncertainties have the most significant pulls in the
fit, however, the impact of the corresponding systematic uncertainties on the observables is small (up to
\SI{1}{\percent} for cross-sections and negligible for \Rc). These parameters are
constrained by the observed width of the \Dmeson peaks in the data. The NP shifts depend on the charge of the \Dmeson meson and are
therefore treated with independent parameters for each charge. They account for the small residual resolution degradation that is not accounted
for in the MC simulation.
 
Theoretical predictions of the \WplusDmeson cross-section for a variety of state-of-the-art PDF sets are obtained using the signal
\aMGNLO\ samples with the configuration described in \Sect{\ref{sec:vjetsMC}}. A finite charm quark
mass of $m_c=\SI{1.55}{\GeV}$ is used to regularize the cross-section and a full CKM matrix is used to calculate
the hard-scattering amplitudes. For each PDF set, the uncertainty is obtained from the alternative generator weights
using the LHAPDF prescription~\cite{Buckley:2014ana}.
Uncertainties due to the choice of \PYTHIA[8] tune are assessed by replacing the A14 tune with the Monash tune~\cite{Skands:2014pea}.
Uncertainties associated with the choice of parton shower model are estimated from a comparison of events generated with the
baseline configuration and events generated with \HERWIG[7.2]~\cite{Bellm:2019zci} using its default tune.
Differences between predictions associated with the choice of NLO matching algorithm are assessed by comparing the \aMGNLO\
cross-sections with those obtained using the calculation described in \Refn{\cite{Bevilacqua:2021ovq}}. This calculation is
based on the \Powhel event generator, which uses the \POWHEGBOX[v2] interface to implement \POWHEG NLO matching. A charm quark
mass $m_c=\SI{1.5}{\GeV}$ is used to regularize the cross-section. Effects of nondiagonal CKM matrix elements and off-shell
\Wboson boson decays including spin correlations are taken into account
in both the \aMGNLO and \Powhel calculations. For these comparisons, the renormalization and factorization
scales are set to one half of the transverse mass calculated using all final-state partons and leptons, and the ABMP16\_3\_NLO PDF set with $\alphas = 0.118$ and Monash
\PYTHIA[8.2] tune are used for both samples. The uncertainty in the direct charm production fractions is assessed
using the results from \Refn{\cite{Lisovyi:2015uqa}}.
 
\begin{figure}[htbp]
\centering
\subfloat[]{
\includegraphics[width=0.50\textwidth]{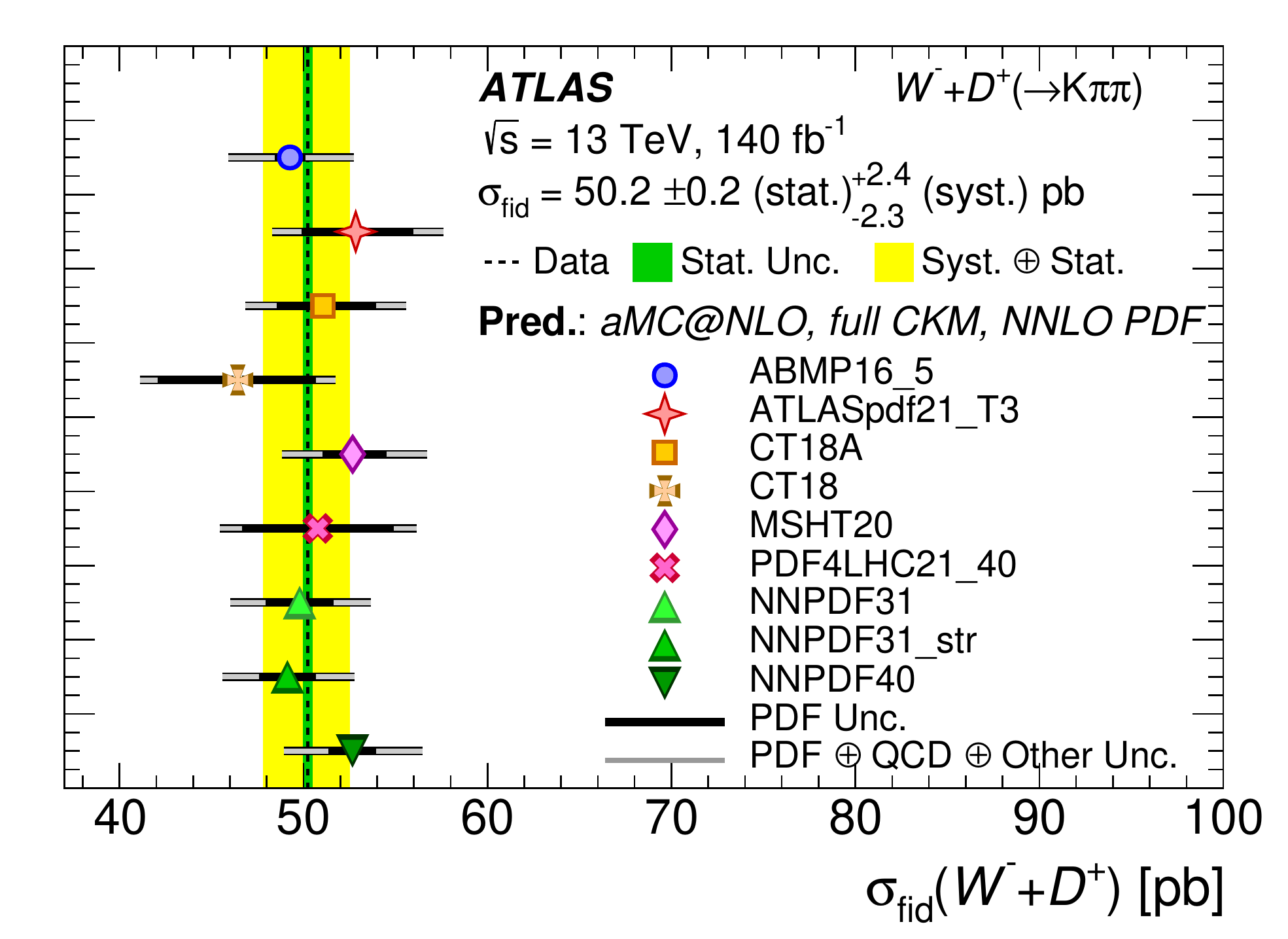}
\label{fig:results:fit_results_dplus:Wminus_tot_pt}
}
\subfloat[]{
\includegraphics[width=0.50\textwidth]{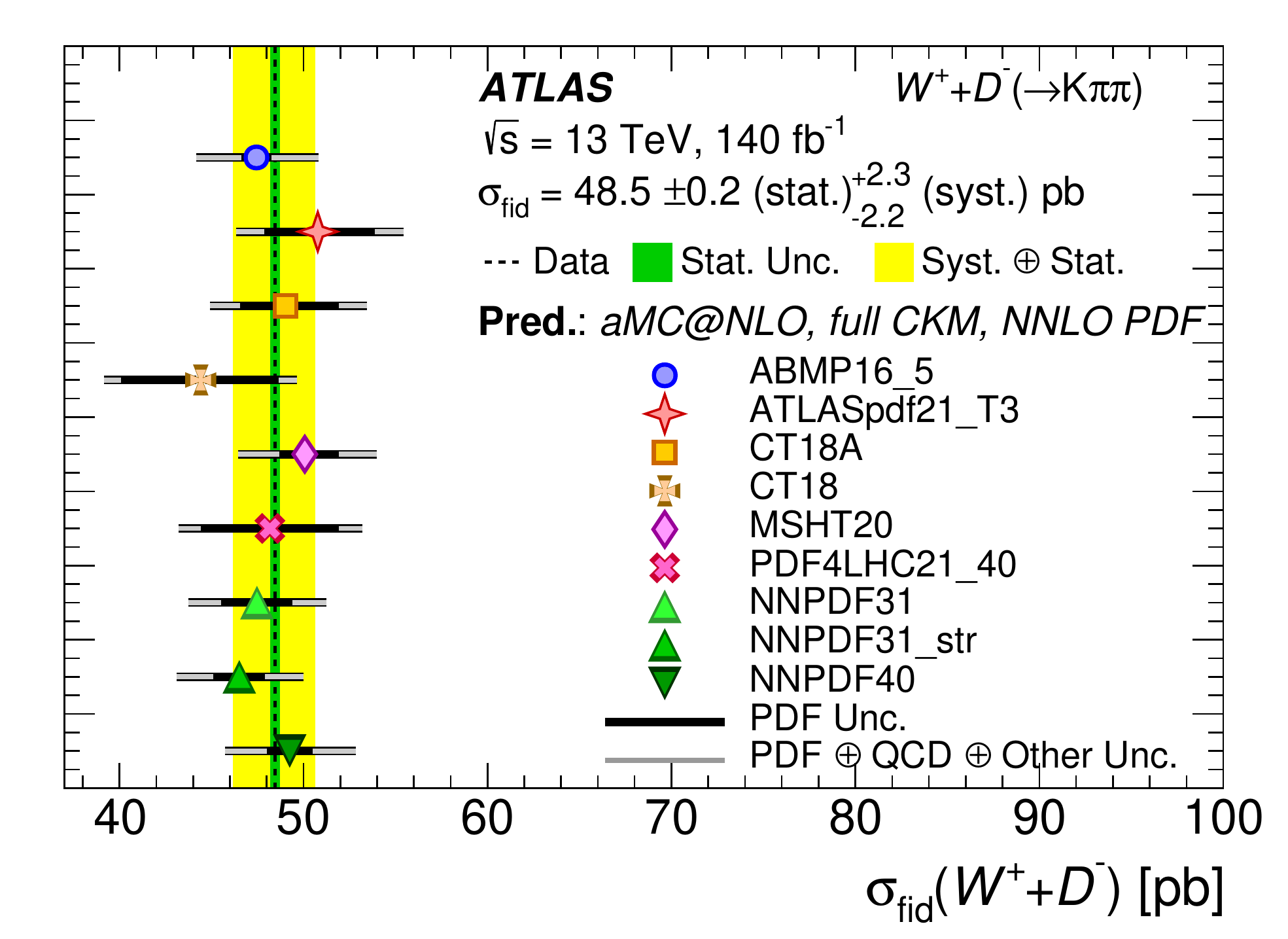}
\label{fig:results:fit_results_dplus:Wplus_tot_pt}
}
\\
\subfloat[]{
\includegraphics[width=0.50\textwidth]{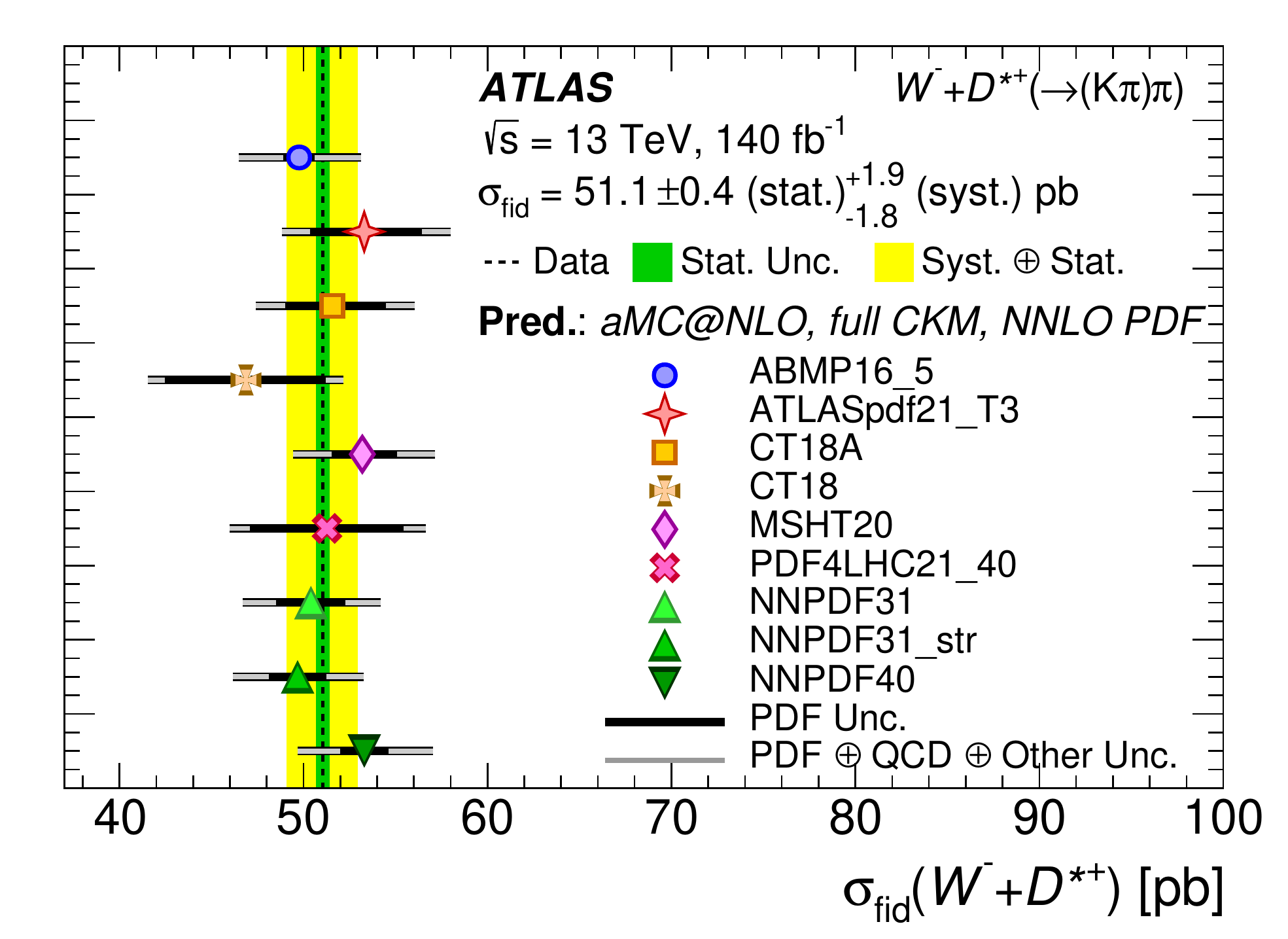}
\label{fig:results:fit_results_dstar:Wminus_tot_pt}
}
\subfloat[]{
\includegraphics[width=0.50\textwidth]{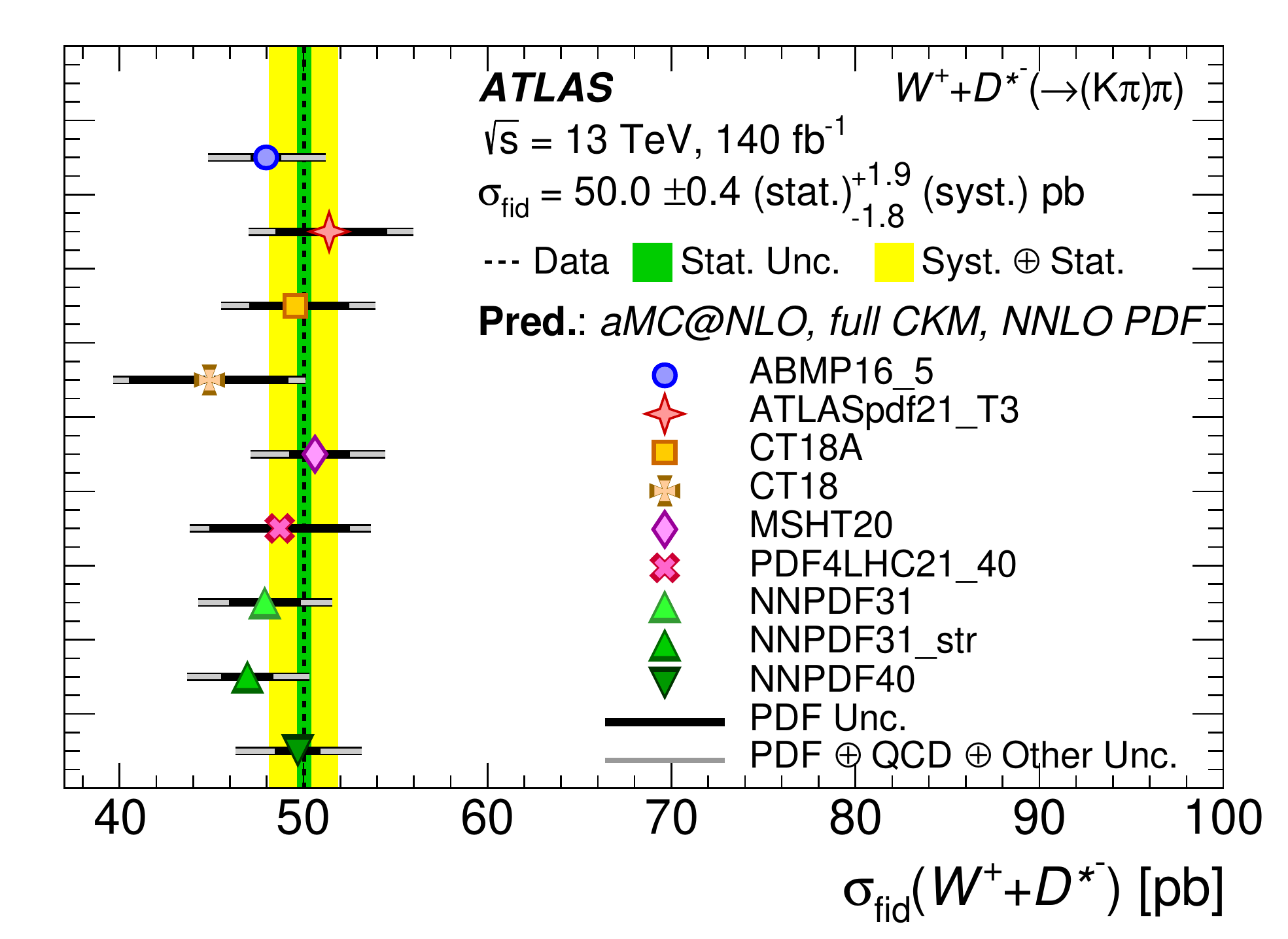}
\label{fig:results:fit_results_dstar:Wplus_tot_pt}
}
\caption{
Measured fiducial cross-section times the single-lepton-flavor $W$ branching ratio
compared with different NNLO PDF predictions for
\protect\subref{fig:results:fit_results_dplus:Wminus_tot_pt} \WmplusD,
\protect\subref{fig:results:fit_results_dplus:Wplus_tot_pt} \WpplusD,
\protect\subref{fig:results:fit_results_dstar:Wminus_tot_pt} \WmplusDstar, and
\protect\subref{fig:results:fit_results_dstar:Wplus_tot_pt} \WpplusDstar.
The dotted vertical line shows the central value of the measurement,
the green band shows the statistical uncertainty and the yellow band
shows the combined statistical and systematic uncertainty.  The PDF predictions are designated
by markers. The inner error bars on the theoretical predictions show the 68\%\ CL
uncertainties obtained from the error sets provided with each PDF set, while the outer error
bar represents the quadrature sum of the 68\% CL PDF, scale, hadronization, and matching uncertainties.
The PDF predictions are based on NLO calculations performed using \AMCatNLO and a full CKM matrix:
ABMP16\_5~\cite{Alekhin:2017kpj}, ATLASpdf21\_T3~\cite{STDM-2020-32}, CT18A, CT18~\cite{Hou:2019efy}, MSHT20~\cite{Bailey:2020ooq},
PDF4LHC21\_40~\cite{PDF4LHCWorkingGroup:2022cjn}, NNPDF31~\cite{NNPDF:2017mvq}, NNPDF31\_str~\cite{Faura:2020oom}, NNPDF40~\cite{NNPDF:2021njg}.
ABMP16\_5, ATLASpdf21\_T3, CT18A, and CT18 impose symmetric strange-sea PDFs.
}
\label{fig:results:sigmaWD}
\end{figure}

\Fig{\ref{fig:results:sigmaWD}} shows the measured fiducial cross-sections for each of the four channels compared with the theoretical
predictions obtained using different NNLO PDF sets, including a PDF set tailored to describe the strangeness of the proton --
NNPDF3.1\_strange~\cite{Faura:2020oom}. Results for all four channels show a consistent pattern. The experimental precision
is comparable to the PDF uncertainties and smaller than the total NLO theory uncertainty. All PDF sets are consistent with the measured
cross-sections once the combined theory and PDF set uncertainties are considered.
 
\begin{figure}[htbp]
\centering
\includegraphics[width=0.60\textwidth]{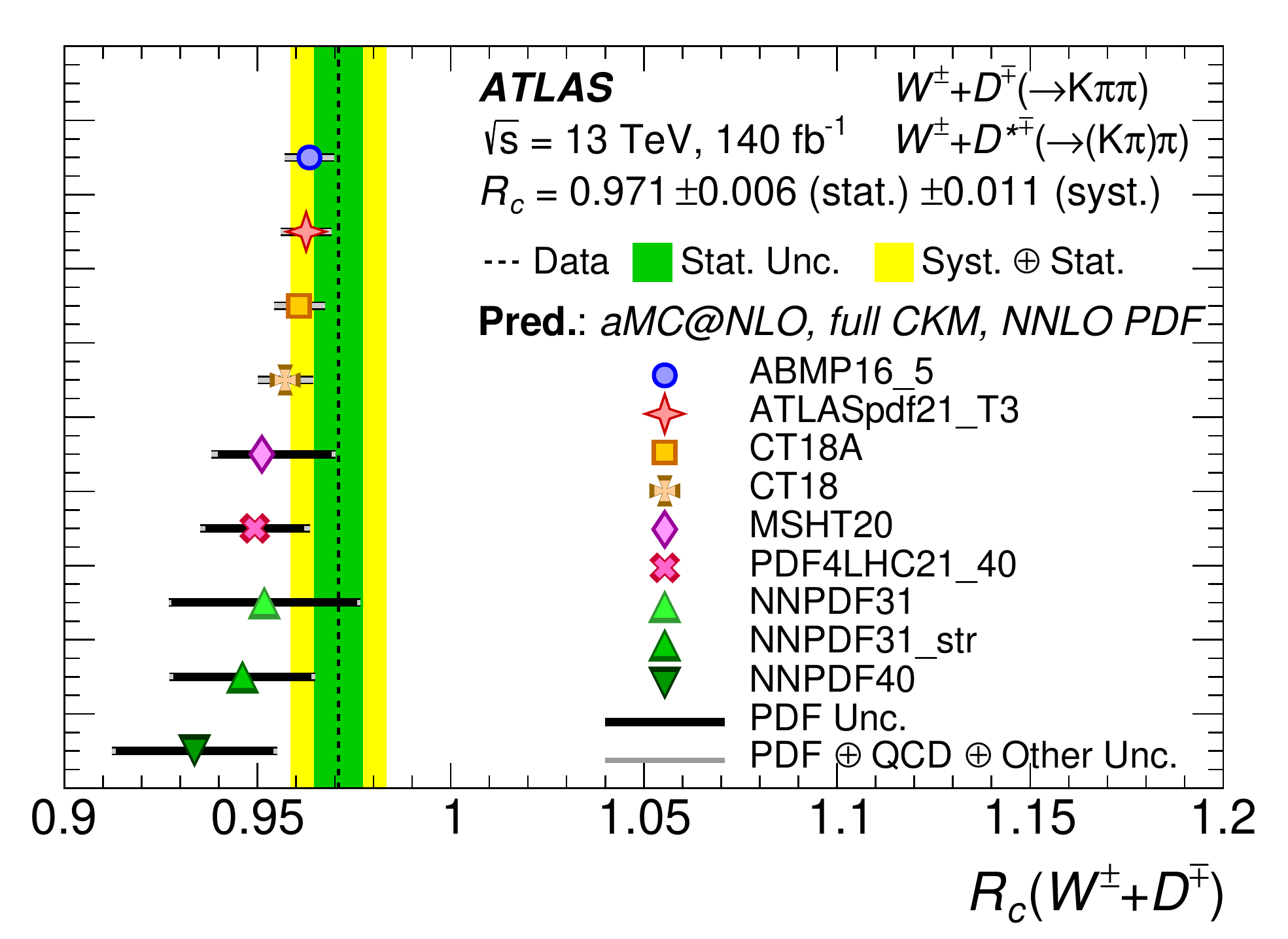}
\caption{
Measured fiducial cross-section ratio, \Rc, compared with different PDF predictions. The data
are a combination of the separate \WplusD and \WplusDstar channel measurements.
The dotted vertical line shows the central value of the measurement,
the green band shows the statistical uncertainty and the yellow band
shows the combined statistical and systematic uncertainty.  The PDF predictions are designated
by markers. The inner error bars on the theoretical predictions show the 68\%\ CL
uncertainties obtained from the error sets provided with each PDF set, while the outer error
bar represents the quadrature sum of the 68\% CL PDF, scale, hadronization, and matching uncertainties.
The PDF predictions are based on NLO calculations performed using \AMCatNLO and a full CKM matrix:
ABMP16\_5~\cite{Alekhin:2017kpj}, ATLASpdf21\_T3~\cite{STDM-2020-32}, CT18A, CT18~\cite{Hou:2019efy}, MSHT20~\cite{Bailey:2020ooq},
PDF4LHC21\_40~\cite{PDF4LHCWorkingGroup:2022cjn}, NNPDF31~\cite{NNPDF:2017mvq}, NNPDF31\_str~\cite{Faura:2020oom}, NNPDF40~\cite{NNPDF:2021njg}.
ABMP16\_5, ATLASpdf21\_T3, CT18A, and CT18 impose symmetric strange-sea PDFs.
}
\label{fig:results:Rc:combined}
\end{figure}

The cross-section ratio, \Rc,
is shown for the combined \Dplus and \Dstarp channel measurements in \Fig{\ref{fig:results:Rc:combined}}.
This combined result is consistent with theoretical predictions
for all PDF sets, although the prediction obtained using \NNPDF[4.0nnlo] shows some tension with the measurement.
Unlike the cross-section measurements, which are dominated by
systematic uncertainties, the measurements of \Rc have comparable statistical and systematic uncertainties. PDF set uncertainties for \Rc fall
into two categories. Those sets that impose the restriction that the strange-sea be symmetric ($s=\bar{s}$), such as CT18 and AMBP16, predict \Rc
with high precision while PDF fits that allow the $s$ and $\bar{s}$ distributions to differ, such as NNPDF or MSHT, have larger uncertainties.
These measurements are consistent with the predictions obtained with  PDF sets that impose a symmetric $s$-$\bar{s}$ sea, suggesting that
any $s$-$\bar{s}$ asymmetry is small in the Bjorken-$x$ region probed by this measurement.  Reference~\cite{Czakon:2020coa} presents a detailed study of the NLO and NNLO fiducial cross-sections for different charm-jet selections.  That study uses the same lepton fiducial definition as this paper.
While  $\Wplusc$-jet cross-section calculations cannot be compared with \sigmaWplusD measurements, they provide insight into the behavior of \Rc.
The $\Wplusc$-jet \Rc value calculated at NLO using an OS--SS selection is consistent within statistical uncertainties with that obtained for \WplusDmeson using \MGNLO and the same PDF set (\NNPDF[3.1]).  The NNLO+EW(NLO) value of the $\Wplusc$-jet \Rc is smaller than the NLO value by ${\sim}1\%$, but the two are consistent within the quoted 1\%\ statistical uncertainty.  The effects of NNLO scale uncertainties on \Rc are below 0.3\%.  These results suggest that the PDF comparisons presented in \Fig{\ref{fig:results:Rc:combined}} are likely to look similar for an NNLO+EW(NLO) calculation.

\begin{figure}[htbp]
\centering
\subfloat[]{
\includegraphics[width=0.50\textwidth]{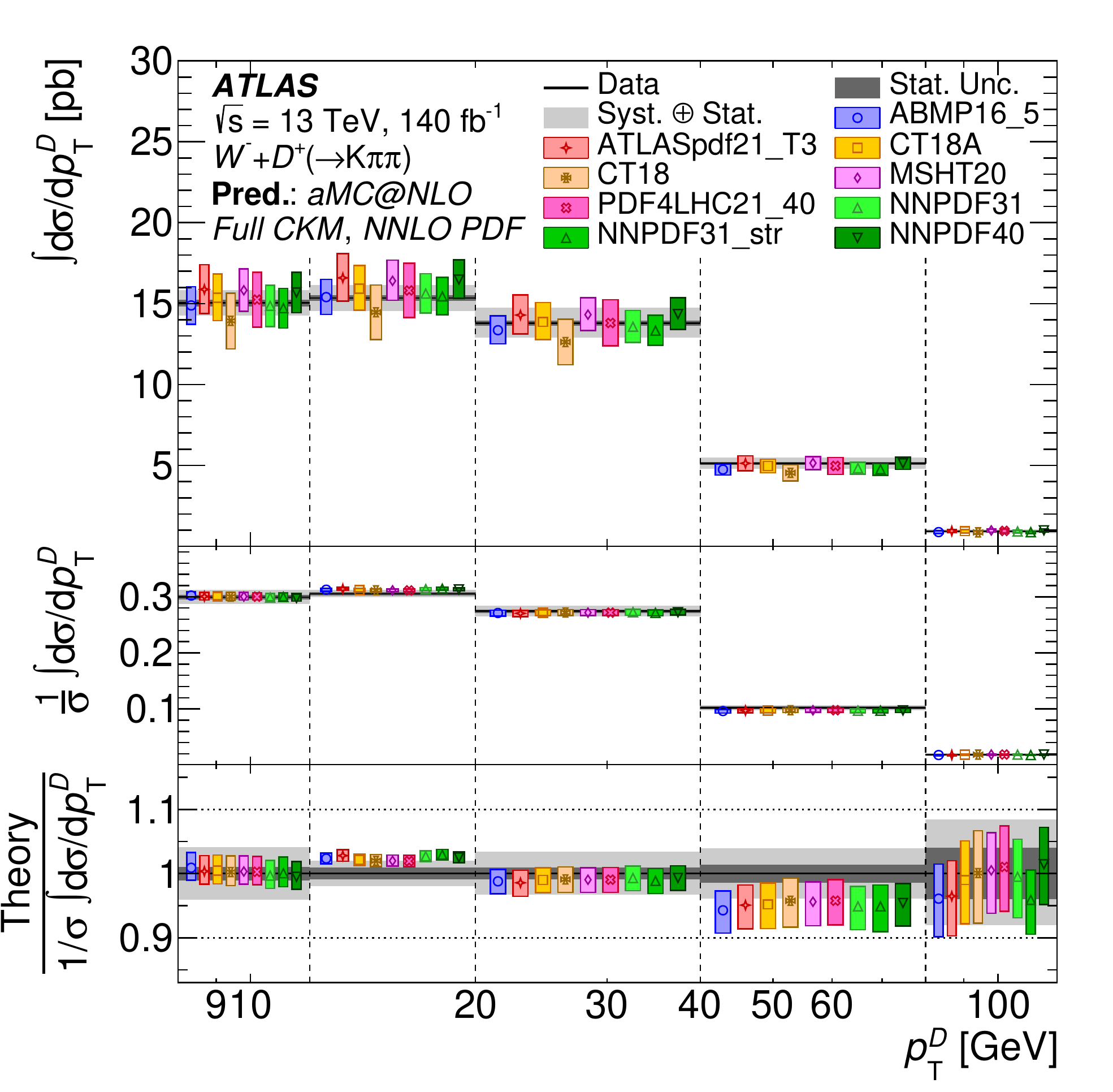}
\label{fig:results:fit_results_dplus:pdf:Wminus_pt}
}
\subfloat[]{
\includegraphics[width=0.50\textwidth]{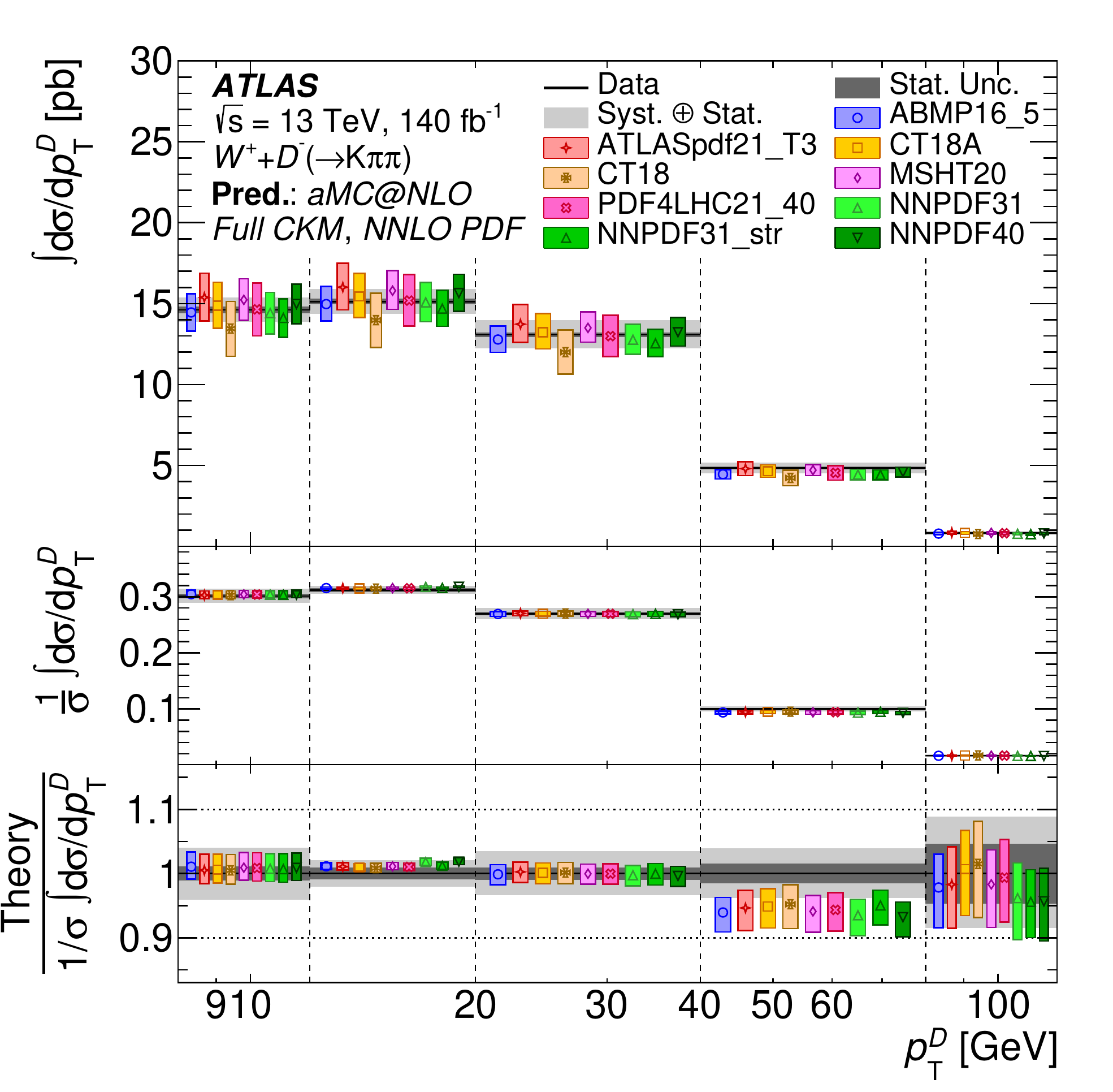}
\label{fig:results:fit_results_dplus:pdf:Wplus_pt}
}
\\
\subfloat[]{
\includegraphics[width=0.50\textwidth]{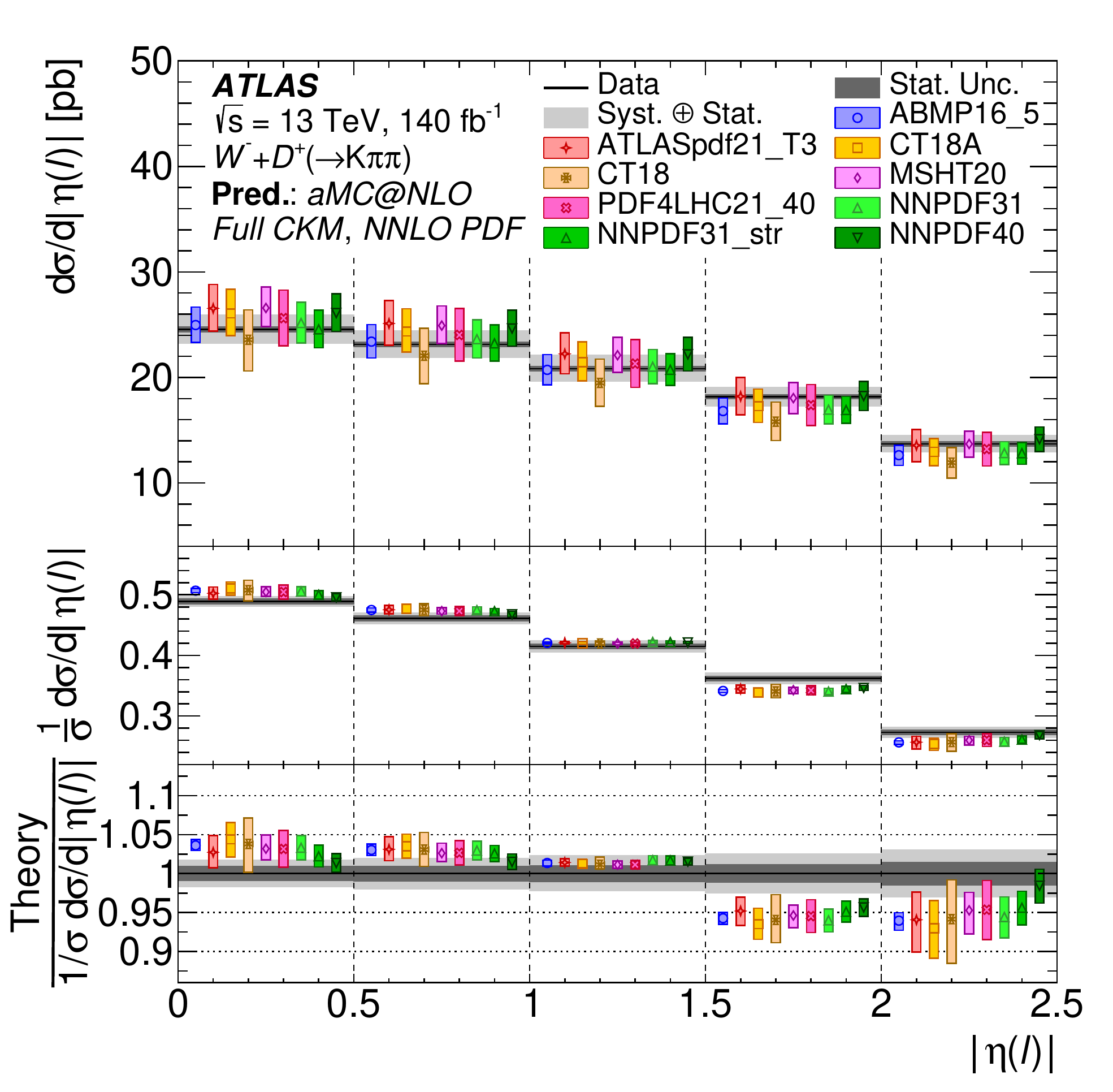}
\label{fig:results:fit_results_dplus:pdf:Wminus_eta}
}
\subfloat[]{
\includegraphics[width=0.50\textwidth]{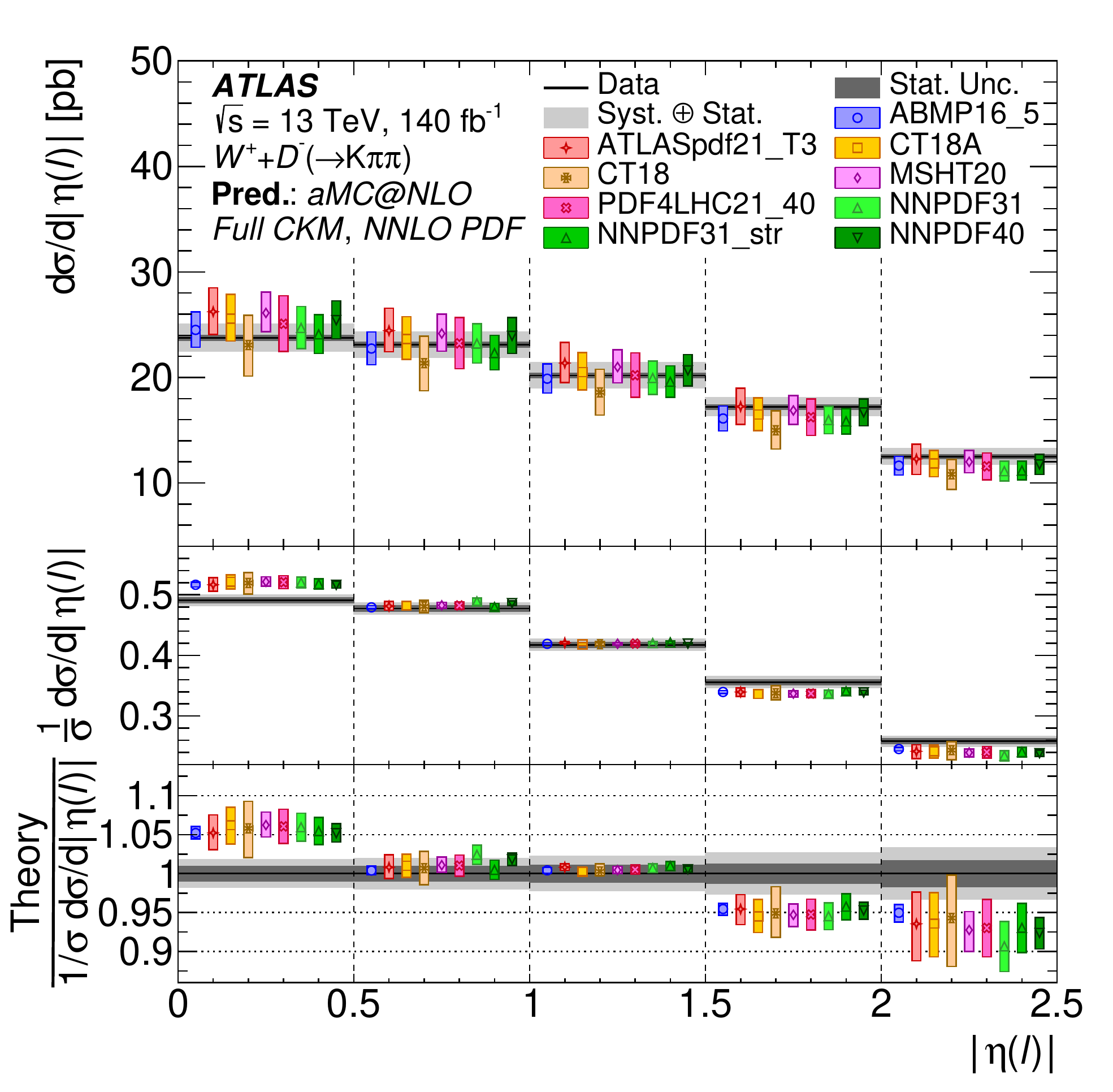}
\label{fig:results:fit_results_dplus:pdf:Wplus_eta}
}
\caption{
Measured differential fiducial cross-section times the single-lepton-flavor $W$ branching ratio
compared with different NNLO PDF predictions in the \Dplus channel:
\protect\subref{fig:results:fit_results_dplus:pdf:Wminus_pt} \WmplusD $\pt(\Dplus)$,
\protect\subref{fig:results:fit_results_dplus:pdf:Wplus_pt} \WpplusD $\pt(\Dplus)$,
\protect\subref{fig:results:fit_results_dplus:pdf:Wminus_eta} \WmplusD \diffeta, and
\protect\subref{fig:results:fit_results_dplus:pdf:Wplus_eta} \WpplusD \diffeta.
The displayed cross sections in \pt(\Dplus) plots are integrated over each differential bin.
Error bars on the MC predictions are the quadrature sum of the QCD scale uncertainty, PDF uncertainties,
hadronization uncertainties, and matching uncertainty. The PDF predictions are based on NLO calculations
performed using \AMCatNLO and a full CKM matrix: ABMP16\_5~\cite{Alekhin:2017kpj}, ATLASpdf21\_T3~\cite{STDM-2020-32},
CT18A, CT18~\cite{Hou:2019efy}, MSHT20~\cite{Bailey:2020ooq}, PDF4LHC21\_40~\cite{PDF4LHCWorkingGroup:2022cjn}, NNPDF31~\cite{NNPDF:2017mvq},
NNPDF31\_str~\cite{Faura:2020oom}, NNPDF40~\cite{NNPDF:2021njg}.
ABMP16\_5, ATLASpdf21\_T3, CT18A, and CT18 impose symmetric strange-sea PDFs.
}
\label{fig:results:pdf:Dplus_diff}
\end{figure}

\begin{figure}[htbp]
\centering
\subfloat[]{
\includegraphics[width=0.50\textwidth]{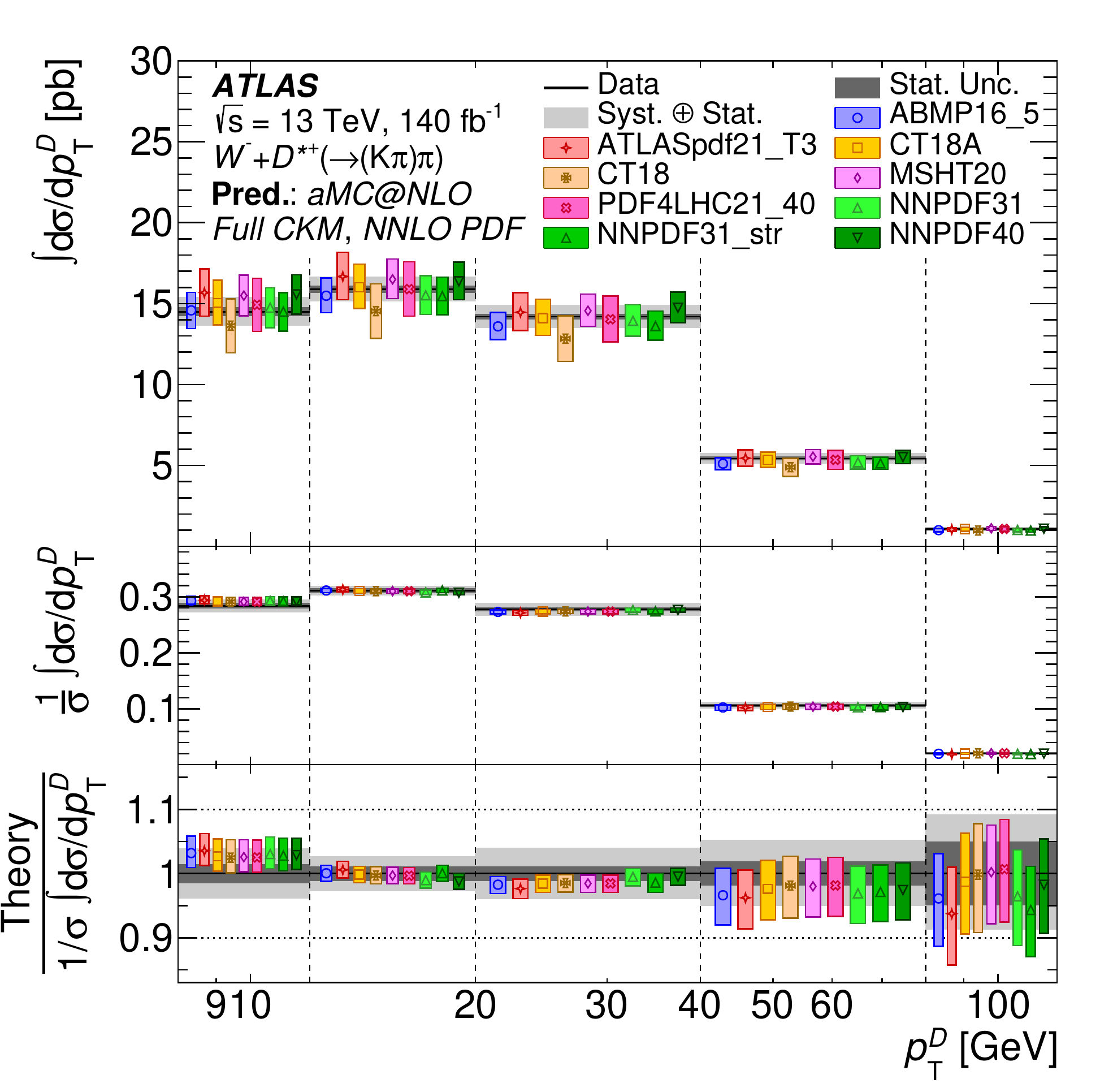}
\label{fig:results:fit_results_dstar:pdf:Wminus_pt}
}
\subfloat[]{
\includegraphics[width=0.50\textwidth]{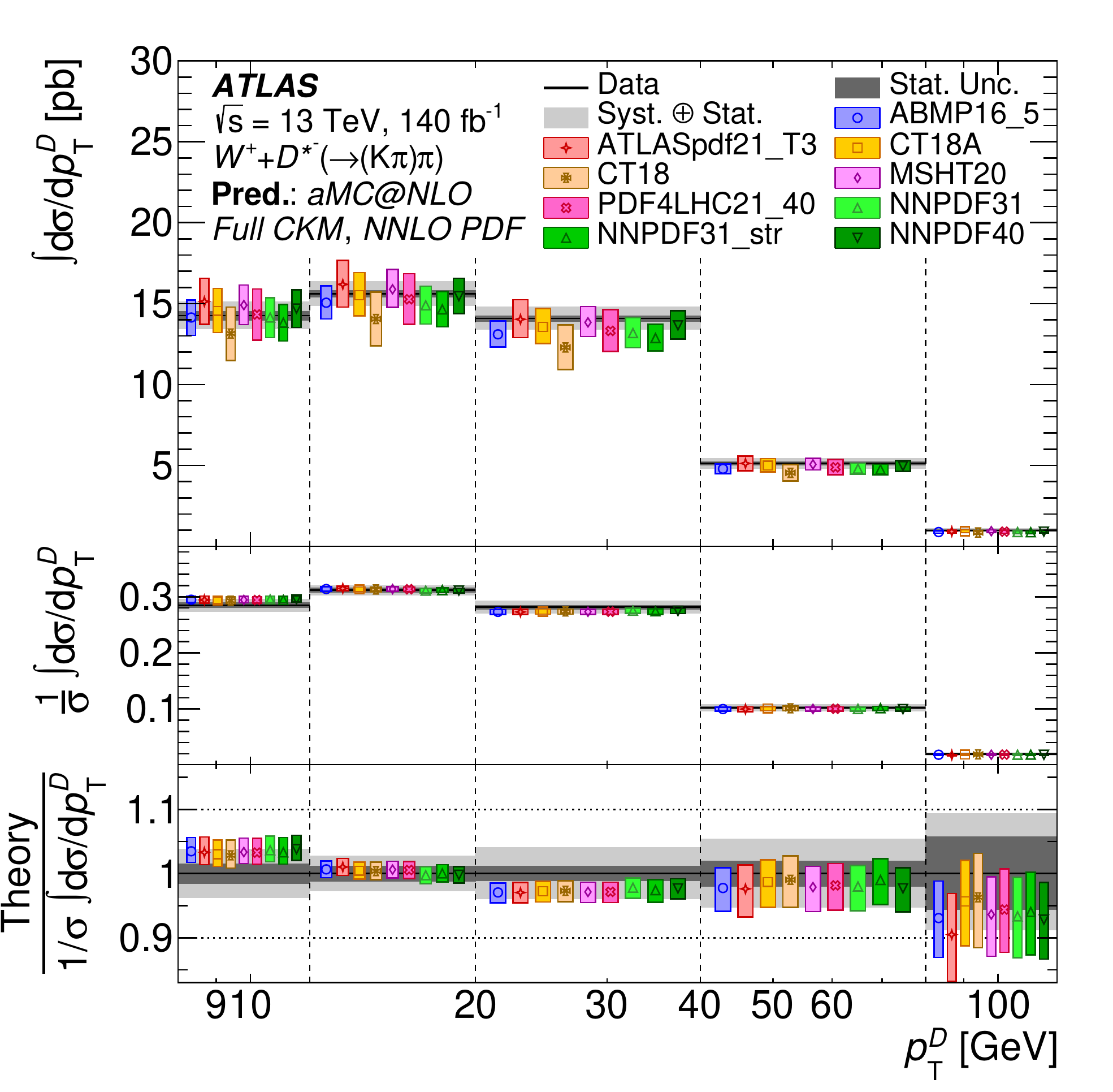}
\label{fig:results:fit_results_dstar:pdf:Wplus_pt}
}
\\
\subfloat[]{
\includegraphics[width=0.50\textwidth]{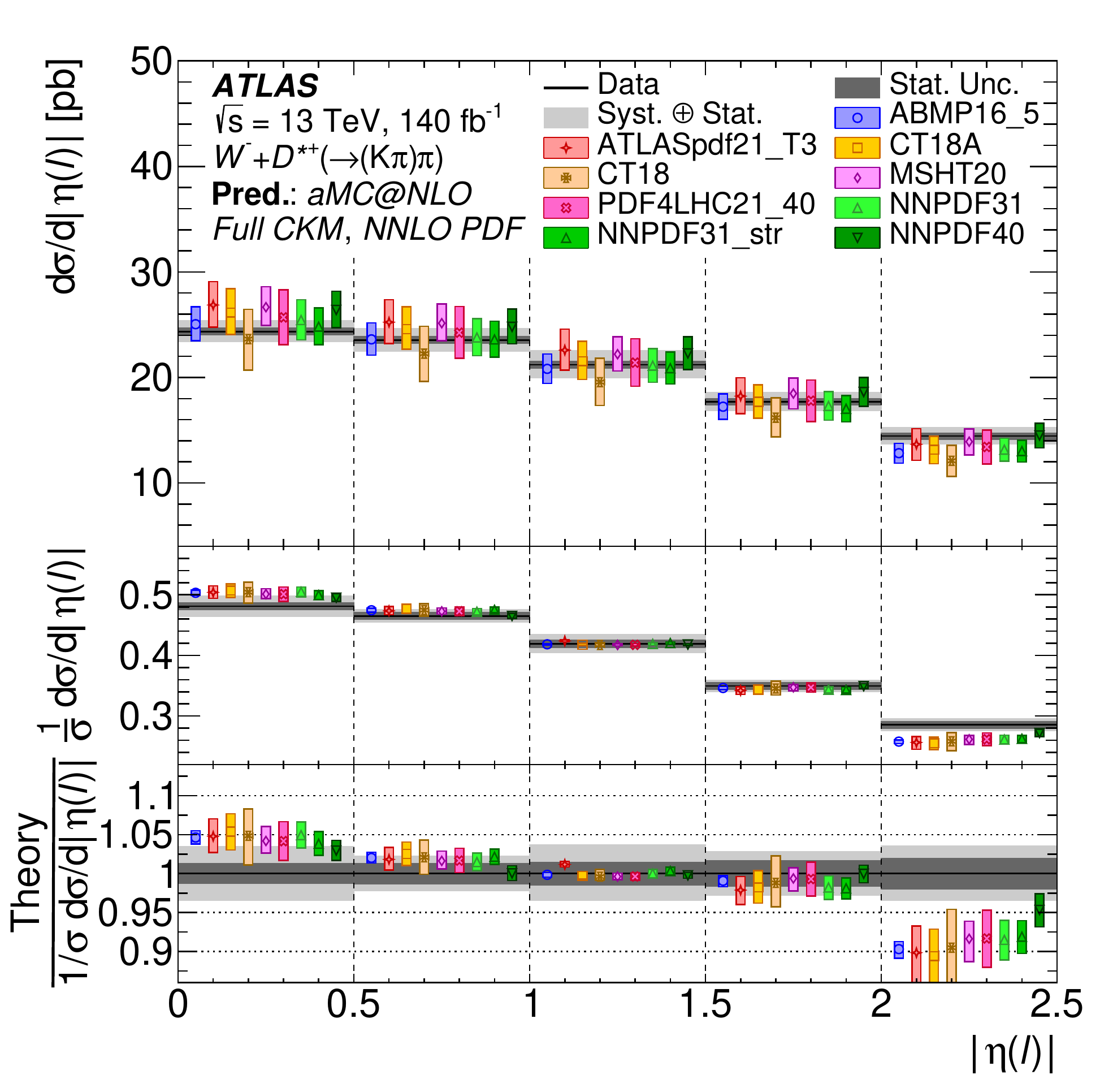}
\label{fig:results:fit_results_dstar:pdf:Wminus_eta}
}
\subfloat[]{
\includegraphics[width=0.50\textwidth]{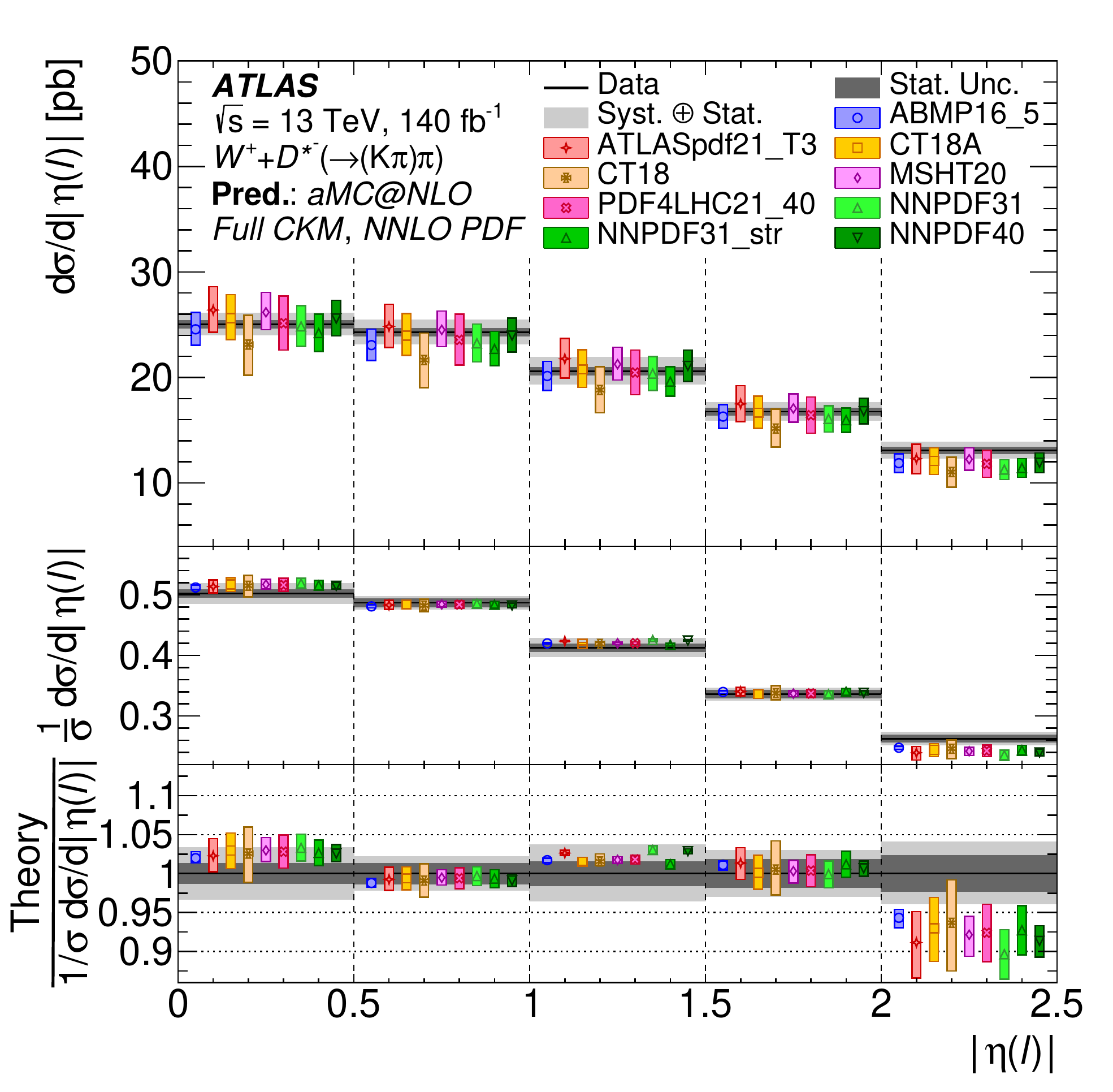}
\label{fig:results:fit_results_dstar:pdf:Wplus_eta}
}
\caption{
Measured differential fiducial cross-section times the single-lepton-flavor $W$ branching ratio
compared with different PDF predictions in the \Dstarp channel:
\protect\subref{fig:results:fit_results_dstar:pdf:Wminus_pt} \WmplusDstar $\pt(\Dstarp)$,
\protect\subref{fig:results:fit_results_dstar:pdf:Wplus_pt} \WpplusDstar $\pt(\Dstarp)$,
\protect\subref{fig:results:fit_results_dstar:pdf:Wminus_eta} \WmplusDstar \diffeta, and
\protect\subref{fig:results:fit_results_dstar:pdf:Wplus_eta} \WpplusDstar \diffeta.
The displayed cross sections in \pt(\Dplus) plots are integrated over each differential bin.
Error bars on the MC predictions are the quadrature sum of the QCD scale uncertainty, PDF uncertainties,
hadronization uncertainties, and matching uncertainty. The PDF predictions are based on NLO calculations
performed using \AMCatNLO and a full CKM matrix: ABMP16\_5~\cite{Alekhin:2017kpj}, ATLASpdf21\_T3~\cite{STDM-2020-32},
CT18A, CT18~\cite{Hou:2019efy}, MSHT20~\cite{Bailey:2020ooq}, PDF4LHC21\_40~\cite{PDF4LHCWorkingGroup:2022cjn}, NNPDF31~\cite{NNPDF:2017mvq},
NNPDF31\_str~\cite{Faura:2020oom}, NNPDF40~\cite{NNPDF:2021njg}.
ABMP16\_5, ATLASpdf21\_T3, CT18A, and CT18 impose symmetric strange-sea PDFs.
}
\label{fig:results:pdf:Dstar_diff}
\end{figure}

The differential cross-sections are shown in \Figs{\ref{fig:results:pdf:Dplus_diff}}{\ref{fig:results:pdf:Dstar_diff}},
together with the predicted cross-sections obtained with different choices of NNLO PDF set. The patterns observed in the \Dplus and \Dstarp
channels are consistent, for both the differential \Dmeson \pT and \diffeta\ distributions. For each \Dmeson species and charge, the differential
distributions are plotted in three separate panels. The top panel compares the measured differential cross-section with theoretical predictions
obtained using the same PDF sets as in \Fig{\ref{fig:results:sigmaWD}}. Systematic uncertainties in the predictions are correlated between
bins and are dominated by uncertainties in the normalization. Differences between PDF sets can be seen more clearly in the middle and lower panels,
which show the normalized differential cross-sections and the ratio of the predictions to the normalized cross-sections, respectively.
Because the integral of the normalized cross-section across all bins is constrained to be unity, the measurements are highly correlated
between bins: if the normalized cross-section in one bin increases, that in another bin must decrease.
 
Variations in the shape of the \diffpt distribution depend only weakly on the choice of PDF. Experimental sensitivity to this
dependence is reduced by the presence of \pT-dependent systematic uncertainties in the \Dmeson\ fiducial efficiency. Thus, while measurements
of the cross-section as a function of \diffpt are an important test of the quality of MC modeling, they do not provide incisive constraints on PDFs.
Systematic uncertainties for \diffeta are small and highly correlated among bins, providing good sensitivity to PDF variations. Measured differential
cross-sections have a broader \diffeta distribution than the central values of the predictions obtained with any of the PDF sets. The significance of
the discrepancy is reduced if the PDF uncertainties are considered.
 
The compatibility of the measurements and predictions is tested with a $\chi^2$ formula using experimental
and theory covariance matrices,
 
$$\chi^2 = \sum_{i,j} (x_i - \mu_i) \left( C^{-1} \right)_{ij} (x_j - \mu_j),$$
 
where $\vec{x}$ are the measured differential cross-sections in the 10 \diffeta bins, and $\vec{\mu}$ are the predicted
cross-sections in the same bin and depend on the choice of  PDF set. The total covariance matrix $C$ is the sum of
the experimental covariance matrix, encoding the measurement error, and the theory covariance matrix describing the
uncertainties in the theory predictions as described below. The $\chi^2$ is then converted to a $p$-value assuming 10
degrees of freedom. Experimental covariance matrices are given in \App{\ref{appendix:cov}}. The theory covariance
matrix corresponding to the PDF uncertainty is calculated following the LHAPDF prescription~\cite{Buckley:2014ana}.
Other theory uncertainties are assumed to be \SI{100}{\percent} correlated across differential bins.
 
The resulting $p$-values for the \AMCatNLO predictions of the \diffeta differential cross-sections with different PDF sets are
given in \Tab{\ref{tab:results:differential:dplus:eta:compatibility}} for the \Dplus channel and in
\Tab{\ref{tab:results:differential:dstar:eta:compatibility}} for the \Dstarp channel. The $p$-values are calculated
with progressively more systematic uncertainties included in the theory covariance matrix, ranging from an \enquote{Exp.\ Only} calculation,
where no systematic uncertainties related to the theory predictions are included, to a calculation including all theory uncertainties:
QCD scale, \enquote{hadronization and matching}, and PDF uncertainties. The hadronization and matching uncertainty is defined to be the quadrature
sum of the uncertainty in the charm production fractions, two-point uncertainties associated with the choice of showering program (\PYTHIA vs.\ \HERWIG),
the tune (A14 vs.\ Monash) and the matching algorithm (\AMCatNLO vs.\ \POWHEG). These uncertainties are treated as fully correlated between the
\WpplusD and \WmplusD channels. Without considering the theory uncertainties (i.e.\ just comparing the PDF central values with the experimental measurements)
the $p$-values are below \SI{10}{\percent} for all PDFs in the \Dplus channel and most of the PDFs in the \Dstarp channel. Adding hadronization
and QCD scale uncertainties increases the probabilities to at most \SI{15}{\percent} in the \Dplus channel and \SI{24}{\percent} in
the \Dstarp channel. Although the QCD scale uncertainty is a large uncertainty in the absolute cross-section, it does not change the
$p$-values significantly because the uncertainty is \SI{100}{\percent} correlated between the \diffeta bins, and it does not have a
large impact on the shape of the differential distribution. Adding the PDF uncertainties greatly increases the $p$-values; the PDF
uncertainty has a significant effect on the shape of the differential \diffeta distribution. This suggests that including these measurements
in a global PDF fit would provide useful constraints on the allowed PDF variations.
 
\begin{table}[htpb]
\caption{The $p$-values for compatibility of the measurement and the predictions, calculated with the $\chi^2$ formula
using experimental and theory covariance matrices. The first column shows the $p$-values for the \diffeta(\Dplus)
differential cross-section using only experimental uncertainties. The next columns show $p$-values when
progressively more theory systematic uncertainties are included.
The PDF predictions are based on NLO calculations performed using \AMCatNLO and a full CKM matrix:
ABMP16\_5~\cite{Alekhin:2017kpj}, ATLASpdf21\_T3~\cite{STDM-2020-32}, CT18A, CT18~\cite{Hou:2019efy}, MSHT20~\cite{Bailey:2020ooq},
PDF4LHC21\_40~\cite{PDF4LHCWorkingGroup:2022cjn}, NNPDF31~\cite{NNPDF:2017mvq}, NNPDF31\_str~\cite{Faura:2020oom}, NNPDF40~\cite{NNPDF:2021njg}.
ABMP16\_5, ATLASpdf21\_T3, CT18A, and CT18 impose symmetric strange-sea PDFs.
}
\label{tab:results:differential:dplus:eta:compatibility}
\centering
\begin{tabular}{
l |
S[table-format=2.1, round-precision=1, round-mode=places] |
S[table-format=2.1, round-precision=1, round-mode=places] |
S[table-format=2.1, round-precision=1, round-mode=places] |
S[table-format=2.1, round-precision=1, round-mode=places]
}
\toprule
Channel & \multicolumn{4}{c}{\Dplus \diffeta} \\
\midrule
$p$-value for PDF [\%]
& \multicolumn{1}{c|}{Exp.\ Only}
& \multicolumn{1}{c|}{$\oplus$ QCD Scale}
& \multicolumn{1}{c|}{$\oplus$ Had.\ and Matching}
& \multicolumn{1}{c}{$\oplus$ PDF} \\
\midrule
ABMP16\_5\_nnlo                   & 7.090085 & 11.808625 & 12.894139 & 19.754568 \\
ATLASpdf21\_T3                    & 9.034615 & 9.663936 & 11.515174 & 84.668574 \\
CT18ANNLO                         & 0.716061 & 0.957653 & 1.125354 & 76.044492 \\
CT18NNLO                          & 1.428913 & 6.057887 & 6.299684 & 87.640329 \\
MSHT20nnlo\_as118                 & 2.712597 & 2.906547 & 3.256016 & 45.601462 \\
PDF4LHC21\_40                     & 3.857017 & 5.283401 & 5.617906 & 75.806332 \\
NNPDF31\_nnlo\_as\_0118\_hessian  & 1.486617 & 2.600123 & 2.757183 & 50.721853 \\
NNPDF31\_nnlo\_as\_0118\_strange  & 9.064739 & 14.732635 & 15.204274 & 59.935306 \\
NNPDF40\_nnlo\_as\_01180\_hessian & 9.856755 & 10.154986 & 10.170689 & 43.735452 \\
\bottomrule
\end{tabular}
\end{table}
 
\begin{table}[htpb]
\caption{The $p$-values for compatibility of the measurement and the predictions, calculated with the $\chi^2$ formula
using experimental and theory covariance matrices. The first column shows the $p$-values for the \diffeta(\Dstarp)
differential cross-section using only experimental uncertainties. The next columns show $p$-values when
progressively more theory systematic uncertainties are included.
The PDF predictions are based on NLO calculations performed using \AMCatNLO and a full CKM matrix:
ABMP16\_5~\cite{Alekhin:2017kpj}, ATLASpdf21\_T3~\cite{STDM-2020-32}, CT18A, CT18~\cite{Hou:2019efy}, MSHT20~\cite{Bailey:2020ooq},
PDF4LHC21\_40~\cite{PDF4LHCWorkingGroup:2022cjn}, NNPDF31~\cite{NNPDF:2017mvq}, NNPDF31\_str~\cite{Faura:2020oom}, NNPDF40~\cite{NNPDF:2021njg}.
ABMP16\_5, ATLASpdf21\_T3, CT18A, and CT18 impose symmetric strange-sea PDFs.
}
\label{tab:results:differential:dstar:eta:compatibility}
\centering
\begin{tabular}{
l |
S[table-format=2.1, round-precision=1, round-mode=places] |
S[table-format=2.1, round-precision=1, round-mode=places] |
S[table-format=2.1, round-precision=1, round-mode=places] |
S[table-format=2.1, round-precision=1, round-mode=places]
}
\toprule
Channel & \multicolumn{4}{c}{\Dstarp \diffeta} \\
\midrule
$p$-value for PDF [\%]
& \multicolumn{1}{c|}{Exp.\ Only}
& \multicolumn{1}{c|}{$\oplus$ QCD Scale}
& \multicolumn{1}{c|}{$\oplus$ Had.\ and Matching}
& \multicolumn{1}{c}{$\oplus$ PDF} \\
\midrule
ABMP16\_5\_nnlo                   & 22.772418 & 23.652713 & 25.015109 & 28.834252 \\
ATLASpdf21\_T3                    & 1.904514 & 2.859969 & 3.390468 & 33.717575 \\
CT18ANNLO                         & 6.539364 & 6.880909 & 7.785994 & 47.278717 \\
CT18NNLO                          & 9.386447 & 19.207494 & 19.681068 & 52.789701 \\
MSHT20nnlo\_as118                 & 7.035638 & 9.430646 & 10.397872 & 31.274221 \\
PDF4LHC21\_40                     & 14.150632 & 14.241338 & 15.185642 & 51.378703 \\
NNPDF31\_nnlo\_as\_0118\_hessian  & 5.028859 & 5.063434 & 5.465402 & 34.859813 \\
NNPDF31\_nnlo\_as\_0118\_strange  & 11.443917 & 12.421560 & 13.222548 & 45.970413 \\
NNPDF40\_nnlo\_as\_01180\_hessian & 4.467649 & 6.073818 & 6.393742 & 36.035473 \\
\bottomrule
\end{tabular}
\end{table}
 
\FloatBarrier


\section{Conclusions}
\label{sec:conclusions}
 
Fiducial cross-sections for $W$ boson production in association with a \Dmeson meson are measured as a function of
\diffpt and \diffeta using \SI{140.1}{\ifb} of $\sqrt{s}=\SI{13}{\TeV}$ $pp$ collision data at
collected with the ATLAS detector at the Large Hadron Collider. A secondary-vertex fit is used to tag events containing
a \Dplus or a \Dstarp meson and a profile likelihood fit is used to extract the \WplusDmeson observables.
The single-lepton-species integrated cross-sections and cross-section ratios for the fiducial region $\pT(\ell) > \SI{30}{\GeV}$, $|\eta(\ell)| < 2.5$,
$\pT(\Dmeson) > \SI{8}{\GeV}$ and $|\eta(\Dmeson)| < 2.2$ are measured to be:
\begin{eqnarray*}
\sigmaWmplusD     & = & 50.2 \pm0.2  \,\mathrm{(stat.)}\,^{+2.4}_{-2.3}\,\mathrm{(syst.)}\,\mathrm{pb} \\
\sigmaWpplusD     & = & 48.5 \pm0.2  \,\mathrm{(stat.)}\,^{+2.3}_{-2.2}\,\mathrm{(syst.)}\,\mathrm{pb} \\
\sigmaWmplusDstar & = & 51.1 \pm0.4  \,\mathrm{(stat.)}\,^{+1.9}_{-1.8}\,\mathrm{(syst.)}\,\mathrm{pb} \\
\sigmaWpplusDstar & = & 50.0 \pm0.4  \,\mathrm{(stat.)}\,^{+1.9}_{-1.8}\,\mathrm{(syst.)}\,\mathrm{pb} \\
\Rc(\Dmeson)      & = & 0.971\pm0.006\,\mathrm{(stat.)}\,\pm0.011\,\mathrm{(syst.)} \\
\end{eqnarray*}
 
The uncertainty in the measured absolute integrated and differential fiducial cross-sections is about \SI{5}{\percent}
and is dominated by the systematic uncertainty. On the other hand, cross-section
ratios and normalized differential cross-sections are measured with percent-level precision and have comparable
contributions from systematic and statistical uncertainties. The experimental precision of these measurements is
comparable to the PDF uncertainties and smaller than the total theory uncertainty.
 
Measured differential cross-sections as a function of \diffeta have a broader distribution than the central values
of the predictions. These measurements are, however, consistent with the predictions if the uncertainties associated with the PDF sets
are included, indicating that these measurements would provide useful constraints for global PDF fits. The measured values of \Rc are
consistent with predictions obtained with a range of PDF sets, including those that constrain the $s$-$\overline s$ sea to be symmetric.


\section*{Acknowledgments}


We thank CERN for the very successful operation of the LHC, as well as the
support staff from our institutions without whom ATLAS could not be
operated efficiently.
 
We acknowledge the support of
ANPCyT, Argentina;
YerPhI, Armenia;
ARC, Australia;
BMWFW and FWF, Austria;
ANAS, Azerbaijan;
CNPq and FAPESP, Brazil;
NSERC, NRC and CFI, Canada;
CERN;
ANID, Chile;
CAS, MOST and NSFC, China;
Minciencias, Colombia;
MEYS CR, Czech Republic;
DNRF and DNSRC, Denmark;
IN2P3-CNRS and CEA-DRF/IRFU, France;
SRNSFG, Georgia;
BMBF, HGF and MPG, Germany;
GSRI, Greece;
RGC and Hong Kong SAR, China;
ISF and Benoziyo Center, Israel;
INFN, Italy;
MEXT and JSPS, Japan;
CNRST, Morocco;
NWO, Netherlands;
RCN, Norway;
MEiN, Poland;
FCT, Portugal;
MNE/IFA, Romania;
MESTD, Serbia;
MSSR, Slovakia;
ARRS and MIZ\v{S}, Slovenia;
DSI/NRF, South Africa;
MICINN, Spain;
SRC and Wallenberg Foundation, Sweden;
SERI, SNSF and Cantons of Bern and Geneva, Switzerland;
MOST, Taiwan;
TENMAK, T\"urkiye;
STFC, United Kingdom;
DOE and NSF, United States of America.
In addition, individual groups and members have received support from
BCKDF, CANARIE, Compute Canada and CRC, Canada;
PRIMUS 21/SCI/017 and UNCE SCI/013, Czech Republic;
COST, ERC, ERDF, Horizon 2020 and Marie Sk{\l}odowska-Curie Actions, European Union;
Investissements d'Avenir Labex, Investissements d'Avenir Idex and ANR, France;
DFG and AvH Foundation, Germany;
Herakleitos, Thales and Aristeia programmes co-financed by EU-ESF and the Greek NSRF, Greece;
BSF-NSF and MINERVA, Israel;
Norwegian Financial Mechanism 2014-2021, Norway;
NCN and NAWA, Poland;
La Caixa Banking Foundation, CERCA Programme Generalitat de Catalunya and PROMETEO and GenT Programmes Generalitat Valenciana, Spain;
G\"{o}ran Gustafssons Stiftelse, Sweden;
The Royal Society and Leverhulme Trust, United Kingdom.
 
The crucial computing support from all WLCG partners is acknowledged gratefully, in particular from CERN, the ATLAS Tier-1 facilities at TRIUMF (Canada), NDGF (Denmark, Norway, Sweden), CC-IN2P3 (France), KIT/GridKA (Germany), INFN-CNAF (Italy), NL-T1 (Netherlands), PIC (Spain), ASGC (Taiwan), RAL (UK) and BNL (USA), the Tier-2 facilities worldwide and large non-WLCG resource providers. Major contributors of computing resources are listed in Ref.~\cite{ATL-SOFT-PUB-2021-003}.


\clearpage
\clearpage
\appendix
 
\section{Breakdown of systematic uncertainties in differential bins}
\label{appendix:sys}
 
The breakdown of uncertainties in the measured differential fiducial cross-sections is summarized in
\Tabrange{\ref{tab:fits:systematics:diff:dplus:pt}}{\ref{tab:fits:systematics:diff:dstar:eta}}.
The uncertainties in the normalized cross-sections are given in parentheses next to the uncertainties
in the corresponding absolute cross-sections.
 
\begin{table}[htpb]
\caption{Summary of the main systematic uncertainties as percentages of the measured observable
for the $\pT(\Dplus)$ differential cross-sections in the \Dplus channel. The uncertainty in the corresponding
normalized cross-section is given in parentheses next to the uncertainty in the absolute differential
cross-section.}
\label{tab:fits:systematics:diff:dplus:pt}
\renewcommand{\arraystretch}{1.0}
\setlength\tabcolsep{3.2pt}
\centering
\resizebox*{1.0\textwidth}{!}{
\begin{tabular}{l | c c c c c | c c c c c
}
\toprule
Uncertainty [\%]              & \multicolumn{5}{c|}{$d\sigmaWmplusD/d(\pT(\Dplus))$ ($1/\sigma d\sigma/d\pT$)}
& \multicolumn{5}{c}{$d\sigmaWpplusD/d(\pT(\Dplus))$ ($1/\sigma d\sigma/d\pT$)} \\
\midrule
$\pT(\Dplus)$ bins [\GeV]       & $[8,\,12]$
& $[12,\,20]$
& $[20,\,40]$
& $[40,\,80]$
& $[80,\,\infty)$
& $[8,\,12]$
& $[12,\,20]$
& $[20,\,40]$
& $[40,\,80]$
& $[80,\,\infty)$ \\
\midrule
SV reconstruction             & 3.1 (1.2) & 2.8 (0.6) & 3.2 (0.7) & 4.7 (2.6) & 5.7 (4.3) & 2.6 (1.0) & 2.5 (0.7) & 3.3 (0.7) & 4.5 (2.5) & 5.8 (3.9) \\
Jets and \MET        & 1.8 (0.8) & 1.9 (0.4) & 1.9 (0.5) & 2.0 (1.2) & 3.4 (2.4) & 2.1 (0.6) & 1.9 (0.6) & 2.1 (0.7) & 2.0 (1.2) & 3.7 (2.7) \\
Luminosity                    & 0.8 (0.0) & 0.8 (0.0) & 0.8 (0.0) & 0.8 (0.0) & 0.8 (0.0) & 0.8 (0.0) & 0.8 (0.0) & 0.8 (0.0) & 0.8 (0.0) & 0.8 (0.0) \\
Muon reconstruction           & 0.8 (0.2) & 0.7 (0.1) & 0.6 (0.1) & 0.5 (0.3) & 0.6 (0.5) & 0.8 (0.2) & 0.7 (0.1) & 0.6 (0.1) & 0.5 (0.3) & 0.5 (0.4) \\
Electron reconstruction       & 0.2 (0.0) & 0.2 (0.1) & 0.3 (0.0) & 0.4 (0.2) & 0.5 (0.4) & 0.2 (0.0) & 0.2 (0.0) & 0.2 (0.0) & 0.4 (0.2) & 0.5 (0.4) \\
Multijet background           & 0.3 (0.2) & 0.3 (0.1) & 0.2 (0.1) & 0.1 (0.3) & 1.1 (1.3) & 0.1 (0.1) & 0.3 (0.1) & 0.2 (0.1) & 0.2 (0.1) & 0.1 (0.2) \\
\midrule
Signal modeling               & 1.5 (3.2) & 2.7 (0.7) & 4.6 (2.7) & 2.4 (0.4) & 3.0 (1.2) & 1.5 (3.2) & 2.7 (0.7) & 4.6 (2.7) & 2.3 (0.4) & 3.0 (1.1) \\
Signal branching ratio        & 1.7 (0.1) & 1.6 (0.0) & 1.5 (0.1) & 1.6 (0.0) & 1.7 (0.1) & 1.7 (0.1) & 1.6 (0.0) & 1.5 (0.1) & 1.6 (0.0) & 1.7 (0.1) \\
Background modeling           & 1.7 (1.4) & 1.5 (0.8) & 1.8 (1.2) & 1.8 (1.6) & 1.8 (1.7) & 1.9 (1.5) & 1.6 (1.0) & 1.8 (1.3) & 1.6 (1.5) & 3.5 (3.2) \\
\midrule
Finite size of MC samples     & 2.3 (1.7) & 1.7 (1.3) & 1.6 (1.3) & 2.1 (1.9) & 4.6 (4.6) & 2.4 (1.8) & 1.7 (1.3) & 1.7 (1.4) & 2.1 (1.9) & 4.8 (4.6) \\
Data statistical uncertainty  & 1.2 (1.0) & 0.9 (0.8) & 0.9 (0.9) & 1.4 (1.4) & 4.0 (4.0) & 1.3 (1.1) & 1.0 (0.9) & 1.0 (0.9) & 1.5 (1.5) & 4.6 (4.6) \\
\midrule
Total                         & 5.1 (4.0) & 5.1 (1.9) & 6.5 (3.3) & 6.5 (3.9) & 9.9 (8.2) & 5.0 (4.0) & 5.0 (2.0) & 6.6 (3.4) & 6.3 (3.8) & 10.6 (8.6) \\
\bottomrule
\end{tabular}
}
\end{table}
 
\begin{table}[htpb]
\caption{Summary of the main systematic uncertainties as percentages of the measured observable
for the \diffeta differential cross-sections in the \Dplus channel. The uncertainty in the corresponding
normalized cross-section is given in parentheses next to the uncertainty in the absolute differential
cross-section.}
\label{tab:fits:systematics:diff:dplus:eta}
\renewcommand{\arraystretch}{1.0}
\setlength\tabcolsep{3.2pt}
\centering
\resizebox*{1.0\textwidth}{!}{
\begin{tabular}{l | c c c c c | c c c c c
}
\toprule
Uncertainty [\%]              & \multicolumn{5}{c|}{$d\sigmaWmplusD/d(\diffeta)$ ($1/\sigma d\sigma/d\eta$)}
& \multicolumn{5}{c}{$d\sigmaWpplusD/d(\diffeta)$ ($1/\sigma d\sigma/d\eta$)} \\
\midrule
\diffeta bins                 & $[0.0,\,0.5]$
& $[0.5,\,1.0]$
& $[1.0,\,1.5]$
& $[1.5,\,2.0]$
& $[2.0,\,2.5]$
& $[0.0,\,0.5]$
& $[0.5,\,1.0]$
& $[1.0,\,1.5]$
& $[1.5,\,2.0]$
& $[2.0,\,2.5]$ \\
\midrule
SV reconstruction             & 3.2 (0.1) & 3.1 (0.2) & 3.2 (0.2) & 3.2 (0.1) & 3.3 (0.2) & 3.1 (0.1) & 3.0 (0.1) & 3.1 (0.2) & 3.0 (0.2) & 3.1 (0.2) \\
Jets and \MET        & 1.6 (0.2) & 1.9 (0.4) & 1.6 (0.2) & 1.5 (0.6) & 1.7 (0.4) & 1.6 (0.2) & 1.8 (0.3) & 1.8 (0.2) & 1.5 (0.4) & 1.9 (0.5) \\
Luminosity                    & 0.8 (0.0) & 0.8 (0.0) & 0.8 (0.0) & 0.8 (0.0) & 0.8 (0.0) & 0.8 (0.0) & 0.8 (0.0) & 0.8 (0.0) & 0.8 (0.0) & 0.8 (0.0) \\
Muon reconstruction           & 0.5 (0.2) & 0.6 (0.1) & 0.8 (0.1) & 0.8 (0.1) & 0.8 (0.2) & 0.5 (0.2) & 0.6 (0.1) & 0.8 (0.1) & 0.8 (0.1) & 0.9 (0.2) \\
Electron reconstruction       & 0.2 (0.2) & 0.3 (0.0) & 0.3 (0.1) & 0.4 (0.1) & 0.4 (0.1) & 0.2 (0.2) & 0.3 (0.0) & 0.3 (0.1) & 0.4 (0.1) & 0.4 (0.2) \\
Multijet background           & 0.2 (0.2) & 0.2 (0.2) & 0.2 (0.2) & 0.3 (0.1) & 0.9 (0.7) & 0.2 (0.3) & 0.1 (0.1) & 0.1 (0.1) & 0.4 (0.3) & 0.7 (0.6) \\
\midrule
Signal modeling               & 3.2 (0.4) & 2.9 (0.3) & 3.9 (1.1) & 1.8 (1.4) & 2.4 (0.7) & 3.2 (0.4) & 2.9 (0.3) & 3.9 (1.2) & 1.9 (1.4) & 2.5 (0.7) \\
Signal branching ratio        & 1.6 (0.0) & 1.6 (0.0) & 1.5 (0.0) & 1.6 (0.0) & 1.5 (0.0) & 1.6 (0.0) & 1.6 (0.0) & 1.6 (0.0) & 1.7 (0.1) & 1.6 (0.0) \\
Background modeling           & 1.5 (0.8) & 2.2 (1.2) & 1.7 (0.7) & 1.2 (0.8) & 2.1 (1.3) & 1.8 (0.7) & 2.0 (1.2) & 1.7 (0.8) & 1.3 (0.9) & 1.9 (1.4) \\
\midrule
Finite size of MC samples     & 1.6 (1.3) & 1.8 (1.4) & 2.1 (1.6) & 1.9 (1.7) & 2.7 (2.4) & 1.7 (1.3) & 1.8 (1.5) & 1.9 (1.5) & 2.2 (1.8) & 3.0 (2.7) \\
Data statistical uncertainty  & 1.0 (0.9) & 1.1 (1.0) & 1.2 (1.1) & 1.2 (1.1) & 1.6 (1.5) & 1.1 (1.0) & 1.1 (1.0) & 1.2 (1.1) & 1.3 (1.2) & 1.8 (1.7) \\
\midrule
Total                         & 5.5 (1.7) & 5.5 (2.0) & 6.0 (2.3) & 5.0 (2.5) & 5.8 (3.0) & 5.4 (1.8) & 5.4 (2.0) & 6.0 (2.3) & 5.1 (2.7) & 6.0 (3.4) \\
\bottomrule
\end{tabular}
}
\end{table}
 
\begin{table}[htpb]
\caption{Summary of the main systematic uncertainties as percentages of the measured observable
for the $\pT(\Dstar)$ differential cross-sections in the \Dstar channel. The uncertainty in the corresponding
normalized cross-section is given in parentheses next to the uncertainty in the absolute differential
cross-section.}
\label{tab:fits:systematics:diff:dstar:pt}
\renewcommand{\arraystretch}{1.0}
\setlength\tabcolsep{3.2pt}
\centering
\resizebox*{1.0\textwidth}{!}{
\begin{tabular}{l | c c c c c | c c c c c
}
\toprule
Uncertainty [\%]              & \multicolumn{5}{c|}{$d\sigmaWmplusD/d(\pT(\Dstar))$ ($1/\sigma d\sigma/d\pT$)}
& \multicolumn{5}{c}{$d\sigmaWpplusD/d(\pT(\Dstar))$ ($1/\sigma d\sigma/d\pT$)} \\
\midrule
$\pT(\Dstar)$ bins [\GeV]       & $[8,\,12]$
& $[12,\,20]$
& $[20,\,40]$
& $[40,\,80]$
& $[80,\,\infty)$
& $[8,\,12]$
& $[12,\,20]$
& $[20,\,40]$
& $[40,\,80]$
& $[80,\,\infty)$ \\
\midrule
SV reconstruction             & 2.4 (0.5) & 2.3 (0.3) & 2.3 (0.3) & 2.4 (1.0) & 4.5 (2.8) & 2.4 (0.5) & 2.3 (0.3) & 2.3 (0.4) & 2.5 (1.0) & 4.8 (2.9) \\
Jets and \MET        & 1.5 (0.6) & 1.6 (0.5) & 1.4 (0.5) & 2.0 (1.3) & 4.3 (3.2) & 1.4 (0.6) & 1.8 (0.6) & 1.5 (0.4) & 1.8 (1.3) & 3.9 (3.2) \\
Luminosity                    & 0.8 (0.0) & 0.8 (0.0) & 0.8 (0.0) & 0.8 (0.0) & 0.8 (0.0) & 0.8 (0.0) & 0.8 (0.0) & 0.8 (0.0) & 0.8 (0.0) & 0.8 (0.0) \\
Muon reconstruction           & 0.8 (0.2) & 0.7 (0.1) & 0.6 (0.1) & 0.5 (0.4) & 0.6 (0.6) & 0.8 (0.2) & 0.7 (0.1) & 0.6 (0.1) & 0.5 (0.3) & 0.5 (0.5) \\
Electron reconstruction       & 0.2 (0.1) & 0.1 (0.2) & 0.3 (0.2) & 0.4 (0.2) & 0.6 (0.4) & 0.1 (0.1) & 0.1 (0.2) & 0.3 (0.2) & 0.4 (0.2) & 0.6 (0.4) \\
Multijet background           & 0.1 (0.1) & 0.1 (0.1) & 0.1 (0.1) & 0.1 (0.1) & 1.0 (1.0) & 0.2 (0.1) & 0.1 (0.1) & 0.1 (0.1) & 0.1 (0.1) & 0.3 (0.4) \\
\midrule
Signal modeling               & 3.7 (2.9) & 2.6 (1.9) & 3.1 (3.5) & 3.5 (3.9) & 1.2 (0.4) & 3.7 (2.8) & 2.6 (1.9) & 3.1 (3.5) & 3.5 (3.9) & 1.2 (0.4) \\
Signal branching ratio        & 1.1 (0.0) & 1.0 (0.0) & 1.1 (0.0) & 1.1 (0.0) & 1.1 (0.0) & 1.0 (0.0) & 1.0 (0.0) & 1.1 (0.0) & 1.1 (0.0) & 1.1 (0.0) \\
Background modeling           & 2.2 (1.3) & 1.3 (0.6) & 1.2 (0.7) & 1.2 (0.9) & 2.7 (2.2) & 1.7 (0.8) & 1.5 (0.5) & 1.3 (0.7) & 1.8 (1.5) & 1.9 (1.8) \\
\midrule
Finite size of MC samples     & 2.6 (1.9) & 1.8 (1.4) & 1.7 (1.4) & 2.6 (2.3) & 7.2 (6.9) & 2.5 (1.8) & 1.9 (1.4) & 1.7 (1.4) & 2.7 (2.4) & 6.3 (6.0) \\
Data statistical uncertainty  & 1.8 (1.4) & 1.2 (1.1) & 1.1 (1.1) & 1.9 (1.8) & 5.0 (4.9) & 1.9 (1.5) & 1.3 (1.1) & 1.2 (1.1) & 2.0 (2.0) & 5.7 (5.7) \\
\midrule
Total                         & 6.0 (3.8) & 4.7 (2.7) & 4.8 (4.0) & 5.8 (5.1) & 10.3 (8.9) & 5.8 (3.8) & 4.8 (2.7) & 4.8 (4.0) & 6.0 (5.3) & 10.4 (9.1) \\
\bottomrule
\end{tabular}
}
\end{table}
 
\begin{table}[htpb]
\caption{Summary of the main systematic uncertainties as percentages of the measured observable
for the \diffeta differential cross-sections in the \Dstar channel. The uncertainty in the corresponding
normalized cross-section is given in parentheses next to the uncertainty in the absolute differential
cross-section.}
\label{tab:fits:systematics:diff:dstar:eta}
\renewcommand{\arraystretch}{1.0}
\setlength\tabcolsep{3.2pt}
\centering
\resizebox*{1.0\textwidth}{!}{
\begin{tabular}{l | c c c c c | c c c c c
}
\toprule
Uncertainty [\%]              & \multicolumn{5}{c|}{$d\sigmaWmplusD/d(\diffeta)$ ($1/\sigma d\sigma/d\eta$)}
& \multicolumn{5}{c}{$d\sigmaWpplusD/d(\diffeta)$ ($1/\sigma d\sigma/d\eta$)} \\
\midrule
\diffeta bins                 & $[0.0,\,0.5]$
& $[0.5,\,1.0]$
& $[1.0,\,1.5]$
& $[1.5,\,2.0]$
& $[2.0,\,2.5]$
& $[0.0,\,0.5]$
& $[0.5,\,1.0]$
& $[1.0,\,1.5]$
& $[1.5,\,2.0]$
& $[2.0,\,2.5]$ \\
\midrule
SV reconstruction             & 2.4 (0.1) & 2.4 (0.0) & 2.4 (0.1) & 2.5 (0.1) & 2.5 (0.2) & 2.4 (0.1) & 2.5 (0.1) & 2.4 (0.2) & 2.4 (0.1) & 2.4 (0.1) \\
Jets and \MET        & 1.4 (0.7) & 1.5 (0.4) & 1.4 (0.4) & 1.6 (0.5) & 1.4 (1.0) & 1.5 (0.2) & 1.5 (0.2) & 1.4 (0.2) & 1.3 (0.3) & 1.1 (0.5) \\
Luminosity                    & 0.8 (0.0) & 0.8 (0.0) & 0.8 (0.0) & 0.8 (0.0) & 0.8 (0.0) & 0.8 (0.0) & 0.8 (0.0) & 0.8 (0.0) & 0.8 (0.0) & 0.8 (0.0) \\
Muon reconstruction           & 0.5 (0.2) & 0.6 (0.1) & 0.8 (0.1) & 0.8 (0.1) & 0.8 (0.2) & 0.5 (0.2) & 0.6 (0.1) & 0.8 (0.1) & 0.8 (0.1) & 0.9 (0.2) \\
Electron reconstruction       & 0.2 (0.1) & 0.2 (0.0) & 0.3 (0.0) & 0.4 (0.1) & 0.3 (0.1) & 0.2 (0.1) & 0.3 (0.0) & 0.3 (0.1) & 0.3 (0.1) & 0.3 (0.1) \\
Multijet background           & 0.1 (0.1) & 0.2 (0.1) & 0.1 (0.1) & 0.2 (0.2) & 0.2 (0.2) & 0.1 (0.1) & 0.1 (0.1) & 0.1 (0.1) & 0.1 (0.1) & 0.2 (0.2) \\
\midrule
Signal modeling               & 1.1 (2.7) & 2.0 (0.2) & 4.6 (2.7) & 1.8 (0.4) & 2.6 (0.7) & 1.1 (2.7) & 2.1 (0.2) & 4.5 (2.7) & 1.8 (0.4) & 2.6 (0.8) \\
Signal branching ratio        & 1.1 (0.0) & 1.0 (0.0) & 1.0 (0.0) & 1.1 (0.0) & 1.1 (0.0) & 1.1 (0.0) & 1.0 (0.0) & 1.0 (0.0) & 1.0 (0.0) & 1.1 (0.0) \\
Background modeling           & 1.4 (0.6) & 1.8 (1.0) & 1.5 (0.8) & 1.7 (1.0) & 1.1 (0.7) & 1.4 (0.7) & 1.8 (1.0) & 1.3 (0.7) & 1.7 (1.1) & 1.6 (0.9) \\
\midrule
Finite size of MC samples     & 1.9 (1.6) & 1.9 (1.6) & 2.2 (1.8) & 2.6 (2.2) & 3.3 (2.9) & 1.8 (1.5) & 1.9 (1.6) & 2.1 (1.8) & 2.7 (2.3) & 3.8 (3.3) \\
Data statistical uncertainty  & 1.4 (1.3) & 1.5 (1.3) & 1.6 (1.5) & 1.8 (1.6) & 2.2 (2.0) & 1.4 (1.3) & 1.5 (1.3) & 1.7 (1.5) & 2.0 (1.8) & 2.5 (2.3) \\
\midrule
Total                         & 4.1 (3.5) & 4.6 (2.2) & 6.2 (3.6) & 5.0 (2.8) & 5.5 (3.5) & 4.1 (3.4) & 4.7 (2.2) & 6.2 (3.6) & 5.0 (3.0) & 6.0 (4.0) \\
\bottomrule
\end{tabular}
}
\end{table}
 
\FloatBarrier
 
\section{Differential cross-section tables}
\label{appendix:results}
 
The measured differential cross sections in bins of \diffpt and \diffeta are
shown in \Tabrange{\ref{tab:results:diff:dplus:pt}}{\ref{tab:results:diff:dstar:eta}} for the \Dplus and \Dstar channels.
 
\begin{table}[htpb]
\caption{Measured \diffptDplus differential fiducial cross-section times the single-lepton-flavor $W$ branching ratio
in the \WplusD channel. The displayed cross sections are integrated over each differential bin.}
\label{tab:results:diff:dplus:pt}
\renewcommand{\arraystretch}{1.2}
\centering
\begin{tabular}{l |
S[table-format=2.2, round-precision=2, round-mode=places] @{$\,\pm\,$}
S[table-format=1.2, round-precision=2, round-mode=places] @{$\,$(stat.)$\,$}
c @{$\,$(syst.)$\,$} |
S[table-format=1.4, round-precision=4, round-mode=places] @{$\,\pm\,$}
S[table-format=1.4, round-precision=4, round-mode=places] @{$\,$(stat.)$\,$}
c @{$\,$(syst.)$\,$}
}
\toprule
\diffpt[\GeV] & \multicolumn{3}{c|}{${\int}d\sigmaWmplusD/d(\pt(\Dplus))$ [pb]} & \multicolumn{3}{c}{$1/\sigma\,{\int}d\sigmaWmplusD/d(\pt(\Dplus))$} \\
\midrule
$[8 ,\,12]$ & 15.0380 & 0.1865 & $^{+0.76}_{-0.72}$ & 0.299442 & 0.003008 & $^{+0.0117}_{-0.0116}$ \\
$[12,\,20]$ & 15.3390 & 0.1437 & $^{+0.78}_{-0.75}$ & 0.305431 & 0.002583 & $^{+0.0052}_{-0.0052}$ \\
$[20,\,40]$ & 13.7823 & 0.1241 & $^{+0.92}_{-0.85}$ & 0.274430 & 0.002357 & $^{+0.0088}_{-0.0085}$ \\
$[40,\,80]$ & 5.1296 & 0.0704 & $^{+0.34}_{-0.31}$ & 0.102139 & 0.001391 & $^{+0.0038}_{-0.0036}$ \\
$[80,\,\infty)$ & 0.9320 & 0.0369 & $^{+0.09}_{-0.08}$ & 0.018558 & 0.000732 & $^{+0.0014}_{-0.0013}$ \\
\midrule
& \multicolumn{3}{c|}{${\int}d\sigmaWpplusD/d(\pt(\Dminus))$ [pb]} & \multicolumn{3}{c}{$1/\sigma\,{\int}d\sigmaWpplusD/d(\pt(\Dminus))$} \\
\midrule
$[8 ,\,12]$ & 14.6085 & 0.1908 & $^{+0.73}_{-0.69}$ & 0.301425 & 0.003177 & $^{+0.0116}_{-0.0115}$ \\
$[12,\,20]$ & 15.1228 & 0.1476 & $^{+0.75}_{-0.72}$ & 0.312033 & 0.002741 & $^{+0.0057}_{-0.0057}$ \\
$[20,\,40]$ & 13.0716 & 0.1246 & $^{+0.89}_{-0.82}$ & 0.269705 & 0.002460 & $^{+0.0089}_{-0.0085}$ \\
$[40,\,80]$ & 4.8427 & 0.0714 & $^{+0.31}_{-0.29}$ & 0.099920 & 0.001465 & $^{+0.0036}_{-0.0035}$ \\
$[80,\,\infty)$ & 0.8199 & 0.0380 & $^{+0.08}_{-0.07}$ & 0.016917 & 0.000780 & $^{+0.0013}_{-0.0012}$ \\
\bottomrule
\end{tabular}
\end{table}
 
\begin{table}[htpb]
\caption{Measured \diffeta differential fiducial cross-section times the single-lepton-flavor $W$ branching ratio
in the \WplusD channel. The displayed cross sections are integrated over each differential bin.}
\label{tab:results:diff:dplus:eta}
\renewcommand{\arraystretch}{1.2}
\centering
\begin{tabular}{l |
S[table-format=2.2, round-precision=2, round-mode=places] @{$\,\pm\,$}
S[table-format=1.2, round-precision=2, round-mode=places] @{$\,$(stat.)$\,$}
c @{$\,$(syst.)$\,$} |
S[table-format=1.4, round-precision=4, round-mode=places] @{$\,\pm\,$}
S[table-format=1.4, round-precision=4, round-mode=places] @{$\,$(stat.)$\,$}
c @{$\,$(syst.)$\,$}
}
\toprule
\diffeta      & \multicolumn{3}{c|}{${\int}d\sigmaWmplusD/d(\diffeta)$ [pb]} & \multicolumn{3}{c}{$1/\sigma\,{\int}d\sigmaWmplusD/d(\diffeta)$} \\
\midrule
$[0.0,\,0.5]$ & 12.2741 & 0.1265 & $^{+0.67}_{-0.64}$ & 0.244551 & 0.002275 & $^{+0.0036}_{-0.0036}$ \\
$[0.5,\,1.0]$ & 11.5677 & 0.1218 & $^{+0.63}_{-0.61}$ & 0.230475 & 0.002214 & $^{+0.0040}_{-0.0040}$ \\
$[1.0,\,1.5]$ & 10.4133 & 0.1198 & $^{+0.64}_{-0.59}$ & 0.207476 & 0.002182 & $^{+0.0042}_{-0.0041}$ \\
$[1.5,\,2.0]$ & 9.0863 & 0.1106 & $^{+0.45}_{-0.43}$ & 0.181036 & 0.002046 & $^{+0.0041}_{-0.0041}$ \\
$[2.0,\,2.5]$ & 6.8491 & 0.1071 & $^{+0.39}_{-0.37}$ & 0.136462 & 0.001995 & $^{+0.0037}_{-0.0036}$ \\
\midrule
& \multicolumn{3}{c|}{${\int}d\sigmaWpplusD/d(\diffeta)$ [pb]} & \multicolumn{3}{c}{$1/\sigma\,{\int}d\sigmaWpplusD/d(\diffeta)$} \\
\midrule
$[0.0,\,0.5]$ & 11.8736 & 0.1263 & $^{+0.65}_{-0.62}$ & 0.245490 & 0.002381 & $^{+0.0037}_{-0.0037}$ \\
$[0.5,\,1.0]$ & 11.5464 & 0.1233 & $^{+0.61}_{-0.60}$ & 0.238726 & 0.002333 & $^{+0.0041}_{-0.0041}$ \\
$[1.0,\,1.5]$ & 10.0929 & 0.1217 & $^{+0.61}_{-0.57}$ & 0.208674 & 0.002304 & $^{+0.0042}_{-0.0040}$ \\
$[1.5,\,2.0]$ & 8.6040 & 0.1157 & $^{+0.43}_{-0.41}$ & 0.177890 & 0.002207 & $^{+0.0042}_{-0.0042}$ \\
$[2.0,\,2.5]$ & 6.2499 & 0.1126 & $^{+0.37}_{-0.35}$ & 0.129220 & 0.002168 & $^{+0.0038}_{-0.0037}$ \\
\bottomrule
\end{tabular}
\end{table}
 
\begin{table}[htpb]
\caption{Measured \diffptDstar differential fiducial cross-section times the single-lepton-flavor $W$ branching ratio
in the \WplusDstar channel. The displayed cross sections are integrated over each differential bin.}
\label{tab:results:diff:dstar:pt}
\renewcommand{\arraystretch}{1.2}
\centering
\begin{tabular}{l |
S[table-format=2.2, round-precision=2, round-mode=places] @{$\,\pm\,$}
S[table-format=1.2, round-precision=2, round-mode=places] @{$\,$(stat.)$\,$}
c @{$\,$(syst.)$\,$} |
S[table-format=1.4, round-precision=4, round-mode=places] @{$\,\pm\,$}
S[table-format=1.4, round-precision=4, round-mode=places] @{$\,$(stat.)$\,$}
c @{$\,$(syst.)$\,$}
}
\toprule
\diffpt[\GeV] & \multicolumn{3}{c|}{${\int}d\sigmaWmplusDstar/d(\pt(\Dstarp))$ [pb]} & \multicolumn{3}{c}{$1/\sigma\,{\int}d\sigmaWmplusDstar/d(\pt(\Dstarp))$} \\
\midrule
$[8 ,\,12]$ & 14.4967 & 0.2635 & $^{+0.85}_{-0.79}$ & 0.283948 & 0.004114 & $^{+0.0102}_{-0.0100}$ \\
$[12,\,20]$ & 15.8788 & 0.1933 & $^{+0.73}_{-0.69}$ & 0.311022 & 0.003415 & $^{+0.0075}_{-0.0075}$ \\
$[20,\,40]$ & 14.1889 & 0.1617 & $^{+0.68}_{-0.64}$ & 0.277923 & 0.003022 & $^{+0.0107}_{-0.0105}$ \\
$[40,\,80]$ & 5.4238 & 0.1010 & $^{+0.31}_{-0.29}$ & 0.106238 & 0.001944 & $^{+0.0052}_{-0.0049}$ \\
$[80,\,\infty)$ & 1.0654 & 0.0526 & $^{+0.10}_{-0.09}$ & 0.020869 & 0.001022 & $^{+0.0016}_{-0.0015}$ \\
\midrule
& \multicolumn{3}{c|}{${\int}d\sigmaWpplusDstar/d(\pt(\Dstarm))$ [pb]} & \multicolumn{3}{c}{$1/\sigma\,{\int}d\sigmaWpplusDstar/d(\pt(\Dstarm))$} \\
\midrule
$[8 ,\,12]$ & 14.2556 & 0.2718 & $^{+0.82}_{-0.76}$ & 0.284899 & 0.004315 & $^{+0.0100}_{-0.0097}$ \\
$[12,\,20]$ & 15.6012 & 0.2002 & $^{+0.74}_{-0.70}$ & 0.311793 & 0.003583 & $^{+0.0076}_{-0.0076}$ \\
$[20,\,40]$ & 14.0799 & 0.1658 & $^{+0.68}_{-0.64}$ & 0.281391 & 0.003179 & $^{+0.0108}_{-0.0107}$ \\
$[40,\,80]$ & 5.1128 & 0.1011 & $^{+0.30}_{-0.28}$ & 0.102179 & 0.001991 & $^{+0.0052}_{-0.0050}$ \\
$[80,\,\infty)$ & 0.9877 & 0.0564 & $^{+0.09}_{-0.08}$ & 0.019739 & 0.001119 & $^{+0.0015}_{-0.0013}$ \\
\bottomrule
\end{tabular}
\end{table}
 
\begin{table}[htpb]
\caption{Measured \diffeta differential fiducial cross-section times the single-lepton-flavor $W$ branching ratio
in the \WplusDstar channel. The displayed cross sections are integrated over each differential bin.}
\label{tab:results:diff:dstar:eta}
\renewcommand{\arraystretch}{1.2}
\centering
\begin{tabular}{l |
S[table-format=2.2, round-precision=2, round-mode=places] @{$\,\pm\,$}
S[table-format=1.2, round-precision=2, round-mode=places] @{$\,$(stat.)$\,$}
c @{$\,$(syst.)$\,$} |
S[table-format=1.4, round-precision=4, round-mode=places] @{$\,\pm\,$}
S[table-format=1.4, round-precision=4, round-mode=places] @{$\,$(stat.)$\,$}
c @{$\,$(syst.)$\,$}
}
\toprule
\diffeta      & \multicolumn{3}{c|}{${\int}d\sigmaWmplusDstar/d(\diffeta)$ [pb]} & \multicolumn{3}{c}{$1/\sigma\,{\int}d\sigmaWmplusDstar/d(\diffeta)$} \\
\midrule
$[0.0,\,0.5]$ & 12.1768 & 0.1762 & $^{+0.48}_{-0.46}$ & 0.240525 & 0.003129 & $^{+0.0078}_{-0.0078}$ \\
$[0.5,\,1.0]$ & 11.7719 & 0.1713 & $^{+0.53}_{-0.50}$ & 0.232530 & 0.003056 & $^{+0.0042}_{-0.0041}$ \\
$[1.0,\,1.5]$ & 10.6051 & 0.1726 & $^{+0.67}_{-0.61}$ & 0.209484 & 0.003072 & $^{+0.0071}_{-0.0066}$ \\
$[1.5,\,2.0]$ & 8.8473 & 0.1584 & $^{+0.42}_{-0.40}$ & 0.174759 & 0.002874 & $^{+0.0040}_{-0.0039}$ \\
$[2.0,\,2.5]$ & 7.2243 & 0.1563 & $^{+0.38}_{-0.36}$ & 0.142702 & 0.002845 & $^{+0.0042}_{-0.0040}$ \\
\midrule
& \multicolumn{3}{c|}{${\int}d\sigmaWpplusDstar/d(\diffeta)$ [pb]} & \multicolumn{3}{c}{$1/\sigma\,{\int}d\sigmaWpplusDstar/d(\diffeta)$} \\
\midrule
$[0.0,\,0.5]$ & 12.5222 & 0.1807 & $^{+0.50}_{-0.48}$ & 0.251041 & 0.003265 & $^{+0.0078}_{-0.0077}$ \\
$[0.5,\,1.0]$ & 12.1392 & 0.1788 & $^{+0.55}_{-0.52}$ & 0.243367 & 0.003232 & $^{+0.0042}_{-0.0042}$ \\
$[1.0,\,1.5]$ & 10.2927 & 0.1762 & $^{+0.64}_{-0.58}$ & 0.206349 & 0.003189 & $^{+0.0070}_{-0.0065}$ \\
$[1.5,\,2.0]$ & 8.3786 & 0.1644 & $^{+0.39}_{-0.37}$ & 0.167974 & 0.003022 & $^{+0.0040}_{-0.0039}$ \\
$[2.0,\,2.5]$ & 6.5478 & 0.1640 & $^{+0.37}_{-0.34}$ & 0.131269 & 0.003025 & $^{+0.0044}_{-0.0042}$ \\
\bottomrule
\end{tabular}
\end{table}
 
\FloatBarrier
 
\section{The measurement covariance matrices}
\label{appendix:cov}
 
\begin{figure}[htbp]
\centering
\subfloat[]{
\includegraphics[width=0.50\textwidth]{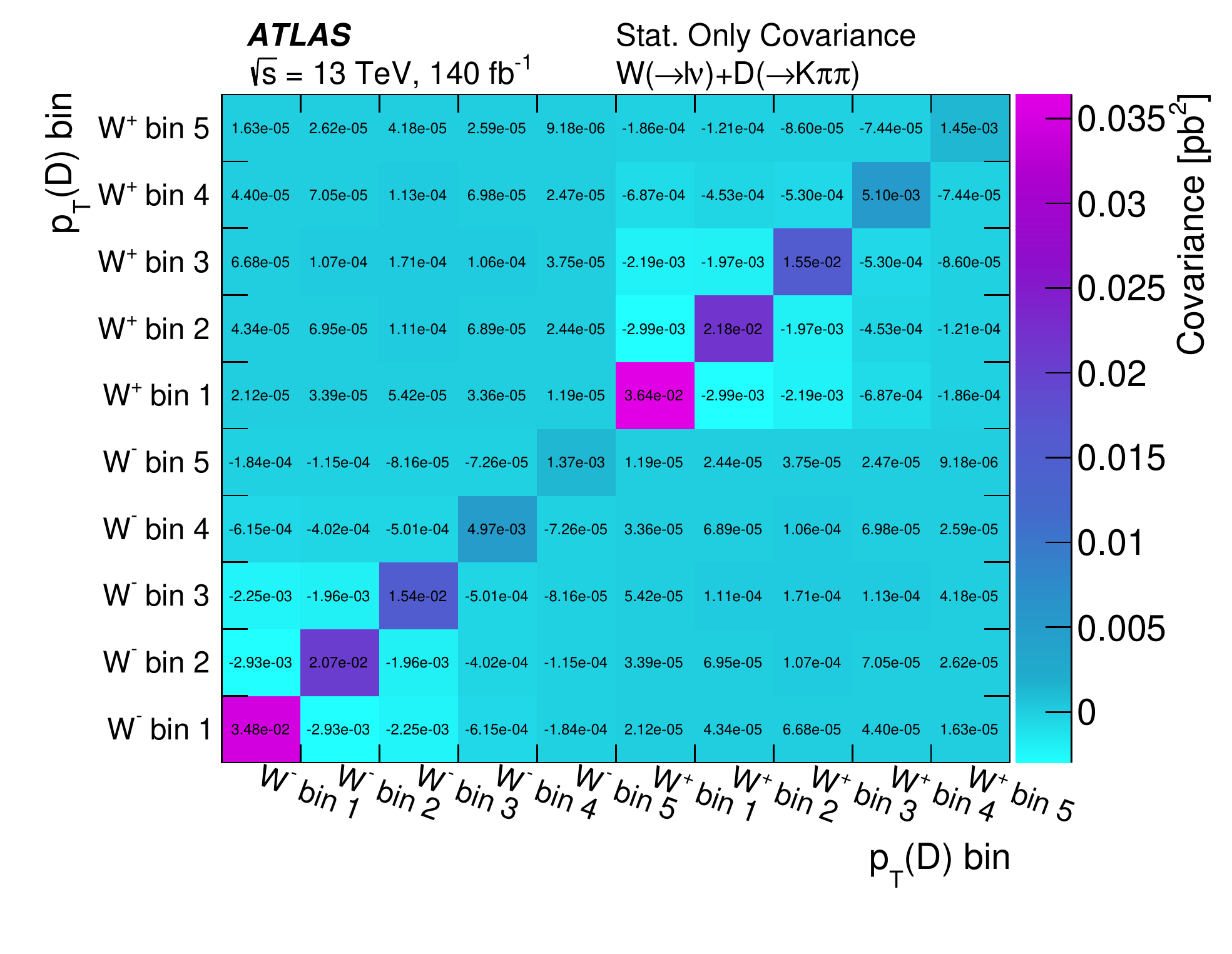}
\label{fig:results:covariance:Covariance_statOnly_dplus_pt}
}
\subfloat[]{
\includegraphics[width=0.50\textwidth]{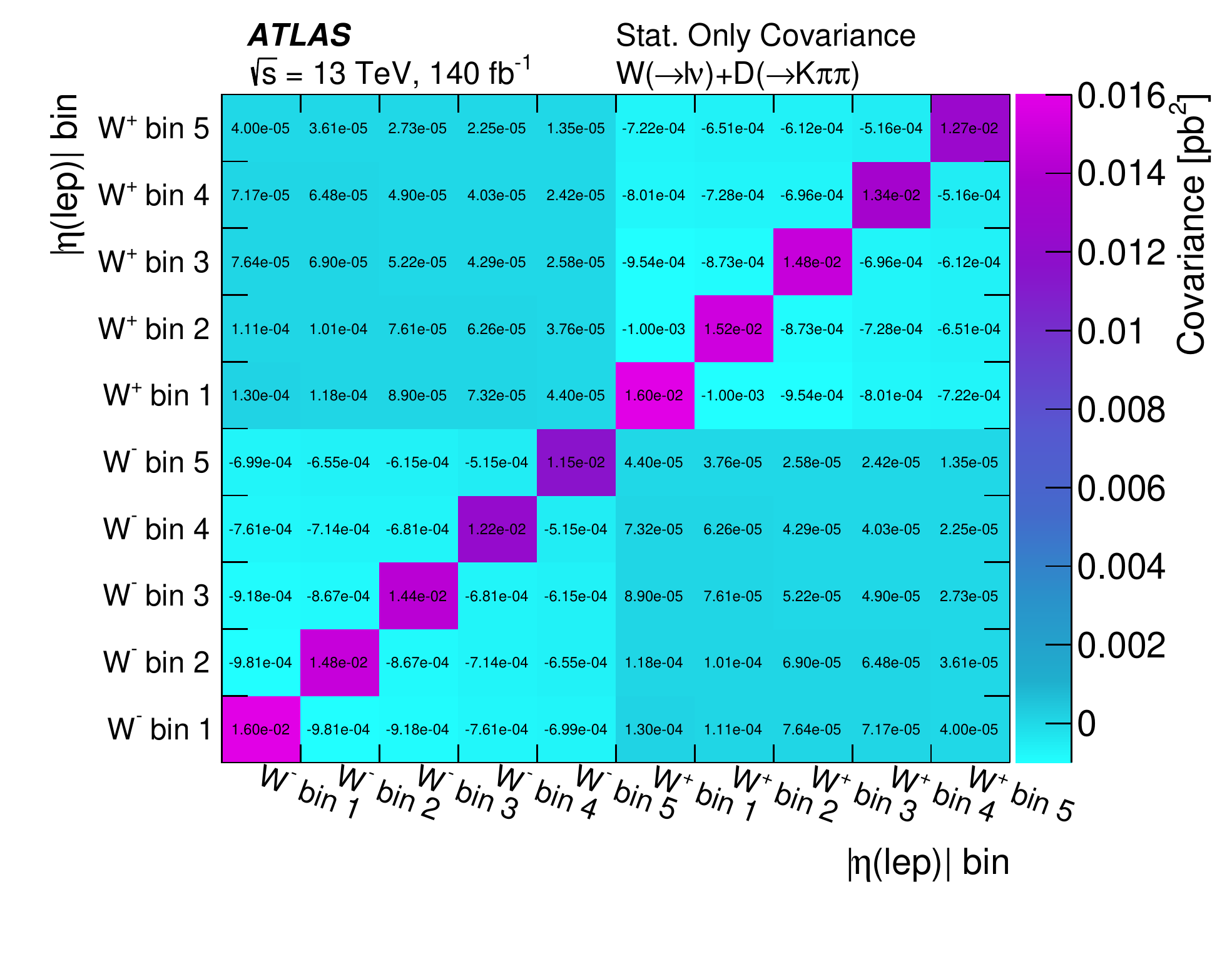}
\label{fig:results:covariance:Covariance_statOnly_dplus_eta}
}
\\
\subfloat[]{
\includegraphics[width=0.50\textwidth]{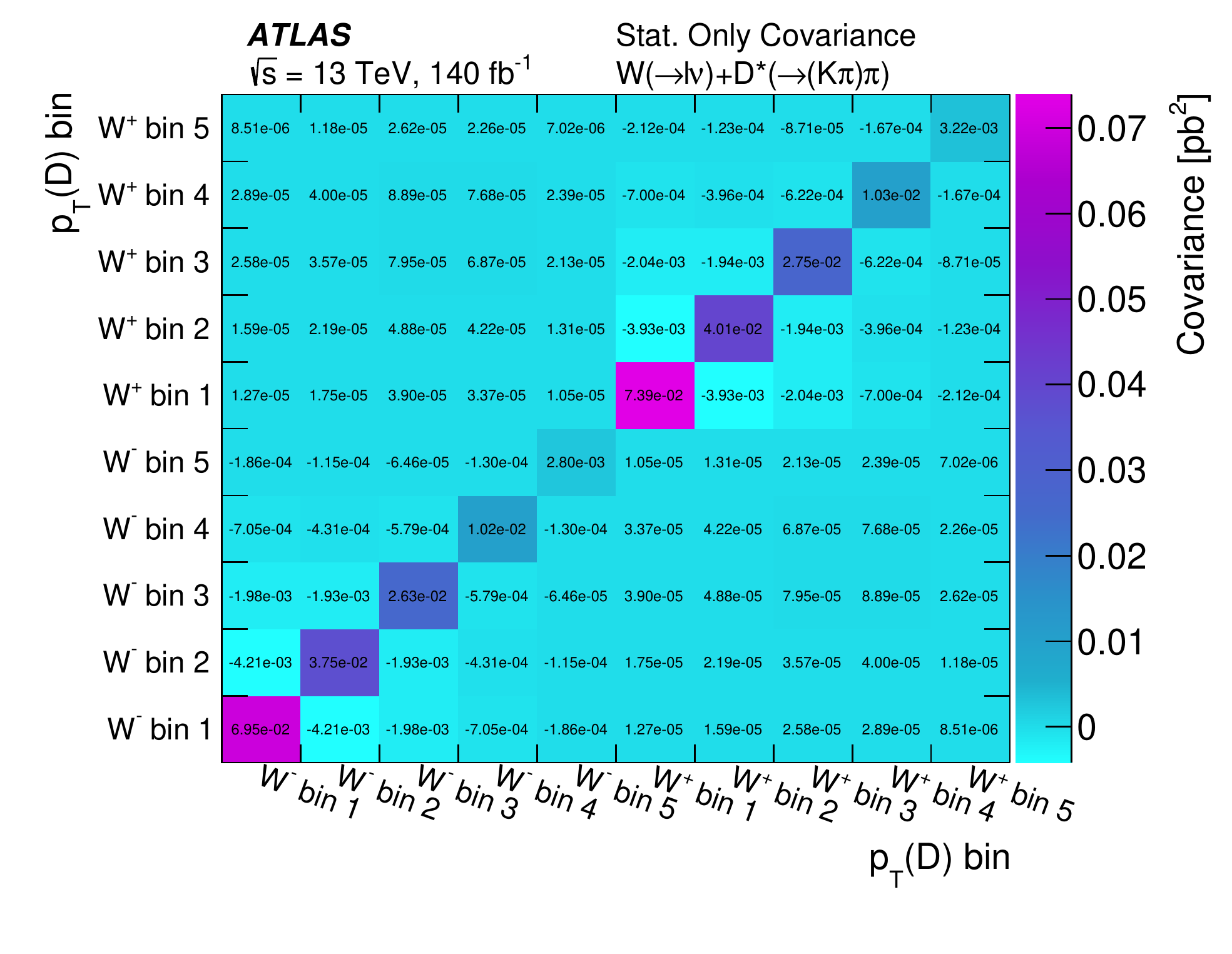}
\label{fig:results:covariance:Covariance_statOnly_dstar_pt}
}
\subfloat[]{
\includegraphics[width=0.50\textwidth]{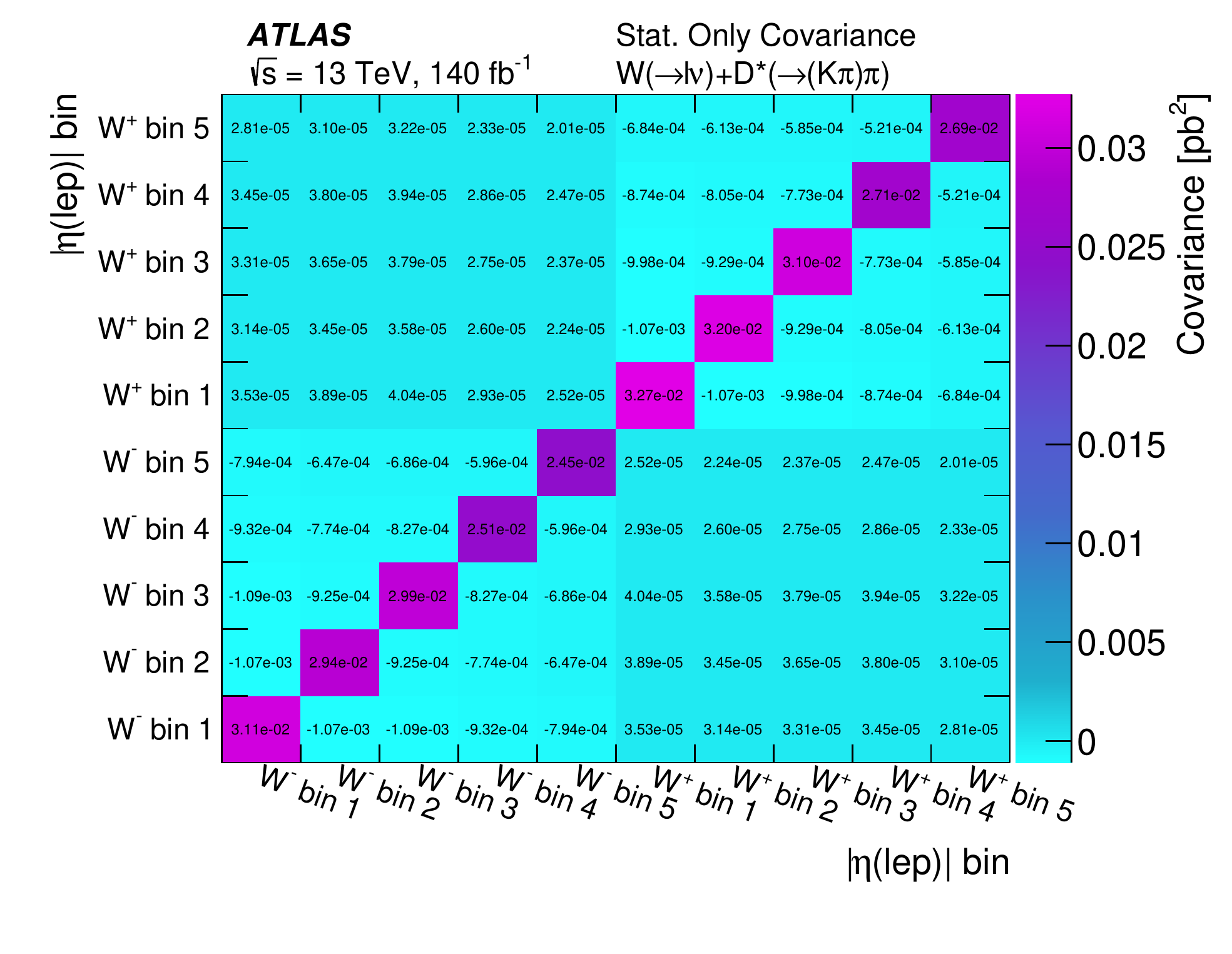}
\label{fig:results:covariance:Covariance_statOnly_dstar_eta}
}
\caption{
The data statistical uncertainty covariance matrix for the differential \WplusDmeson fits:
\protect\subref{fig:results:covariance:Covariance_statOnly_dplus_pt} \Dplus \diffpt fit,
\protect\subref{fig:results:covariance:Covariance_statOnly_dplus_eta} \Dplus \diffeta fit,
\protect\subref{fig:results:covariance:Covariance_statOnly_dstar_pt} \Dstar \diffpt fit, and
\protect\subref{fig:results:covariance:Covariance_statOnly_dstar_eta} \Dstar \diffeta fit.
}
\label{fig:results:covariance:stat}
\end{figure}

\begin{figure}[htbp]
\centering
\subfloat[]{
\includegraphics[width=0.50\textwidth]{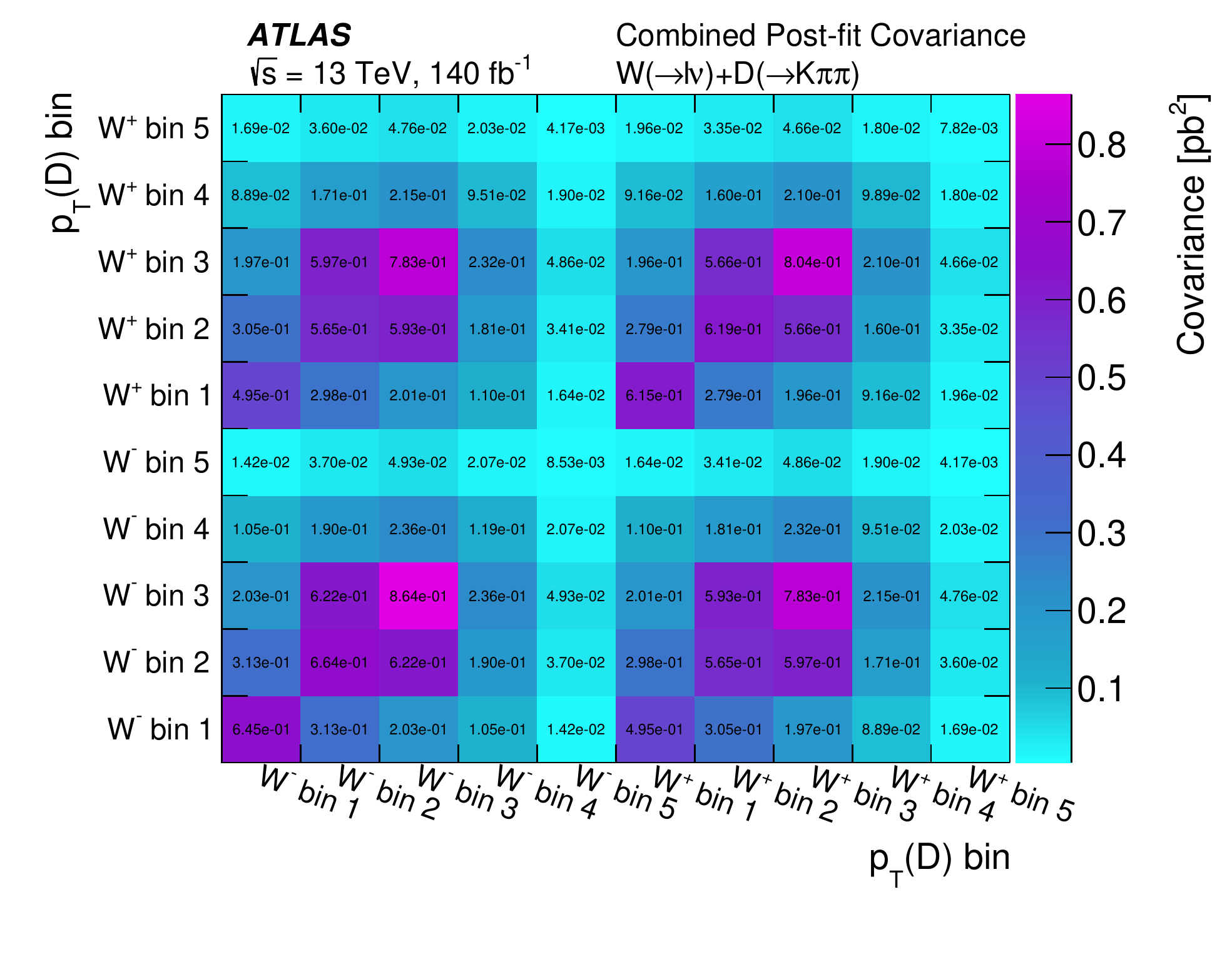}
\label{fig:results:covariance:Combined_Covariance_postfit_dplus_pt}
}
\subfloat[]{
\includegraphics[width=0.50\textwidth]{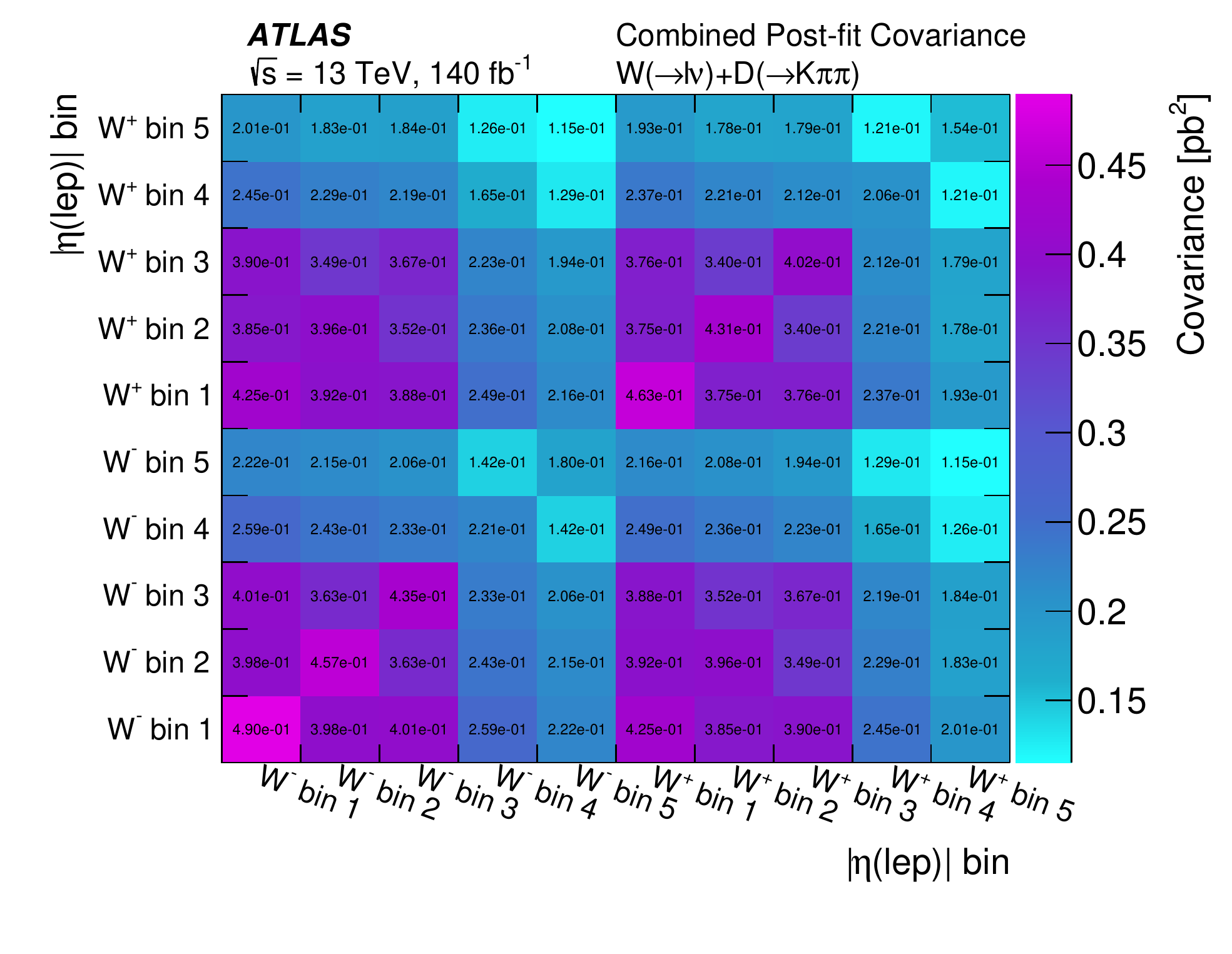}
\label{fig:results:covariance:Combined_Covariance_postfit_dplus_eta}
}
\\
\subfloat[]{
\includegraphics[width=0.50\textwidth]{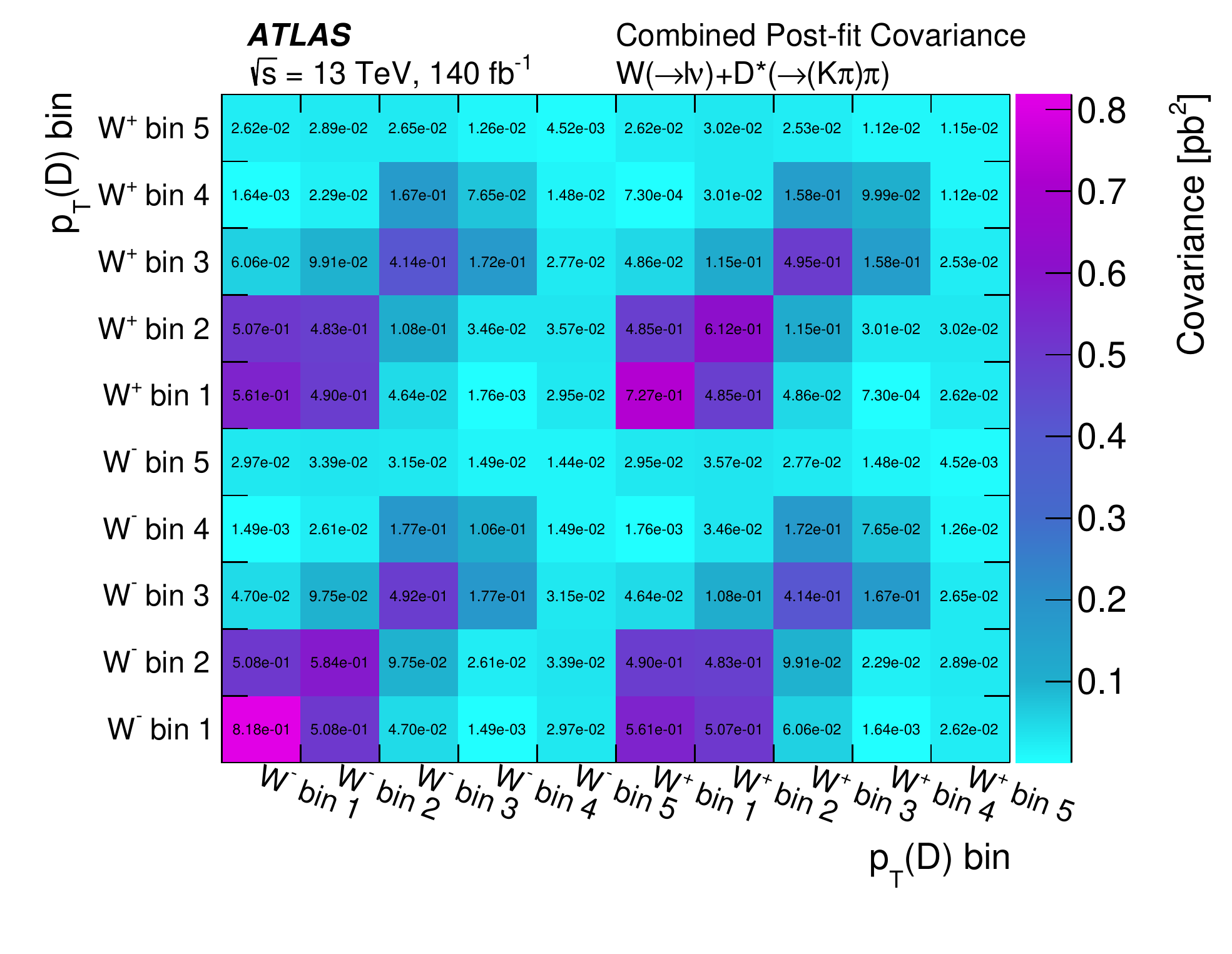}
\label{fig:results:covariance:Combined_Covariance_postfit_dstar_pt}
}
\subfloat[]{
\includegraphics[width=0.50\textwidth]{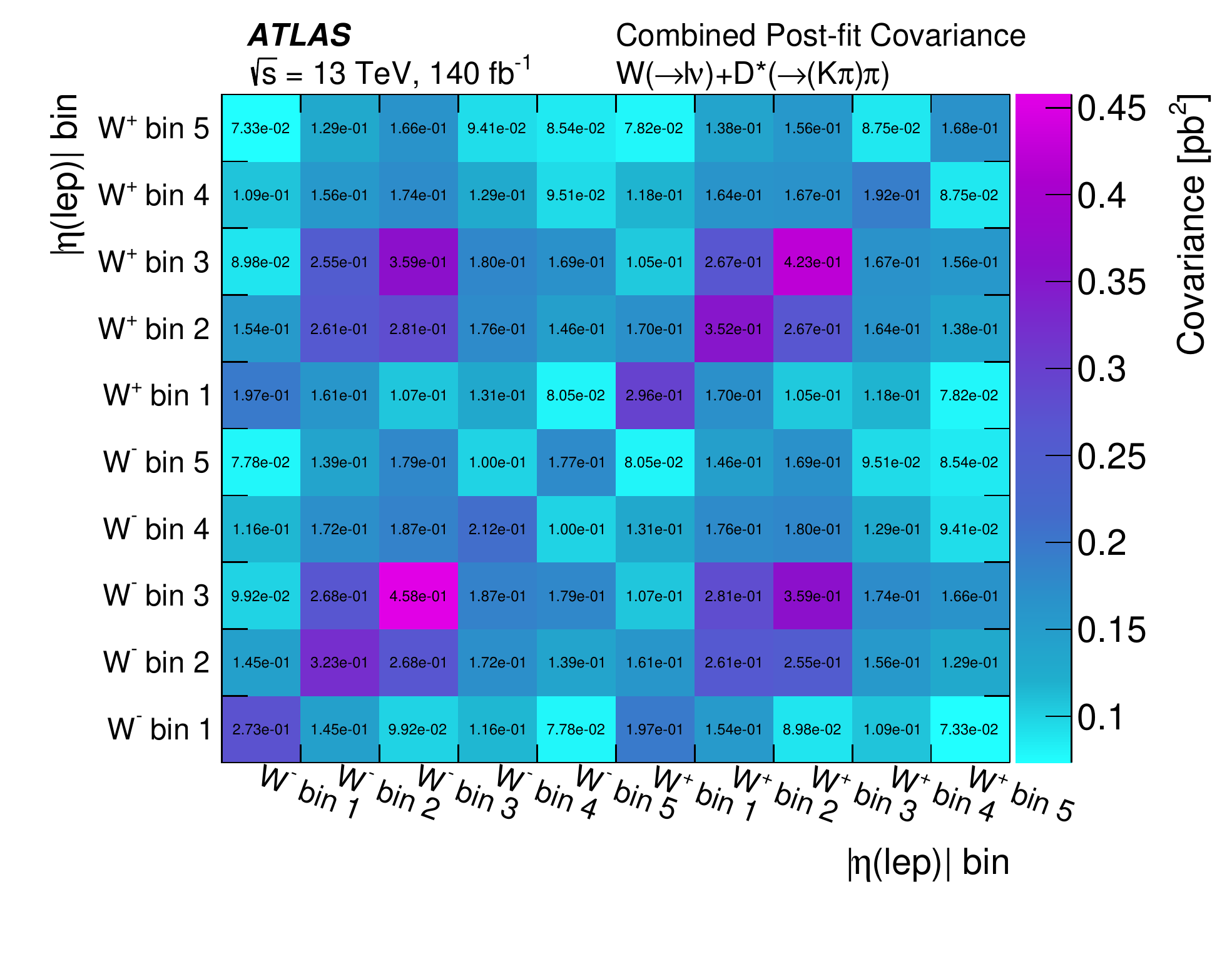}
\label{fig:results:covariance:Combined_Covariance_postfit_dstar_eta}
}
\caption{
The combined statistical and systematic uncertainty covariance matrix for the differential \WplusDmeson fits:
\protect\subref{fig:results:covariance:Combined_Covariance_postfit_dplus_pt} \Dplus \diffpt fit,
\protect\subref{fig:results:covariance:Combined_Covariance_postfit_dplus_eta} \Dplus \diffeta fit,
\protect\subref{fig:results:covariance:Combined_Covariance_postfit_dstar_pt} \Dstar \diffpt fit, and
\protect\subref{fig:results:covariance:Combined_Covariance_postfit_dstar_eta} \Dstar \diffeta fit.
The systematic uncertainties are evaluated with the post-fit values of the nuisance parameters,
corresponding to the measured differential cross-sections.
}
\label{fig:results:covariance:postfit}
\end{figure}

\begin{figure}[htbp]
\centering
\subfloat[]{
\includegraphics[width=0.50\textwidth]{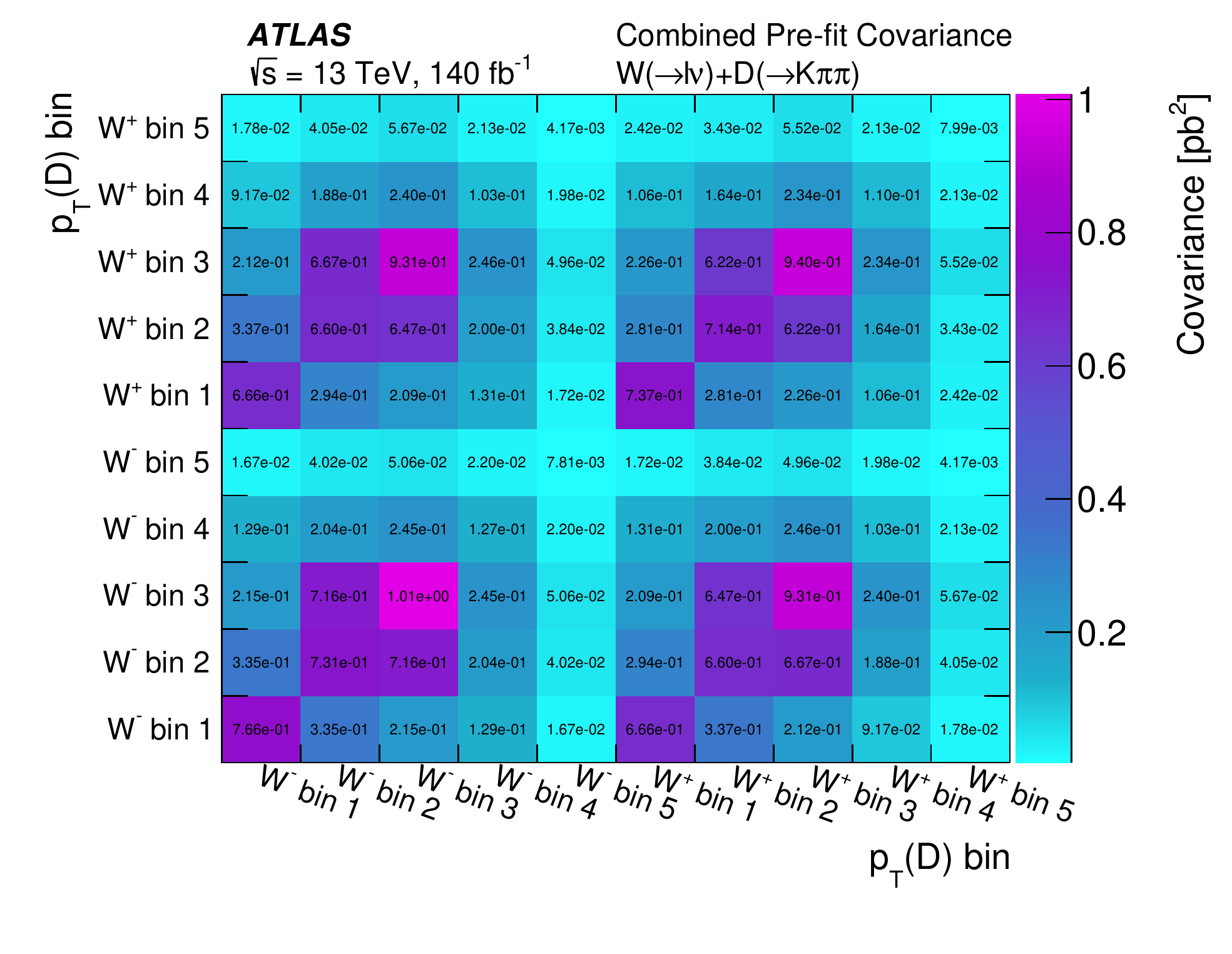}
\label{fig:results:covariance:Combined_Covariance_prefit_dplus_pt}
}
\subfloat[]{
\includegraphics[width=0.50\textwidth]{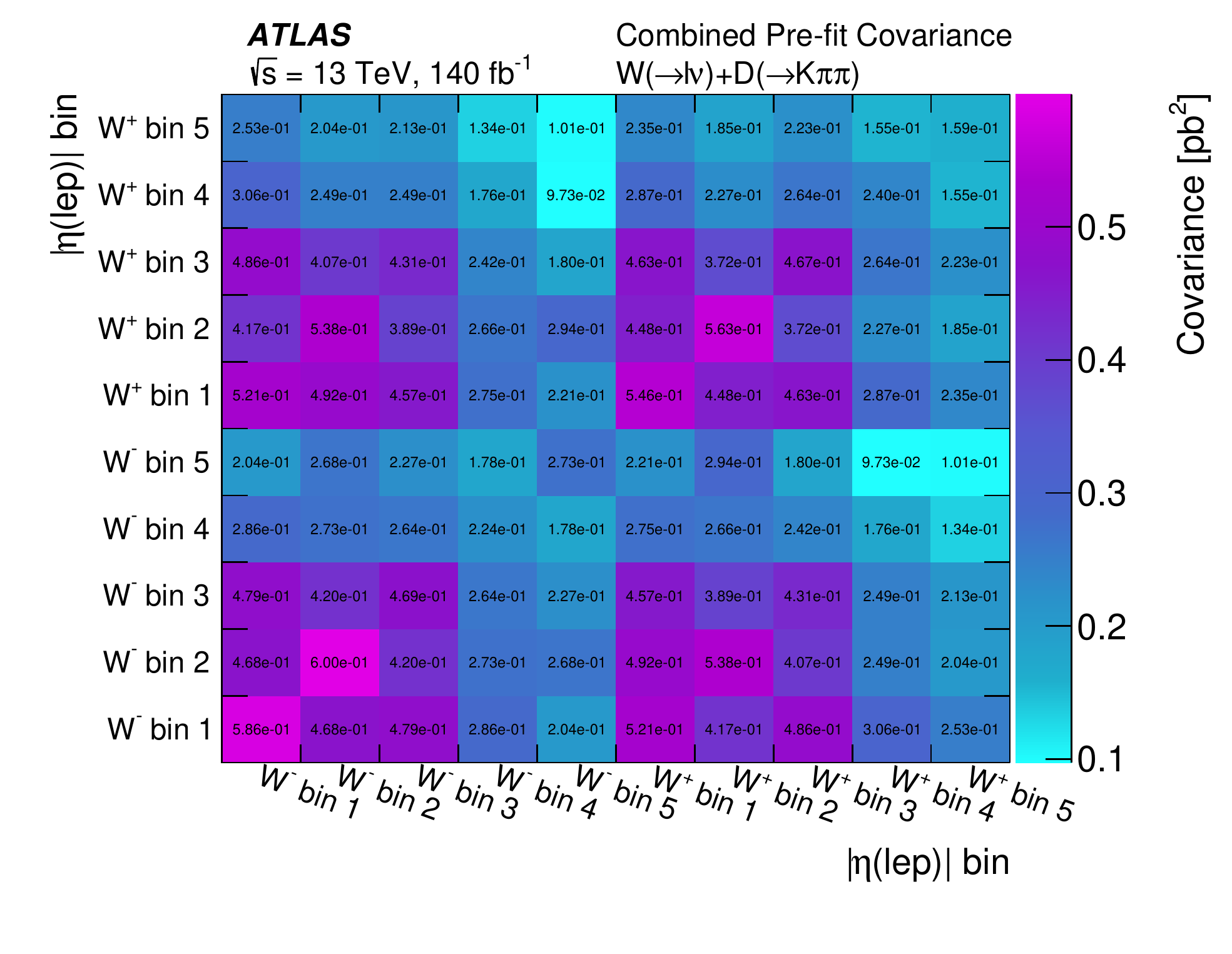}
\label{fig:results:covariance:Combined_Covariance_prefit_dplus_eta}
}
\\
\subfloat[]{
\includegraphics[width=0.50\textwidth]{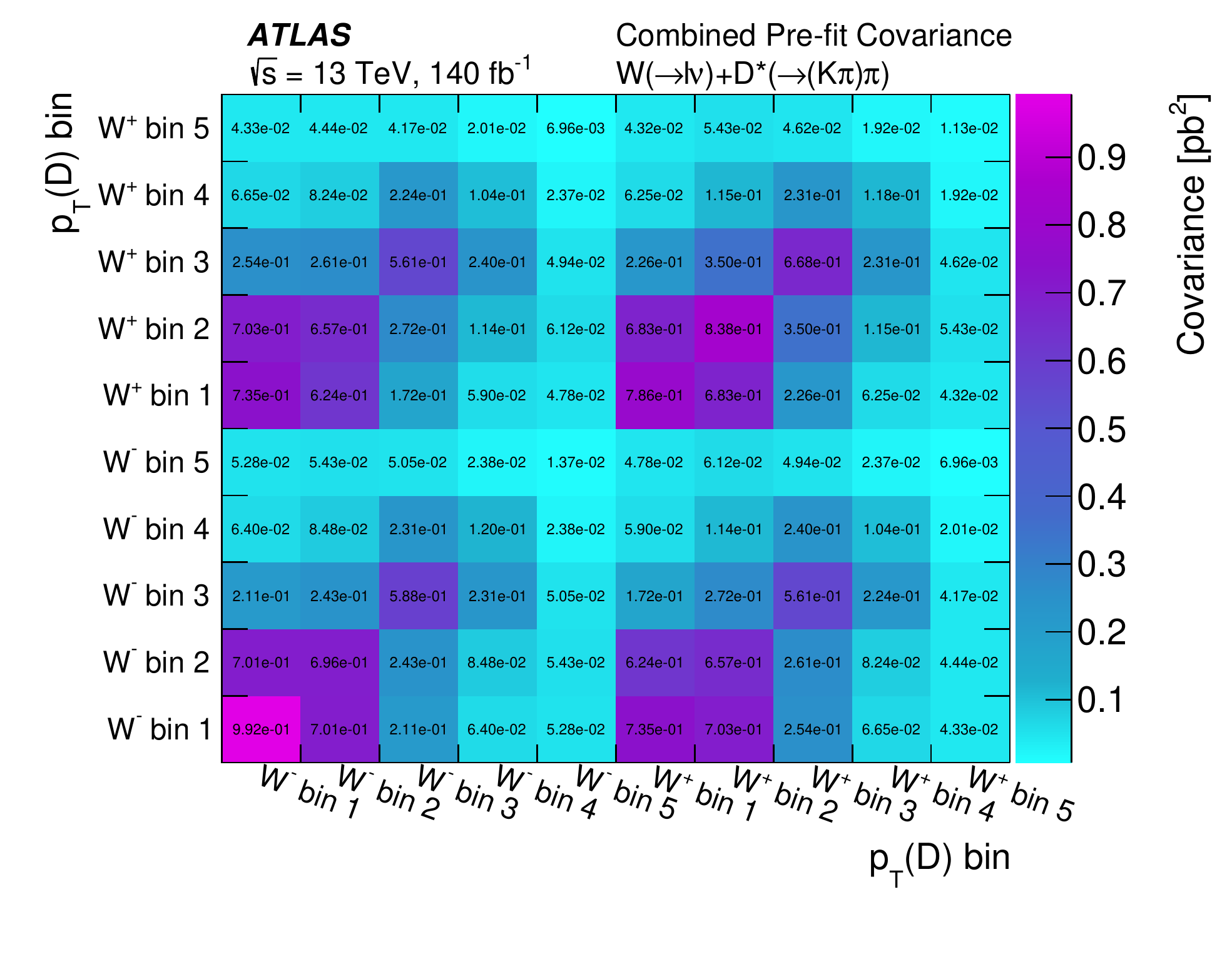}
\label{fig:results:covariance:Combined_Covariance_prefit_dstar_pt}
}
\subfloat[]{
\includegraphics[width=0.50\textwidth]{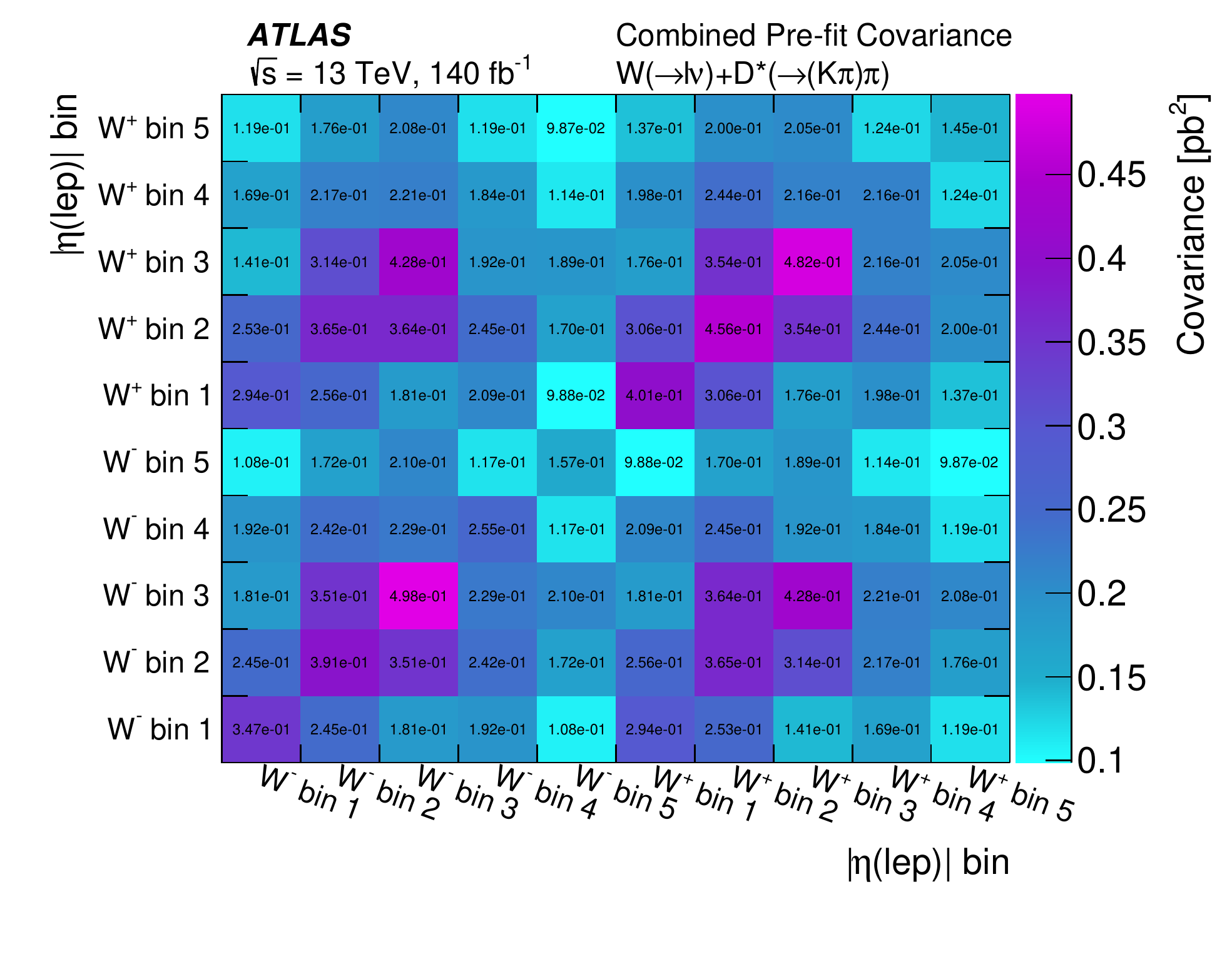}
\label{fig:results:covariance:Combined_Covariance_prefit_dstar_eta}
}
\caption{
The combined statistical and systematic uncertainty covariance matrix for the differential \WplusDmeson fits:
\protect\subref{fig:results:covariance:Combined_Covariance_prefit_dplus_pt} \Dplus \diffpt fit,
\protect\subref{fig:results:covariance:Combined_Covariance_prefit_dplus_eta} \Dplus \diffeta fit,
\protect\subref{fig:results:covariance:Combined_Covariance_prefit_dstar_pt} \Dstar \diffpt fit, and
\protect\subref{fig:results:covariance:Combined_Covariance_prefit_dstar_eta} \Dstar \diffeta fit.
The systematic uncertainties are evaluated with the pre-fit values of the nuisance parameters.
}
\label{fig:results:covariance:prefit}
\end{figure}

Covariance matrices encoding the measurement error associated with the differential
\WplusDmeson cross-section measurement are given in \Figrange{\ref{fig:results:covariance:stat}}{\ref{fig:results:covariance:prefit}}.
Covariance matrices are given separately for the \Dplus and \Dstar channels and separately
for \diffpt and \diffeta differential bins. Covariance matrices encoding only the statistical
uncertainty are given in \Fig{\ref{fig:results:covariance:stat}}. \Fig{\ref{fig:results:covariance:postfit}}
includes the full set of measurement uncertainties with post-fit values of the nuisance parameters
and \Fig{\ref{fig:results:covariance:prefit}} shows the covariance matrix with pre-fit values
of the nuisance parameters.
 
\FloatBarrier

\clearpage

\printbibliography

\clearpage
 
 
\begin{flushleft}
\hypersetup{urlcolor=black}
{\Large The ATLAS Collaboration}

\bigskip

\AtlasOrcid[0000-0002-6665-4934]{G.~Aad}$^\textrm{\scriptsize 102}$,
\AtlasOrcid[0000-0002-5888-2734]{B.~Abbott}$^\textrm{\scriptsize 120}$,
\AtlasOrcid[0000-0002-1002-1652]{K.~Abeling}$^\textrm{\scriptsize 55}$,
\AtlasOrcid[0000-0002-8496-9294]{S.H.~Abidi}$^\textrm{\scriptsize 29}$,
\AtlasOrcid[0000-0002-9987-2292]{A.~Aboulhorma}$^\textrm{\scriptsize 35e}$,
\AtlasOrcid[0000-0001-5329-6640]{H.~Abramowicz}$^\textrm{\scriptsize 151}$,
\AtlasOrcid[0000-0002-1599-2896]{H.~Abreu}$^\textrm{\scriptsize 150}$,
\AtlasOrcid[0000-0003-0403-3697]{Y.~Abulaiti}$^\textrm{\scriptsize 117}$,
\AtlasOrcid[0000-0003-0762-7204]{A.C.~Abusleme~Hoffman}$^\textrm{\scriptsize 137a}$,
\AtlasOrcid[0000-0002-8588-9157]{B.S.~Acharya}$^\textrm{\scriptsize 69a,69b,p}$
% ...
\bigskip
\\

% ...
$^{29}$Physics Department, Brookhaven National Laboratory, Upton NY; United States of America.\\
% ...
$^{35}$$^{(a)}$Facult\'e des Sciences Ain Chock, R\'eseau Universitaire de Physique des Hautes Energies - Universit\'e Hassan II, Casablanca;$^{(b)}$Facult\'{e} des Sciences, Universit\'{e} Ibn-Tofail, K\'{e}nitra;$^{(c)}$Facult\'e des Sciences Semlalia, Universit\'e Cadi Ayyad, LPHEA-Marrakech;$^{(d)}$LPMR, Facult\'e des Sciences, Universit\'e Mohamed Premier, Oujda;$^{(e)}$Facult\'e des sciences, Universit\'e Mohammed V, Rabat;$^{(f)}$Institute of Applied Physics, Mohammed VI Polytechnic University, Ben Guerir; Morocco.\\
% ...
$^{55}$II. Physikalisches Institut, Georg-August-Universit\"{a}t G\"ottingen, G\"ottingen; Germany.\\
% ...
$^{69}$$^{(a)}$INFN Gruppo Collegato di Udine, Sezione di Trieste, Udine;$^{(b)}$ICTP, Trieste;$^{(c)}$Dipartimento Politecnico di Ingegneria e Architettura, Universit\`a di Udine, Udine; Italy.\\
% ...
$^{102}$CPPM, Aix-Marseille Universit\'e, CNRS/IN2P3, Marseille; France.\\
% ...
$^{117}$Department of Physics, New York University, New York NY; United States of America.\\
% ...
$^{120}$Homer L. Dodge Department of Physics and Astronomy, University of Oklahoma, Norman OK; United States of America.\\
% ...
$^{137}$$^{(a)}$Departamento de F\'isica, Pontificia Universidad Cat\'olica de Chile, Santiago;$^{(b)}$Millennium Institute for Subatomic physics at high energy frontier (SAPHIR), Santiago;$^{(c)}$Instituto de Investigaci\'on Multidisciplinario en Ciencia y Tecnolog\'ia, y Departamento de F\'isica, Universidad de La Serena;$^{(d)}$Universidad Andres Bello, Department of Physics, Santiago;$^{(e)}$Instituto de Alta Investigaci\'on, Universidad de Tarapac\'a, Arica;$^{(f)}$Departamento de F\'isica, Universidad T\'ecnica Federico Santa Mar\'ia, Valpara\'iso; Chile.\\
% ...
$^{150}$Department of Physics, Technion, Israel Institute of Technology, Haifa; Israel.\\
$^{151}$Raymond and Beverly Sackler School of Physics and Astronomy, Tel Aviv University, Tel Aviv; Israel.\\
% ...
$^{a}$ Also Affiliated with an institute covered by a cooperation agreement with CERN.\\
$^{b}$ Also at An-Najah National University, Nablus; Palestine.\\
$^{c}$ Also at Borough of Manhattan Community College, City University of New York, New York NY; United States of America.\\
$^{d}$ Also at Bruno Kessler Foundation, Trento; Italy.\\
$^{e}$ Also at Center for High Energy Physics, Peking University; China.\\
$^{f}$ Also at Center for Interdisciplinary Research and Innovation (CIRI-AUTH), Thessaloniki ; Greece.\\
% ...
$^{p}$ Also at Department of Physics, King's College London, London; United Kingdom.\\
% ...

\end{flushleft}


\end{document}